\documentclass[onecolumn,authoryear]{els-mrw}
\usepackage{helvet}
\usepackage{amssymb,amsmath,graphicx,bm,mathrsfs,color,aas_macros,soul}
\usepackage{xcolor}
\makeatletter
\def\@authfoot{}
\makeatother
\graphicspath{ {./figures/}} 
\newcommand{\be}{\begin{equation}} 
\newcommand{\ee}{\end{equation}}
\newcommand{\bea}{\begin{eqnarray}}
\newcommand{\eea}{\end{eqnarray}}

\newcommand{\vecp}{{\bm p}}

\definecolor{red}{rgb}{0.8,0,0}
\definecolor{violet}{rgb}{0.4,0,0.4}
\definecolor{green}{rgb}{0,0.5,0.0}
\definecolor{navy}{rgb}{0.0,0.0,0.6}
\definecolor{orange}{rgb}{0.8,0.2,0.0}
\usepackage[normalem]{ulem}  % \sout{old text} for strikeout

\usepackage[breaklinks,colorlinks,colorlinks=true]{hyperref}
\hypersetup{
  linkcolor=Red,          % color of internal links
  citecolor=Violet,       % color of links to bibliography
  urlcolor=Blue}           % color of the url

\begin{document}

\chapter{Nuclear Physics of Binary Neutron Star Mergers}
\label{chap1}

\author[1,2]{Armen Sedrakian}
\address[1]{\orgname{Institute of Theoretical Physics}, \orgdiv{University of Wroc\l{}aw}, \orgaddress{50-204 Wroc\l{}aw, Poland}}
\address[2]{\orgname{Frankfurt Institute for Advanced Studies}, \orgaddress{60438 Frankfurt am Main, Germany}}
%\articletag{Chapter Article tagline: update of previous edition, reprint.}                                                

\maketitle

\begin{abstract}[Abstract]
Binary neutron star mergers provide a unique laboratory for studying matter under conditions that cannot be reproduced in terrestrial experiments. They probe dense matter at supranuclear density, finite temperature, rapid rotation, strong gravity, and extreme neutron excess, while producing observable signals in gravitational waves, electromagnetic radiation, and, in principle, neutrinos. This review focuses on the nuclear physics of binary neutron star mergers. We discuss the dense-matter equation of state (EoS), the inspiral and merger dynamics, the structure and lifetime of the post-merger remnant, transport and dissipative processes, weak interactions and neutrino transport, and the production of heavy elements through $r$-process nucleosynthesis. Particular emphasis is placed on the connection between microscopic physics and multimessenger observables, including tidal deformability, post-merger gravitational-wave spectra, kilonova light curves, short gamma-ray bursts, and afterglows. We also review how observations of events such as GW170817, together with neutron star mass and radius measurements, laboratory nuclear experiments, and theoretical many-body calculations, constrain the EoS and the composition of dense matter. The goal is to summarize the current understanding of how nuclear physics controls the dynamics and observable signatures of binary neutron star mergers, and to identify the open questions that future multimessenger observations and improved nuclear theory will address.
\end{abstract}

\begin{BoxTypeA}[chap:bns:keypoints]{Key points}
\begin{itemize}
    \item Binary neutron-star mergers probe dense matter at supranuclear densities, finite temperature, rapid rotation, strong gravity, and extreme neutron excess, providing conditions inaccessible in terrestrial experiments.

    \item The dense-matter equation of state determines neutron-star radii, maximum masses, tidal deformabilities, the threshold for prompt collapse, and the lifetime and stability of the post-merger remnant.

    \item Gravitational waves connect the equation of state to observations: tidal deformability affects the late-inspiral waveform, while post-merger oscillation frequencies and damping encode the finite-temperature, composition-dependent properties of dense matter.

    \item Multimessenger constraints from GW170817, massive pulsars, NICER radius measurements, laboratory nuclear data, and nuclear many-body theory already restrict the range of viable equations of state, but the high-density composition remains uncertain.

    \item Additional degrees of freedom, such as hyperons, $\Delta$ resonances, or deconfined quark matter, can soften the equation of state, modify the maximum mass and radius, and may leave observable signatures in post-merger gravitational-wave spectra.

    \item Transport coefficients, which include shear viscosity, bulk viscosity, thermal conductivity, electrical conductivity, and diffusion coefficients, encode the microscopic scattering and weak-interaction rates that control dissipative relativistic hydrodynamics and MHD.

    \item Bulk viscosity is especially important in the post-merger remnant because weak interactions restore chemical equilibrium during density oscillations, potentially damping gravitational-wave emission and affecting the remnant lifetime.

    \item Electrical and thermal conductivities regulate magnetic-field evolution, heat transport, and current dissipation; in strong magnetic fields these coefficients become tensorial and can generate Hall and thermoelectric effects.

    \item A self-consistent description of mergers requires the equation of state, composition, neutrino transport, and transport coefficients to be treated together, since each feeds back on the others and on the observable multimessenger signals.
\end{itemize}
\end{BoxTypeA}

%\begin{keyword}  Dense Matter\sep 
%  Neutron Stars\sep Binaries\sep Transport processes\sep Weak Interactions
% \PACS 21.65.+f   \sep  97.60.Jd.  \sep     97.80.-d    \sep    66.10.-x   \sep 13.15.+g
%\end{keyword}

%\tableofcontents

\section{Introduction}
\label{sec:Chapter1}

Binary neutron star (BNS) mergers represent one of the most extreme astrophysical transients in the universe. They involve an intimate interplay between nuclear physics, neutrino physics, relativistic hydrodynamics, general relativity and other related fields. These events serve as unique cosmic laboratories: within a region spanning tens of kilometers and over timescales ranging from milliseconds to tens of seconds, matter is compressed to densities five or more times greater than those found in atomic nuclei, while temperatures reach tens to hundreds of MeV. During these events the spacetime itself undergoes violent dynamical evolution.  The physics of these systems spans an enormous range of scales: from the femtometer-scale interactions of quarks and gluons, to the kilometer-scale structure of neutron stars, with multimessenger signals propagating over mega- to giga-parsec distances to the electromagnetic and gravitational wave (GW) detectors on Earth.  Understanding the physics of BNS mergers therefore requires a multi-scale analysis ranging from microphysics based on nuclear theory and experiment to astronomical observations of distant extragalactic objects.

This review focuses on the nuclear physics aspects of this problem. We discuss the microphysics of dense matter underlying neutron stars, as well as the hot, dense matter created during the merger process, including its dynamics, neutrino interactions, nuclear reactions, and related phenomena that ultimately determine the observable signatures. The review is intended both for researchers already working on BNS physics who wish to deepen their understanding of the nuclear physics aspects of the problem, and as an entry point for those wishing to enter the field. Our aim is to review the methods that are used, current open questions and the exciting opportunities that BNS mergers offer for addressing them; for related reviews covering various aspects of BNS physics, see~\cite{Faber2012LRR,Rosswog2015,Baiotti2017,Baiotti2019}.

BNS mergers are excellent multimessenger astrophysical events, in the sense that information is carried by multiple distinct messengers: specifically GWs, electromagnetic radiation across a broad spectrum, and neutrinos. These messengers probe different aspects of the merger and carry complementary and largely independent information. A central challenge -- but also an opportunity -- is therefore to develop theoretical and data-analysis frameworks capable of combining these signals into a coherent picture, thereby connecting astrophysical observations to the underlying microphysics.

{\it Gravitational waves} provide a direct probe of the bulk dynamics of the system. They trace all the phases from inspiral, merger, to post-merger evolution. Their amplitude scales inversely with distance, $h \sim D^{-1}$, while their phase evolution encodes the masses, spins, and internal structure of the merging objects. During the inspiral, tidal interactions — which are manifestations of finite-size effects — modify the waveform, thereby relating the EoS of dense matter to the observed signal through the {\it tidal deformability} $\Lambda$. In the post-merger phase, if a remnant neutron star survives, GWs carry information about its oscillation modes, which are sensitive to the finite-temperature, composition-dependent EoS at moderate to high densities. Extracting this information requires accurate theoretical templates based on both perturbative approaches and numerical simulations of the merger and post-merger phases, ultimately enabling constraints on dense matter.

{\it Electromagnetic radiation} from BNS mergers spans wavelengths from gamma rays to optical, infrared, and radio, and probes the physics of relativistic jets, disk formation, radiation transport in the ejecta, and the interaction with the surrounding medium. Of particular relevance for nuclear physics is kilonova emission, which arises from the radioactive decay of freshly synthesized heavy nuclei in the neutron-rich ejecta~\citep{Metzger2017}. The luminosity, energetics, and temporal evolution of a kilonova depend sensitively on the ejecta mass and composition, the opacity (absorption and scattering of radiation) of $r$-process elements, and the nuclear heating rates. These, in turn, require input from low-energy nuclear physics, in particular nuclear structure and reaction rates that can be studied in terrestrial laboratories. Short gamma-ray bursts (sGRBs), most likely powered by relativistic jets launched from the post-merger remnant, may also carry indirect information about the remnant structure and the properties of dense matter.

{\it Neutrinos} are emitted with luminosities that can reach $10^{53}$--$10^{54}$~erg~s$^{-1}$ during the hot post-merger phase, with the total emitted energy depending sensitively on the remnant lifetime.
These luminosities are comparable to those in core-collapse supernovae, although the physical setting is different: in BNS mergers the matter is already at nuclear density before contact, and the neutrino emission is powered by shock heating, compression, viscous/turbulent dissipation, and weak equilibration in the hot remnant. The key difference between these transients lies in their timescales: in supernovae the emission extends over $\sim 10$~s, whereas in BNS mergers it is concentrated over tens of milliseconds to seconds, depending on the remnant lifetime.
 Despite their weak coupling, the high temperatures reached in the remnant make neutrinos dynamically important, with their interaction rates set by weak processes in dense matter. Electron neutrinos and antineutrinos drive out-of-equilibrium $\beta$-processes, thereby controlling the electron fraction $Y_e$ and setting the composition of the outflows, with direct consequences for transport and dissipation. More generally, neutrino transport and diffusion in a strongly gravitating, dynamically evolving medium remain essential ingredients in merger simulations. Neutrinos from extragalactic BNS mergers are currently undetectable. However, their physics is manifest through their influence on the dynamics and nucleosynthesis. Future nearby events may nonetheless allow direct detection.

Decades of theoretical work on various aspects of BNS mergers were validated observationally by the detection of GW170817 in August 2017, the first multimessenger observation of a BNS merger~\citep{Abbott2017}. The coincident detection of sGRB\,170817A confirmed the association between BNS mergers and short gamma-ray bursts, while the kilonova AT2017gfo provided the first direct evidence for $r$-process nucleosynthesis in such events~\citep{Cowperthwaite2017,Pian2017}. These observations firmly established BNS mergers as multimessenger probes of fundamental physics opening the prospect of using these events to study dense matter under conditions unattainable elsewhere in the Universe. The impact of this event can hardly be overstated.

\begin{figure}[t]
\centering
\includegraphics[width=0.95\linewidth]{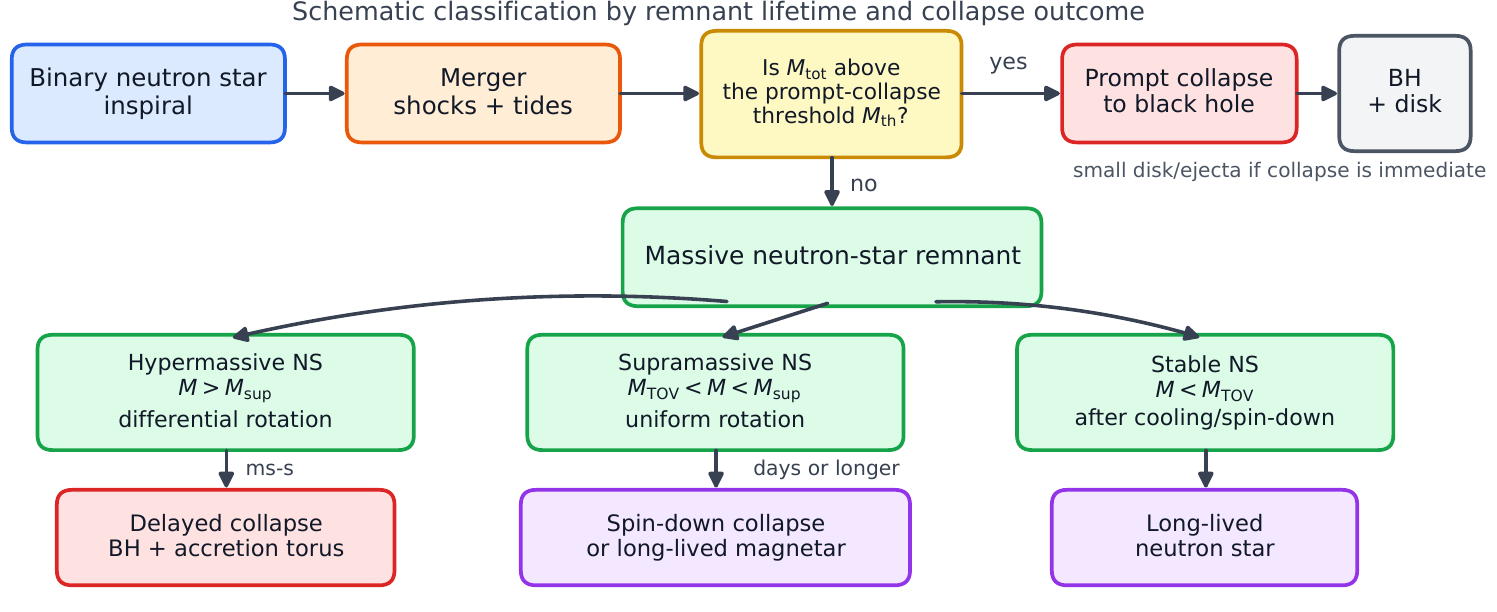}
\caption{Schematic classification of possible outcomes of a binary
neutron-star merger. If the total mass exceeds the prompt-collapse
threshold $M_{\rm th}$, the merger forms a black hole promptly, typically
with a smaller disk and ejecta mass. Otherwise, a massive neutron-star
remnant forms. Depending on its mass relative to the nonrotating maximum
mass $M_{\rm TOV}$ and the maximum mass supportable by uniform rotation
$M_{\rm sup}$, the remnant may be hypermassive, supramassive, or stable.
These outcomes are associated with different collapse timescales, ejecta
channels, post-merger gravitational-wave emission, neutrino emission,
kilonovae, and possible short gamma-ray bursts.}
\label{fig:fig1}
\end{figure}

The qualitative outcome of a BNS merger is primarily controlled by several factors: the total mass of the system, the angular-momentum distribution, thermal support, magnetic-field amplification, and the high-density EoS. If the total mass exceeds the prompt-collapse threshold, the merger remnant collapses to a black hole on a dynamical timescale. For lower masses, the system forms a massive neutron-star remnant whose subsequent evolution may involve delayed collapse or long-term survival, depending on whether differential rotation, uniform rotation, and thermal pressure are sufficient to support it, see Fig.~\ref{fig:fig1} for illustration.

Let us now focus briefly, for the purposes of this review, on several regimes where nuclear physics plays an important role in BNS mergers. The aim is to provide a concise overview of topics that will be discussed in detail in later sections. The importance of nuclear physics stems from the fact that the dynamics of BNS mergers are ultimately governed by the properties of dense matter, as described by quantum chromodynamics (QCD) and nuclear many-body theory under extreme conditions

{\it EoS of dense matter} is a key input in numerical simulations of BNS mergers and a central open question in nuclear physics~\citep{Lattimer2001,Oertel2017,Tews2018,Sedrakian2023}. At densities exceeding nuclear saturation density $n_0 \simeq 0.16$ fm$^{-3}$, matter may consist of nucleons interacting via the strong force, hyperons (strangeness-carrying heavy baryons), and nucleonic resonances. At the highest densities, deconfinement into quarks may give rise to phases characteristic of condensed matter systems, such as color-superconducting matter~\citep{Alford2008}.  The EoS, which specifies the pressure as a function of density, temperature (or entropy), and composition, determines integral properties of neutron stars, such as the maximum mass, radius, moment of inertia and tidal deformability. It also governs the dynamical stability of the post-merger remnant, with additional contributions to the pressure arising from finite-temperature and a trapped neutrino component. At suprasaturation densities, experimental constraints are extremely limited, making this regime weakly constrained and particularly relevant for BNS merger exploration. By contrast, at sub-saturation densities the EoS is better constrained by nuclear experiments, including measurements of nuclear masses and radii and heavy-ion collisions, although uncertainties remain large at the large isospin asymmetries relevant to BNS mergers. The symmetry energy and its density dependence, which control the behavior of neutron-rich matter, are being constrained by experiments such as PREX and CREX, but significant uncertainties remain in the regimes relevant for mergers~\citep{Lattimer2023}.

{\it Transport properties and dissipation are another important aspect of the nuclear physics of BNS mergers.} The dynamical evolution of the remnant, described by relativistic magneto-hydrodynamics (MHD), depends not only on the EoS but also on transport coefficients when dissipation is included. Shear and bulk viscosity, thermal conductivity, and diffusion coefficients cause deviations from ideal hydrodynamics. Therefore, information
beyond the EoS and gravity are need to treat realistically the evolution of matter in the remnant. For example, shear viscosity influences the damping of oscillations and the development of instabilities such as the Kelvin--Helmholtz instability during merger. Bulk viscosity, arising from out-of-equilibrium weak interactions, may be particularly important in the post-merger phase, affecting both GW emission and angular momentum transport. In the presence of electromagnetic fields, electrical conductivity controls current dissipation and the evolution of magnetic fields. Calculations of these coefficients from the underlying microphysics remain in their early stages and will continue to be an important task in the coming years.

\textit{Weak interactions} mediated by $W$ and $Z$ bosons play a crucial role in the hot, dense environment of the BNS merger remnant. Acting on dynamical timescales, they drive matter toward or away from beta equilibrium. Electron captures, beta decays, and neutrino emission control bulk viscosity, neutrino opacity, and energy transport. Through these processes, weak interactions also determine the electron fraction $Y_e$ of the ejecta, which sets the conditions for $r$-process nucleosynthesis. Accurate modeling of weak interaction rates -- especially the effects of many-body correlations -- is therefore an essential ingredient for controlled account of the dissipative dynamics that directly affect the observable signatures.

\textit{Nuclear reactions and nucleosynthesis} occur in the neutron-rich ejecta expelled during the merger. As the ejecta expands and cools, nuclear statistical equilibrium breaks down, and a network of reactions, which include neutron captures, beta decays, and nuclear fission, determines the final abundance pattern. The $r$-process requires a sufficiently large neutron-to-seed ratio to produce very neutron-rich nuclei before beta decay intervenes~\citep{Kajino2019}. The key inputs are nuclear masses, decay rates, capture cross sections, and fission yields. While these are well constrained near stability, they become uncertain for the neutron-rich nuclei relevant here, requiring extrapolation using well-tuned nuclear density functionals, among other theoretical tools. These uncertainties propagate directly into predictions of kilonova light curves and spectra.

Nuclear interactions also determine the radioactive energy release that powers the kilonova. The heating rate depends on the full $\beta$-decay and fission network and is sensitive to nuclear masses and fission properties. Connecting observations to theory therefore provides a direct probe of nuclear physics far from the valley of stability. Laboratory measurements, particularly at facilities such as FRIB and RIKEN, are essential for constraining these inputs. In this sense, BNS mergers (along with the structure of cold neutron stars)
form a bridge between astrophysics and nuclear physics.

\subsection{Key Scientific Questions}

We now briefly list the key questions that motivate the field of nuclear physics of BNS mergers that are addressed in this review.

\begin{itemize}

\item \textit{What is the EoS of dense matter?}
The pressure--density relation above $n_0$ remains a central open problem. Observations of neutron stars, including mass measurements, radius and moment of inertia constraints, and tidal deformability from GWs, provide increasingly strong constraints. Connecting these observations to underlying nuclear interactions and possible phase structure of dense matter is a primary goal.

\item \textit{What is the composition of dense matter?}
Whether neutron stars contain only nucleons or also hyperons, $\Delta$ baryons, or deconfined quarks remains uncertain. The presence of additional degrees of freedom affects the EoS and the maximum mass, and may leave observable signatures in mergers.

\item \textit{What is the maximum neutron star mass?} The maximum mass $M_\mathrm{TOV}$
   of a spherically symmetric (non-rotating, unmagnetized) neutron star determines the fate of merger remnants and has direct implications for GW and electromagnetic signals.
  
 \item\textit{What is the contribution of BNS mergers to heavy-element production?} While GW170817 provides strong evidence for $r$-process nucleosynthesis in BNS mergers, it remains unclear whether mergers alone account for the observed cosmic abundances of heavy elements, or whether additional sources -- such as collapsars or core-collapse supernovae -- are required.

\item \textit{How do transport properties determine observables?} Understanding how transport coefficients, weak interactions, and electromagnetic effects influence observable signals requires connecting microphysics to macroscopic dynamics through the dissipative MHD framework applied to BNS merger simulations.

\end{itemize}

Addressing these questions requires a combination of theoretical modeling, numerical simulations, nuclear physics input, and multimessenger observations. The following sections review the current state of understanding, with emphasis on nuclear physics, and outline the fiuture directions.

%\newpage

\section{Dense Matter EoS}
\label{sec:Chapter2}
The EoS of dense matter provides the relation between thermodynamic quantities such as the pressure $P$, energy density $\epsilon$, temperature $T$ (or entropy $S$) in thermodynamic equilibrium. In microscopic or physics-based models, it also determines the composition of matter, usually expressed in terms of particle fractions $Y_i \equiv n_i/n_B$, where $n_i$ is the number density of species $i$ and $n_B$ is the total baryon density. The EoS is therefore a central input for modeling the structure, dynamics, and observable signatures of neutron stars and, in particular, BNS mergers. In practice, it encodes the nuclear and particle physics relevant to macroscopic stellar properties into a relation of the form $P(\epsilon,T,Y_i)$, which determines the mass--radius relation, the stability of the merger remnant, the composition of the ejecta, and, indirectly, the GW and electromagnetic signals. Constraining this relation over the wide range of densities, temperatures, and compositions encountered in a merger is one of the central goals of the nuclear physics of BNS mergers.

\subsection{Constraints from Astrophysics and Laboratory Experiments}

Constraining the EoS of dense matter requires combining information from complementary sources, since the densities relevant to mature long-lived neutron stars, as well as transients such
as BNS mergers, and core-collapse supernovae span a wide range in density and temperature.
The modern approach therefore combines the astrophysical observations, nuclear theory, and laboratory experiments, typically within Bayesian inference frameworks which
incorporate systematic propagation of uncertainties.

\paragraph{Neutron star mass measurements}
Precise mass measurements of radio pulsars in binary systems, obtained through Shapiro delay or relativistic orbital effects, provide the most direct lower bound on the maximum mass $M_{\rm TOV}$. The measurements of massive pulsars such as PSR~J1614-2230, PSR~J0740+6620, and PSR~J0952-0607 require any viable EoS to support neutron stars with masses at least around $2\,M_\odot$~\citep{Freire2024}. This translates into a lower limit on the pressure at densities of several times $n_0$ and rules out EoS models that soften too strongly at high density~\citep{Oertel2017,Tews2018,Sedrakian2023}.

\paragraph{Radius and compactness measurements}
The NICER mission constrains neutron star radii through pulse-profile modeling of X-ray emission from rotating hot spots. Measurements of PSR~J0030+0451 and PSR~J0740+6620 favor radii of order $R\simeq 11$--$13\,\mathrm{km}$ for canonical-mass neutron stars. These measurements primarily constrain the pressure at intermediate densities, roughly $1$--$2\,n_0$. Although they are subject to systematic uncertainties associated with the surface emission model and geometry, independent analyses have broadly converged toward a consistent range~\citep{Miller2019,Miller2021,Riley2019,Riley2021}.

\paragraph{Gravitational-wave tidal constraints}
The event GW170817 provided the first direct constraint on the tidal deformability of a BNS system. Its inferred value disfavors very stiff EoS with large radii, while the existence of massive pulsars disfavors EoS that are too soft. Together, these observations point toward an EoS that is moderately soft at intermediate densities but sufficiently stiff at higher densities to support massive neutron stars. 

Qunatitatively, the (dimensionless) tidal deformability of a compact binary, is defined as ~\citep{Hinderer2008,Flanagan2008}
\bea
\label{eq:Lambda}
\Lambda = \frac{2}{3}k_2 C^{-5}, \qquad C=\frac{GM}{Rc^2},
\eea
where $k_2$ is the quadrupolar tidal Love number, $M$ and $R$ are the stellar gravitational mass and circumferential radius,  $C$ is the compactness, and $c$ is the speed of light. This quantity describes how the gravitational multipole moments of the star respond to the tidal field of its companion. During the late inspiral, tidal effects accelerate the phase evolution relative to that of two point masses, producing a contribution to the gravitational-wave phase governed by a mass-weighted combination $\tilde{\Lambda}$ formed from individual deformabilities and masses of
both components according to 
%-------------------------------------------
\begin{equation}
 \label{eq:Lambda_tilde}
\tilde{\Lambda} = \frac{16}{13} \frac{(M_1 + 12M_2) M_1^4\Lambda_1 +
(M_2 + 12M_1)M_2^4\Lambda_2}{(M_1 + M_2)^5}.
\end{equation}
%-------------------------------------------
For GW170817, the total binary mass was determined to be $2.73^{+0.04}_{-0.01}\,M_{\odot}$, with component masses in the range $1.16$--$1.60\,M_{\odot}$, yielding $\tilde{\Lambda} = 300^{+420}_{-230}$ at $90\%$ confidence~\citep{Abbott2019}; for GW190425, the total mass of $3.4^{+0.3}_{-0.1}\,M_{\odot}$ and component masses in the range $1.46$--$1.87\,M_{\odot}$ led to an upper limit of $\tilde{\Lambda} \leq 600$~\citep{LVC_GW190814}. Gravitational-wave data alone cannot exclude the possibility that one or both components of this binary were black holes.

\paragraph{Laboratory nuclear physics}
Laboratory experiments provide complementary information near and below saturation density. Nuclear masses, charge radii, dipole polarizabilities, and giant resonances constrain the nuclear energy functional, while heavy-ion collisions probe the EoS through collective flow and particle production. Measurements of neutron skin thicknesses, such as PREX and CREX, constrain the symmetry energy and its slope, although some tension remains between different experimental and theoretical extractions~\citep{Lattimer2023}. Future measurements at radioactive beam facilities, including FRIB and RIKEN-RIBF, will extend nuclear data toward the neutron-rich region relevant for both neutron star matter and $r$-process nucleosynthesis.

The combination of multimessenger astrophysical observations and laboratory nuclear physics--particularly with improved gravitational-wave detections from next-generation detectors, potential observations of post-merger signals, and new data from rare-isotope facilities--is expected to significantly sharpen our understanding of the dense matter EoS in the coming decade.

\subsection{Nuclear Matter at Sub-Saturation Density}

At densities below nuclear saturation (number) density, $n_0$, corresponding to a mass density of roughly $2.7\times 10^{14}\,\mathrm{g\,cm}^{-3}$, the EoS is relatively well constrained by nuclear physics. The binding energies of finite nuclei, nuclear saturation properties, charge radii, and scattering (nucleon-nucleon phase-shifts) and bound state data together constrain several key parameters of the nuclear energy functional in this regime.

The most systematic theoretical framework for describing nuclear matter at sub-saturation densities is chiral effective field theory (chiral EFT)~\citep{Epelbaum2009,Machleidt2011,Drischler2021}. It provides a controlled expansion of nuclear forces in powers of momenta and the pion mass, consistent with the symmetries of QCD and the pattern of spontaneous chiral symmetry breaking. Nuclear interactions are organized hierarchically, starting from two-nucleon forces and supplemented by higher-order corrections, including three-nucleon interactions that appear at next-to-next-to-leading order and beyond. The low-energy constants of chiral EFT encode short-distance physics and are fixed by experimental data, most importantly nucleon--nucleon scattering phase shifts and few-body observables. Once these constants are determined, the theory allows predictive calculations with systematic uncertainty estimates. Truncation errors at a given chiral order can then be propagated to many-body observables and ultimately to the neutron star EoS.

Three-nucleon forces are especially important in neutron-rich matter. They generate additional repulsion with increasing density and are essential for reproducing the empirical saturation properties of nuclear matter. As a result, chiral EFT provides robust predictions for pure neutron matter and symmetric nuclear matter up to densities of order $n_0$, and in some cases moderately beyond. The expansion is expected to become unreliable when the Fermi momentum approaches the breakdown scale, $\Lambda_b\sim 500\,\mathrm{MeV}$, corresponding roughly to densities of order $1$--$2\,n_0$. At higher densities, one must rely on phenomenological potential models in combination with non-perturbative many-body approaches~\citep{Lattimer2001,Sedrakian2007PrPNP}.

Thus, the chiral EFT results are directly relevant for the neutron star crust and outer core. It should be kept in mind, however, that most chiral EFT calculations describe homogeneous neutron or nucleonic matter, whereas the actual crust is highly inhomogeneous and contains nuclear clusters embedded in a neutron gas. Connecting uniform-matter calculations to crust physics therefore requires additional modeling. Furthermore, these results can be used to  constrain phenomenological density functional models, both non-relativistic (Skyrme and Gogny type) or relativistic.

The neutron star crust itself exhibits a rich sequence of phases arising from the competition between the nuclear force and the long-range Coulomb interaction~\citep{Chamel2008}. At low densities, the outer crust consists of a Coulomb lattice of neutron-rich nuclei immersed in a degenerate electron gas. When the neutron chemical potential reaches the neutron rest mass, neutrons begin to drip out of nuclei, at a density $\rho\simeq 4\times 10^{11}\,\mathrm{g\,cm}^{-3}$. In the inner crust, this dripped neutron gas coexists with the nuclear lattice and is expected to become superfluid, with implications for pulsar glitches and thermal evolution~\citep{Page2006}. Near the crust--core transition, at densities $n\sim 0.3$--$0.8\,n_0$, spherical nuclei become unstable because of the competition between surface and Coulomb energies. This leads to the formation of nuclear pasta phases, such as rod-like, slab-like, and more complex geometries. These structures have distinct material properties
that affect neutrino transport, electrical conductivity, etc.~\citep{Horowitz2004,Horowitz2008}.

\subsection{Nuclear Matter at Near-Saturation Density}

The EoS of nuclear matter in the vicinity of saturation density and isospin symmetry can be parameterized by a double expansion in density and isospin asymmetry. Introducing $\chi=(n-n_{0})/3n_{0}$ and $\delta=(n_n-n_p)/n$, where $n_{n/p}$ are the densities of neutrons ($n$) and protons ($p$), one may write
\begin{eqnarray}
  \label{eq:Expansion}
E(\chi,\delta) \simeq E_{\rm sat} + \frac{1}{2}K_{\rm sat}\chi^2 + \frac{1}{6}Q_{\rm sat}\chi^3
+ \left(J_{\rm sym} + L_{\rm sym}\chi + \frac{1}{2}K_{\rm sym}\chi^2 + \frac{1}{6}Q_{\rm sym}\chi^3 \right)\delta^2 ,
\end{eqnarray}
where terms of fourth and higher order in the small parameters have been omitted.
A useful thermodynamic identity relates the pressure to the energy per particle $E$,
\begin{equation}
P = n^2 \frac{\partial E}{\partial n},
\end{equation}
which makes explicit that the zero-temperature pressure is governed by the density dependence of the microscopic energy functional.

The coefficients in this expansion encode the bulk properties of nuclear matter at saturation. In the isoscalar sector, $E_{\rm sat} = E(n_0)$ is the binding energy per nucleon, $K_{\rm sat}$ the incompressibility, and $Q_{\rm sat}$ the skewness parameter. In the isovector sector, $J_{\rm sym}$ defines the symmetry energy at saturation, while $L_{\rm sym}$, $K_{\rm sym}$, and $Q_{\rm sym}$ describe its density dependence. At low order, several of these quantities are reasonably well constrained by nuclear data: $E_{\rm sat}\simeq -16$ MeV, $J_{\rm sym}\simeq 32$ MeV, and $n_{0}\simeq 0.15$--$0.16\,\mathrm{fm}^{-3}$. The incompressibility is typically inferred to lie in the range $K_{\rm sat}\sim 200$--$300$ MeV from giant-resonance measurements. By contrast, the higher-order coefficients, especially $Q_{\rm sat}$ and $K_{\rm sym}$, remain poorly constrained and become a major source of uncertainty when the EoS is extrapolated to suprasaturation densities~\citep{Margueron2018,Sedrakian2023}.
The range of validity of this expansion is limited to densities not too far from saturation, typically up to a few times $n_0$, with the precise range determined by the convergence properties of the series. Within this domain, however, the expansion provides a transparent way to identify how individual nuclear matter properties influence the EoS and how their uncertainties propagate to neutron star observables.

The stiffness of the EoS is conveniently characterized by the speed of sound,
\bea
c_s^2 = \frac{dP}{d\epsilon},
\eea
which measures how efficiently pressure responds to changes in energy density. A stiffer EoS, corresponding to a larger $c_s^2$, supports more massive and generally larger neutron stars. Causality requires $c_s\leq c$, while a non-interacting gas of massless particles -- relevant to the asymptotically free quarks at very high (energy) densities -- gives $c_s^2=1/3$, known as the conformal value. Current theoretical and observational evidence suggests that $c_s^2$  exceeds $1/3$ in hadronic matter at densities relevant to neutron stars and BNS mergers, reflecting strong repulsive interactions and nontrivial phase structure of matter at such densities.
In addition to microscopic models, modern studies often employ phenomenological parameterizations of the EoS directly in terms of $c_s^2(\epsilon)$ or $c_s^2(n)$~\citep{Alford2013,Zdunik2013}. In such constructions, the EoS is obtained by specifying the density dependence of the speed of sound and integrating the relevant thermodynamic relations. This provides a flexible and relatively model-independent way to explore rapid stiffening, phase transitions, or non-monotonic behavior of $c_s^2$ in dense matter. It is particularly useful in Bayesian inference, where neutron star masses, radii, and GW data are used to reconstruct the EoS with minimal theoretical bias. Speed-of-sound parameterizations complement the widely used piecewise-polytropic parameterizations, in which the EoS is approximated by segments $P\propto \rho^\Gamma$, where $\Gamma$ is the adiabatic index~\citep{Read2009}. While polytropes are simple and computationally efficient and have been used from the inception of compact stars physics, notably for white dwarfs, more recent speed-of-sound parameterizations offer a physically more transparent measure of the stiffness of matter.

\subsection{Supra-Nuclear Density Regime}

At densities above 1.5--2\,$n_0$, theoretical uncertainties increase rapidly. The chiral expansion breaks down, many-body perturbation theory becomes less reliable, and non-perturbative aspects of the strong interaction become increasingly important. In the density regime where nucleon--nucleon scattering data still provide useful constraints, microscopic many-body approaches can be used to extend the nuclear interaction into the medium. Self-consistent Green's function theory~\citep{Rios2009,Dickhoff2008} and Brueckner theory~\citep{Muether2017,Bombaci2018}, for example, resum in-medium correlations using renormalized interactions, often formulated in terms of in-medium $T$ matrices~\citep{Sedrakian2007PrPNP}. These approaches go beyond simple Hartree or Hartree--Fock descriptions by including short-range correlations and medium modifications. Three-nucleon forces can be incorporated approximately by averaging over the phase space of the spectator nucleon, thereby generating density-dependent effective two-body interactions.

Quantum Monte Carlo methods, including auxiliary-field diffusion Monte Carlo approaches provide complementary non-perturbative treatments of dense matter~\citep{Carlson2015,Lynn2019}. Their advantage is that they are not restricted to specific diagrammatic resummations, although this comes at the cost of substantial computational complexity. Their empirical input still ultimately relies on nucleon--nucleon phase shifts and few-body observables, which determine the underlying Hamiltonian. These methods therefore provide valuable benchmarks in the density range where they overlap with perturbative and effective approaches. Nuclear lattice simulations offer another complementary ab initio framework, in which nucleons interact via chiral EFT on a discretized spacetime lattice~\citep{Lee2025}. Although currently limited by lattice artifacts and computational cost, such simulations form an important bridge between effective field theory and many-body methods, especially at low densities.

For practical neutron star modeling, covariant density functionals (CDFs) provide a flexible and thermodynamically consistent framework for constructing the EoS~\citep{Oertel2017,Sedrakian2023}. In these models, nucleons are treated as Dirac quasiparticles interacting through effective meson fields, closely related to relativistic mean-field theory. A particularly useful class consists of density-dependent CDFs, in which the effective couplings are promoted to functions of density~\citep{Typel2018}. This allows microscopic input, especially from Dirac--Brueckner--Hartree--Fock calculations, to be incorporated into the functional form of the interaction. In this way, information encoded in nucleon--nucleon phase shifts can be propagated to the EoS through the density dependence of the couplings. Non-relativistic density-functional approaches based on Skyrme or Gogny interactions provide an alternative realization of the same idea, with parameters calibrated to nuclear data and, in some cases, guided by microscopic calculations~\citep{Pearson2018,Colo2020}.

Despite these developments, a fundamental limitation remains: all such models are calibrated near saturation density and must be extrapolated to densities where empirical constraints are sparse. This extrapolation leads to substantial model dependence in the high-density stiffness of the EoS, and therefore in predictions of radii and maximum masses. Observations of neutron stars with masses around $2\,M_\odot$ and above impose a robust lower bound on the maximum mass, requiring sufficient pressure at several times $n_0$. Models that soften too strongly, for example through the early appearance of additional degrees of freedom, are therefore strongly constrained.

\subsection{Stellar models}

Given an EoS of cold nuclear matter, one imposes the conditions of $\beta$-equilibrium and charge neutrality, including both electrons and muons. The core EoS of dense homogeneous matter is then matched smoothly to the inner and outer crust EoS containing nuclear clusters. With this input, the integral properties of compact stars --- in particular the mass $M$ and radius $R$ --- can then be  computed from the Tolman-Oppenheimer-Volkoff (TOV) equations~\citep{Tolman1939,Oppenheimer1939}.

\begin{figure}[tb]
  \begin{center}
     \includegraphics[width=0.45\hsize]{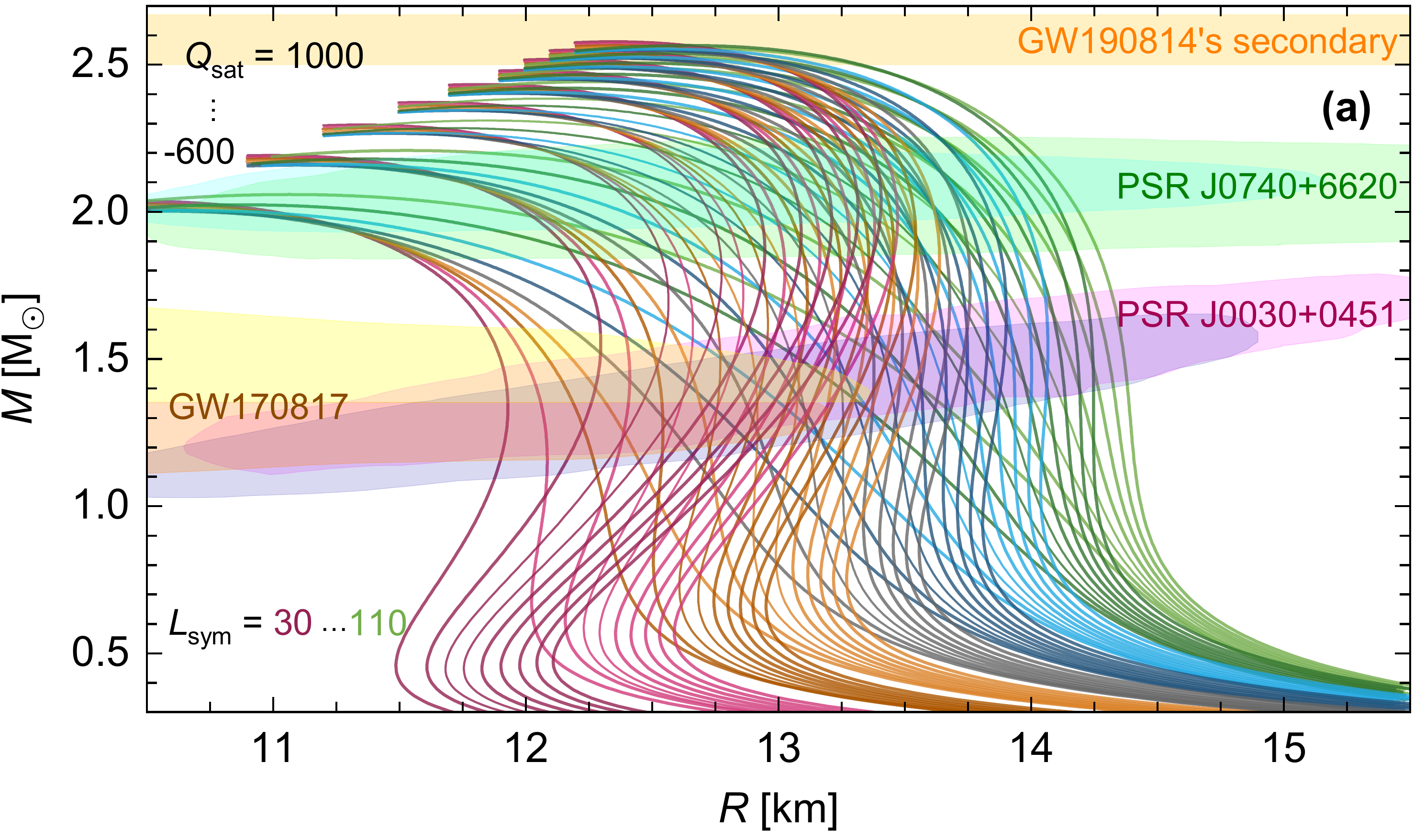}
  \includegraphics[width=0.45\hsize]{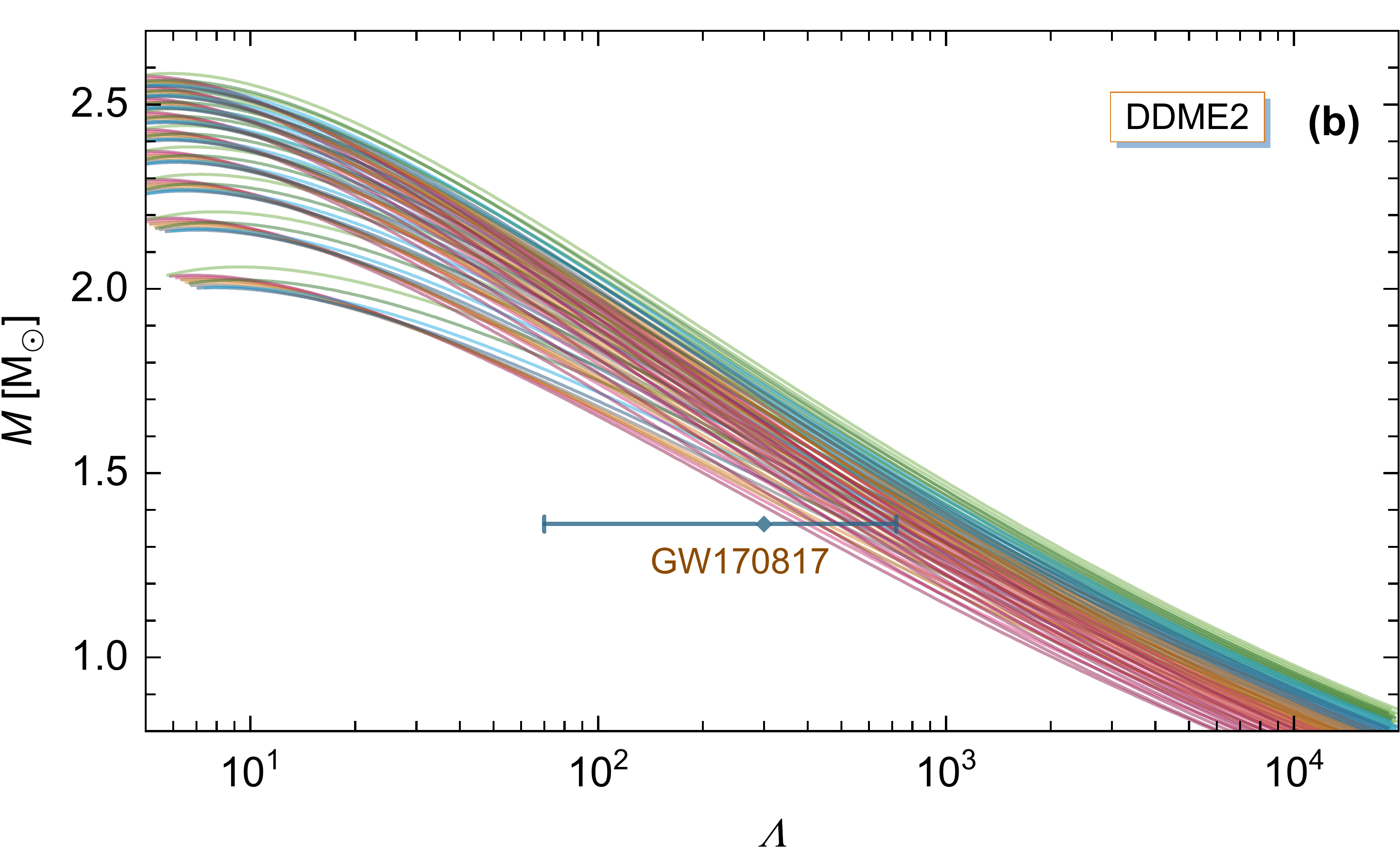}
\caption{
Mass-radius [panel (a)] and mass-tidal deformability [panel (b)]
relations for nucleonic EoS models with different pairs of values of
$Q_{\rm sat}$ and $L_{\rm sym}$ (in MeV)~\citep{Li2023}. In panel (a) the color regions
show the 90\% confidence interval (CI) ellipses from each of the two NICER modeling groups for
PSR~J0030+0451 and J0740+6620~\citep{Miller2019,Miller2021,Riley2019,Riley2021},
the 90\% CI regions for each of the two compact stars that merged in the
gravitational-wave event GW170817~\citep{Abbott2017}, and finally
the 90\% CI for the mass of the secondary component of
GW190814~\citep{LVC_GW190814}. In panel (b), the constraint for a
$1.362\,M_{\odot}$ star deduced from the analysis of GW170817
event~\citep{Abbott2019} is shown too.
}
\label{fig:fig2}
\end{center}
\end{figure}

Mass--radius relations provide a direct visualization of the global stellar structure implied by a given EoS and serve as a standard diagnostic for comparing models with observations. The function $\Lambda(M)$ connects microscopic nuclear physics to a GW observable and is particularly useful for Bayesian inference. As an illustration, we show in Fig.~\ref{fig:fig2} the mass-radius relations [panel (a)] and mass-tidal deformability relations [panel (b)] for static stellar configurations computed from the 81 EoS models of the DDME2 family, with varying $L_{\rm sym}$ and $Q_{\rm sat}$ parameters, overlaid with current multimessenger observational constraints~\citep{Li2023}. This establishes a contact between the observational bounds and empirical parameter of the expansion given by Eq.~\eqref{eq:Expansion}.  The softening of the EoS at intermediate densities, as required by the tidal deformability constraint from GW170817, and the stiffening at high densities, as demanded by the most massive observed compact objects, can be simultaneously accommodated through appropriate choices of the nuclear empirical parameters $L_{\rm sym}$ and $Q_{\rm sat}$, which govern the density dependence of the symmetry energy and the isoscalar skewness, respectively.

%------------------------------------- Fig. 2
%---------------------------------------------------
Above densities of a few times nuclear saturation density, additional degrees of freedom beyond nucleons may become energetically favored~\citep{Weber2007,Tolos2020}. Their appearance modifies the composition of matter, changes the stiffness of the EoS, and may leave observable imprints on neutron star structure and merger dynamics~\citep{Oertel2017,Sedrakian2023}.

\paragraph{Hyperons}
Hyperons, baryons containing strange quarks such as $\Lambda$, $\Sigma$, $\Xi$, and $\Omega$, may appear when the baryon chemical potential exceeds their in-medium threshold. This typically occurs at densities of order $2$--$3\,n_0$. The basic mechanism is simple: as the density increases, the nucleon Fermi energy grows, and it can become energetically favorable to convert a high-energy neutron into a heavier strange baryon. The hyperon--nucleon and hyperon--hyperon interactions are constrained by hypernuclear data and by hyperon scattering, but the uncertainties are considerably larger than in the purely nucleonic sector~\citep{Vidana2018}.

The appearance of hyperons generally softens the EoS, because baryon number is redistributed among additional species and the nucleon Fermi pressure is reduced. In many models this softening lowers the maximum mass below the observed $2\,M_\odot$ scale, leading to the so-called {\it hyperon puzzle}. Possible resolutions include strong short-range hyperon--nucleon repulsion, mediated for example by vector meson exchange, or three-body forces involving hyperons.
The description of dense hypernuclear matter which is based on the CDF framework, provides sufficient flexibility to accommodate both astrophysical observations and laboratory data on hypernuclei.
Lagrangian-based relativistic CDFs — also known as relativistic mean-field models — work with baryons and mesons as effective degrees of freedom and offer a physically transparent and tractable approach to determining the energy density of dense matter. The parameters of these functionals are calibrated to available nuclear data, for reviews see \cite{Oertel2017,Sedrakian2023}. The hyperon puzzle remains an open problem because it lies precisely at the interface between poorly constrained strange-baryon interactions and high-density neutron star observations.
%------------------------------------- Fig. 2
\begin{figure}[tb]
  \begin{center}
    \includegraphics[width=0.45\hsize]{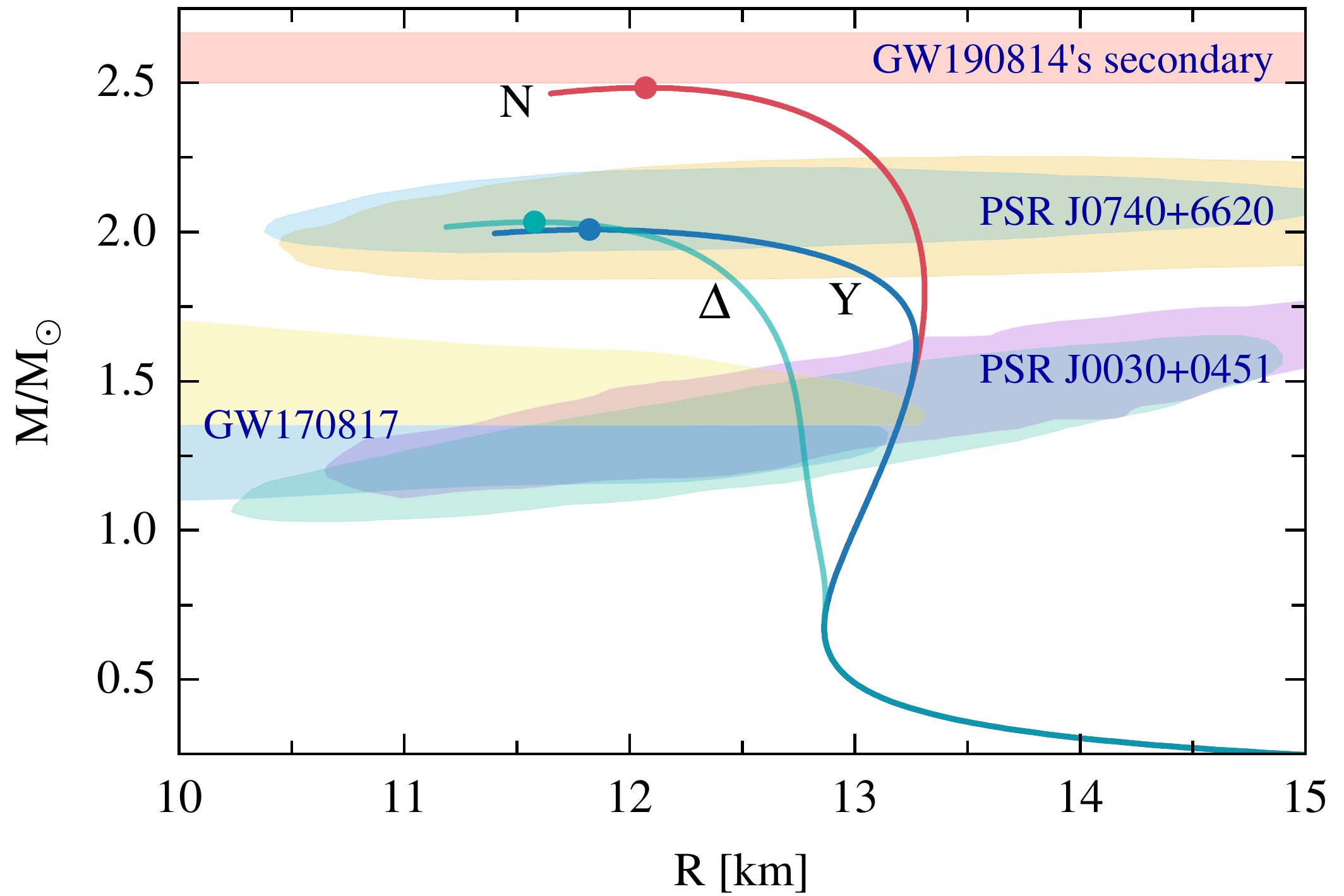} \caption{ Mass-radius relations for cold EoS models with nucleonic ($N$) hyperonic ($Y$) and hyperonic with admixture of $\Delta$-resonances ($\Delta$) compositions~\citep{Sedrakian2023}. The observational shaded ellipses are the same as in Fig.~\ref{fig:fig2}.  Filled circles mark the maximum-mass configuration of each sequence, beyond which the stars are unstable.  }
\label{fig:fig3}
\end{center}
\end{figure}
%-----------------------------------------
\paragraph{$\Delta$ Baryons}
The $\Delta(1232)$ resonance has also received attention as a possible constituent of dense neutron star matter. Its threshold density is uncertain and may be comparable to, or even lower than, the hyperon threshold. Unlike hyperons, the $\Delta$--nucleon interaction is constrained by pion--nucleon scattering data, although its in-medium behavior remains uncertain. CDF based calculations indicate that their inclusion tends make stars more compact. For canonical masses, $M\simeq 1.4\,M_\odot$, the radii may be reduced by about $1\,\mathrm{km}$ relative to purely nucleonic models, reflecting moderate softening at intermediate densities. These results suggest that $\Delta$ degrees of freedom are a viable component of dense matter, although their role remains model dependent~\citep{Drago2014,Li2018, Sedrakian2020,Sedrakian2022}.

As illustrated in Fig.~\ref{fig:fig3}, the progressive inclusion of strangeness degrees of freedom and an admixture of $\Delta$-resonance lowers the maximum masses.   The main effect of the inclusion of $\Delta$-resonances is the additional softening  at intermediate densities, with the net effect of reducing the radius. As seen from the figure, all three compositions can be brought into broad agreement with the multimessenger constraints. The stiffness required at high density to support the most massive observed pulsars places the tightest restrictions on models with exotic degrees of freedom.

\paragraph{Deconfined Quark Matter and Hybrid Stars}
At asymptotically high densities, QCD predicts that quarks become deconfined. In cold dense matter the relevant state is not a thermal quark--gluon plasma~\citep{Collins1975} but a degenerate Fermi liquid of quarks, which may be unstable to pairing and the formation of color-superconducting phases~\citep{Alford2008}. In the color-flavor locked phase, expected at asymptotically high densities, all quark flavors participate in a symmetric pairing pattern, with gaps of order $\Delta\sim 10$--$100\,\mathrm{MeV}$. At the lower densities relevant to neutron star interiors, the pairing pattern is much less certain, and phases such as two-flavor
two-color (2SC) superconductivity, crystalline color superconductivity, or other less symmetric states may occur. These phases can affect the stiffness of the EoS, neutrino emission, and transport properties~\citep{Page2006}.

From the point of view of the high-density EoS, perturbative QCD (pQCD) provides theoretical control only at asymptotically large chemical potentials~\citep{Fraga2014,Gorda2021}. Conservative estimates place the onset of quantitatively reliable weak-coupling behavior at densities far above those realized in neutron stars~\citep{Lattimer2005}. Thus, although pQCD provides an important asymptotic anchor, the density regime relevant for compact stars remains intrinsically non-perturbative and must be modeled with effective descriptions constrained by astrophysical data.

The transition from hadronic to quark matter may occur as a sharp first-order phase transition or as a smooth crossover, depending on the structure of the QCD phase diagram at the relevant temperature and baryon density~\citep{Alford2005,Baym2018}. In either case, the transition modifies the EoS and can leave characteristic signatures in neutron star observables.

A strong first-order transition is signaled thermodynamically by a discontinuity in energy density at fixed pressure in a Maxwell construction, or by a mixed phase when finite-size and Coulomb effects are included. Such a transition softens the EoS over the transition region and produces a corresponding feature in the relation $P(\epsilon)$. Neutron stars with quark cores, commonly called {\it hybrid stars}, may therefore have smaller radii than purely hadronic stars of the same mass, provided the transition occurs in the density range realized in stable configurations.
%------------------------------------- Fig. 4
\begin{figure}[tb]
  \begin{center}
    \includegraphics[width=0.45\hsize]{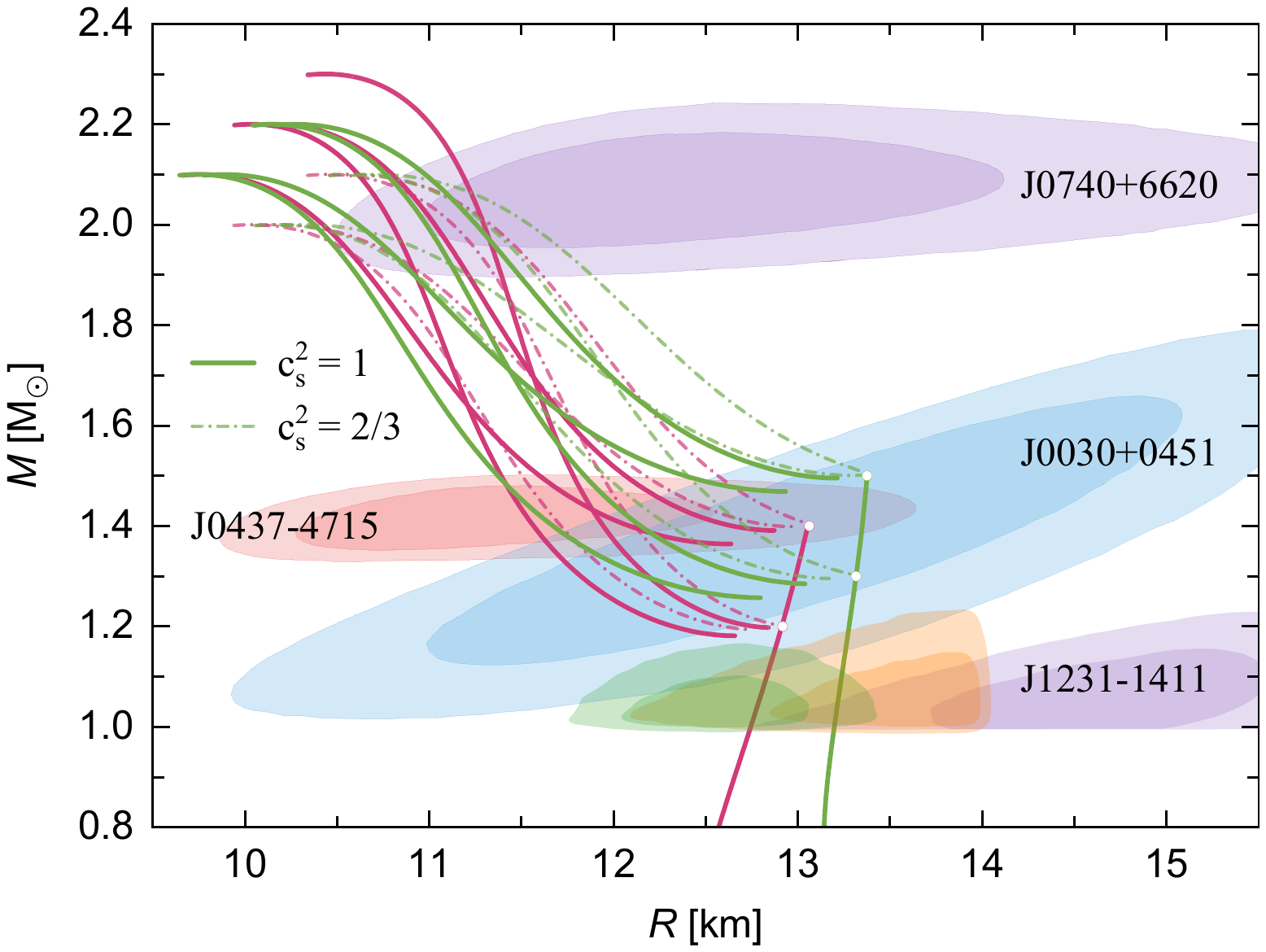}
    \includegraphics[width=0.45\hsize]{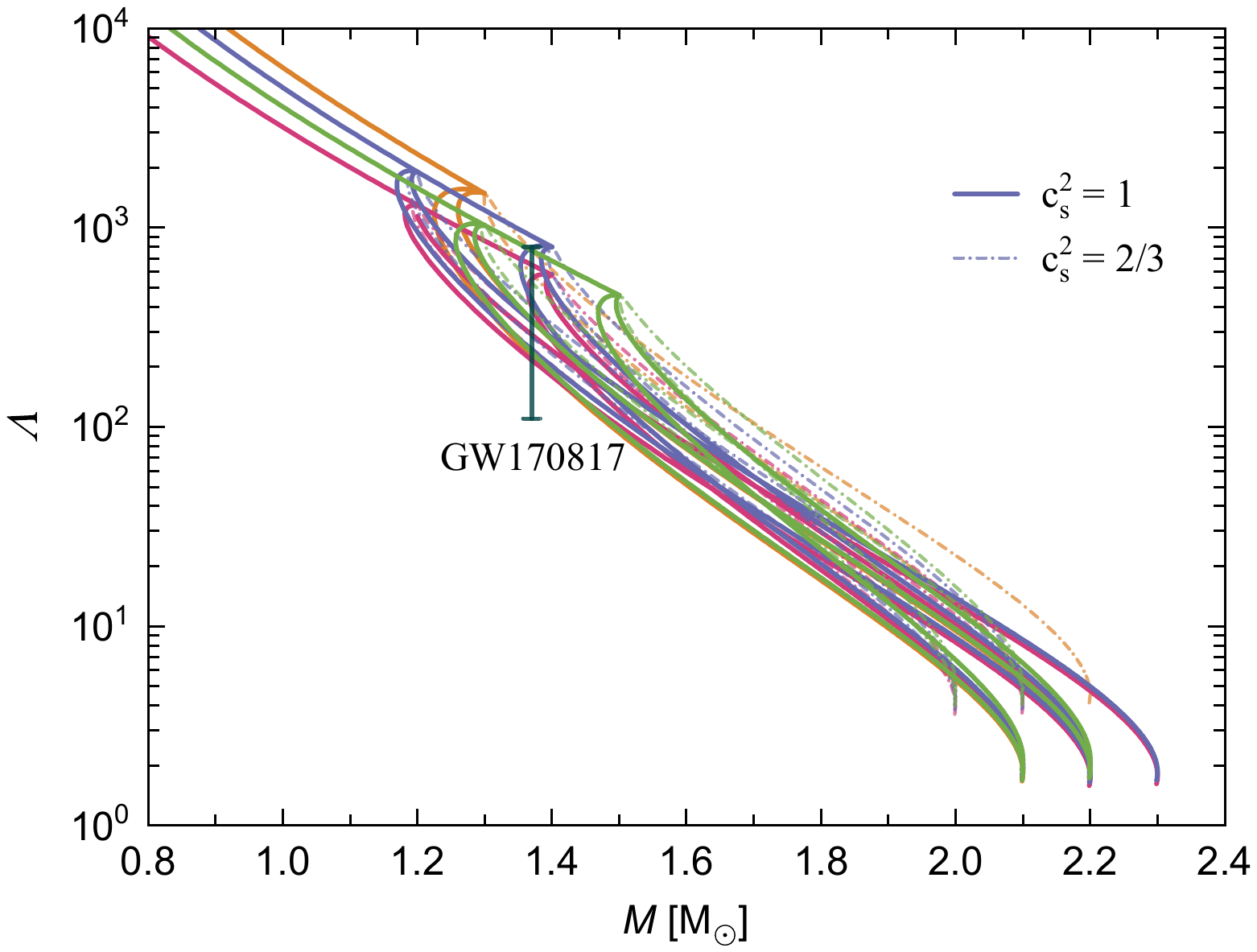}
\caption{
  Mass–radius relations (left panel) for hybrid EoS models with transition masses $M^{ H}_{\rm max}=1.2\text{–}1.5M_\odot$ and hybrid maximum masses $M^{ Q}_{\rm max}=2.0\text{–}2.3M_\odot$, constructed using two representative nucleonic EoSs with $L_{\rm sym}=40$ and 60 MeV~\citep{Li2025}. The ellipses indicate observational constraints at 68\% and 95\% credible levels from recent NICER analyses~\citep{Salmi2024}. The corresponding mass–tidal deformability relations are shown in the right panel, with vertical error bars denoting the 90\% confidence interval for a $1.362\,M_\odot$ star inferred from GW170817~\citep{Abbott2019}.
This panel also includes tidal deformabilities with large $L_{\rm sym}=80$ and 100 MeV (blue and orange curves, respectively). 
}
\label{fig:fig4}
\end{center}
\end{figure}
%-----------------------------------------
One of the most striking possible consequences is the existence of twin-star or third-family solutions. These are pairs of stable stars with the same gravitational mass but different radii and internal compositions~\citep{Glendenning2000,Alford2013}. They occur when the transition is strong enough to destabilize the ordinary branch and generate a new stable branch at higher central density. In more complex scenarios, sequential phase transitions may even produce multiple branches~\citep{Alford2017,Rau2023,Sedrakian2023a}. Such features would provide a direct observational signature of a strong phase transition in dense matter and could be tested through precise mass--radius measurements or, in mergers, through post-merger gravitational-wave spectroscopy.
The occurrence of twin configurations depends primarily on three quantities: the transition density, the magnitude of the energy-density jump across the transition, and the stiffness of quark matter at high density. A sufficiently strong first-order transition is required to destabilize the hadronic branch and generate a disconnected hybrid branch. This condition is closely related to the Seidov criterion, which specifies the critical energy-density discontinuity necessary for destabilization at the onset of the phase transition. If the latent heat is too small, the hybrid branch connects continuously to the hadronic sequence and no twins appear. Conversely, if the energy-density jump is too large and the quark phase is not sufficiently stiff, the star becomes unstable immediately after deconfinement and no stable hybrid configurations survive.

As an illustration Fig.~\ref{fig:fig4} shows the mass-radius and tidal-deformability-mass plots~\citep{Li2025}.
Here, the low-density nucleonic EoS belong to the same family as shown in Fig.~\ref{fig:fig2}
with varying the symmetry energy slope $L_{\rm sym}$ at fixed $Q_{\rm sat}$.
The transition to quark matter is modeled as a first-order phase transition, specified by the transition density and jump in the energy density, and the speed of sound in the quark phase, using the constant speed-of-sound parameterization. The representative values $c_s^2=2/3$ and the causal limit $c_s^2=1$ are implemented. Instead of microscopic parameters, it more transparent to characterize the EoS of a hybrid star in terms of macroscopic observables: the maximum mass of the hadronic branch, and the minimum and maximum masses of the hybrid branch ~\citep{Li2025}.  For suitable parameter choices, this setup yields disconnected mass–radius branches, corresponding to purely hadronic and hybrid stars separated by an unstable region, and allows for the existence of twin configurations. 
All models considered satisfy the tidal deformability constraint for a $1.362\,M_\odot$ star from GW170817. Softer EoSs lie well within the allowed $\Lambda_{1.362}$ range, and in hybrid models with $M^{Q}_{\min} \lesssim 1.4M_\odot$, the onset of quark matter further improves agreement with the data.

BNS mergers provide a particularly important environment for probing hybrid-star EoSs. During inspiral, the tidal deformabilities inferred from gravitational-wave observations constrain the stellar compactness and therefore the stiffness of dense matter. Following merger, the remnant can reach temperatures of tens of MeV and densities several times nuclear saturation density, potentially driving the formation of deconfined quark matter even if the original stars were purely hadronic. The emergence of quark matter can soften the remnant EoS, accelerate collapse to a black hole, and modify the dominant post-merger oscillation frequencies~\citep{Most2019,Most2020,Weih2020}. Simulations incorporating hadron--quark phase transitions have reported shifts in the post-merger gravitational-wave spectrum, changes in the collapse time, and modifications of the ejecta and neutrino emission.

The interpretation of present observations remains model dependent because uncertainties in the hadronic EoS, the phase-transition construction, and the properties of strongly interacting quark matter are still substantial. Nevertheless, the combined constraints from massive pulsars, NICER radius measurements, and the tidal deformability extracted from GW170817 already restrict the range of viable hybrid-star models. In particular, viable models generally require a sufficiently stiff high-density quark phase to maintain stability up to at least $2\,M_{\odot}$~\citep{}. Future detections of post-merger GWs, together with improved multimessenger constraints and advances in QCD-based modeling, may determine whether the dense cores of neutron stars contain deconfined quark matter and whether twin-star configurations are realized in nature.

\subsection{Thermodynamics of Hot Dense Matter}

In dynamical environments such as neutron star mergers, finite-temperature effects are essential. During merger and in the post-merger remnant, temperatures can reach $T\sim 30$--$80\,\mathrm{MeV}$ in the densest regions, with even higher values possible locally in shock-heated layers, shear interfaces, and disk material. Under these conditions the EoS is no longer barotropic, as in cold catalyzed neutron stars, but depends on density, temperature, and composition,
%--------------------------
\bea
P = P(\epsilon,T,Y_i),
\eea
%--------------------------
or equivalently $P=P(n_B,T,Y_e)$, commonly provided in a tabulated form~\citep{Typel2022}.

The composition of matter in this regime is richer than in cold neutron stars. At low temperature and in beta equilibrium, matter is dominated by neutrons, with a smaller fraction of protons and leptons. At finite temperature and away from equilibrium, however, the particle content depends sensitively on both $T$ and $Y_e$. For example, in nucleonic matter at $T\sim 10\,\mathrm{MeV}$ and low electron fraction, $Y_e\sim 0.1$, the system remains strongly neutron-rich, with a modest proton fraction and an electron--positron component. At higher temperatures, $T\sim 50$--$80\,\mathrm{MeV}$, thermal pair production becomes important, positrons are abundant, and the proton fraction can increase as the neutron--proton chemical potential difference is reduced. The~case $Y_{L,e} = 0.1$ is
  characteristic of a BNS merger remnant, whereas
  $Y_{L,e} = 0.2 - 0.4$ correspond to supernovae.

%------------------------------------- Fig. 5
\begin{figure}[tb]
  \begin{center}
    \includegraphics[width=0.8\hsize]{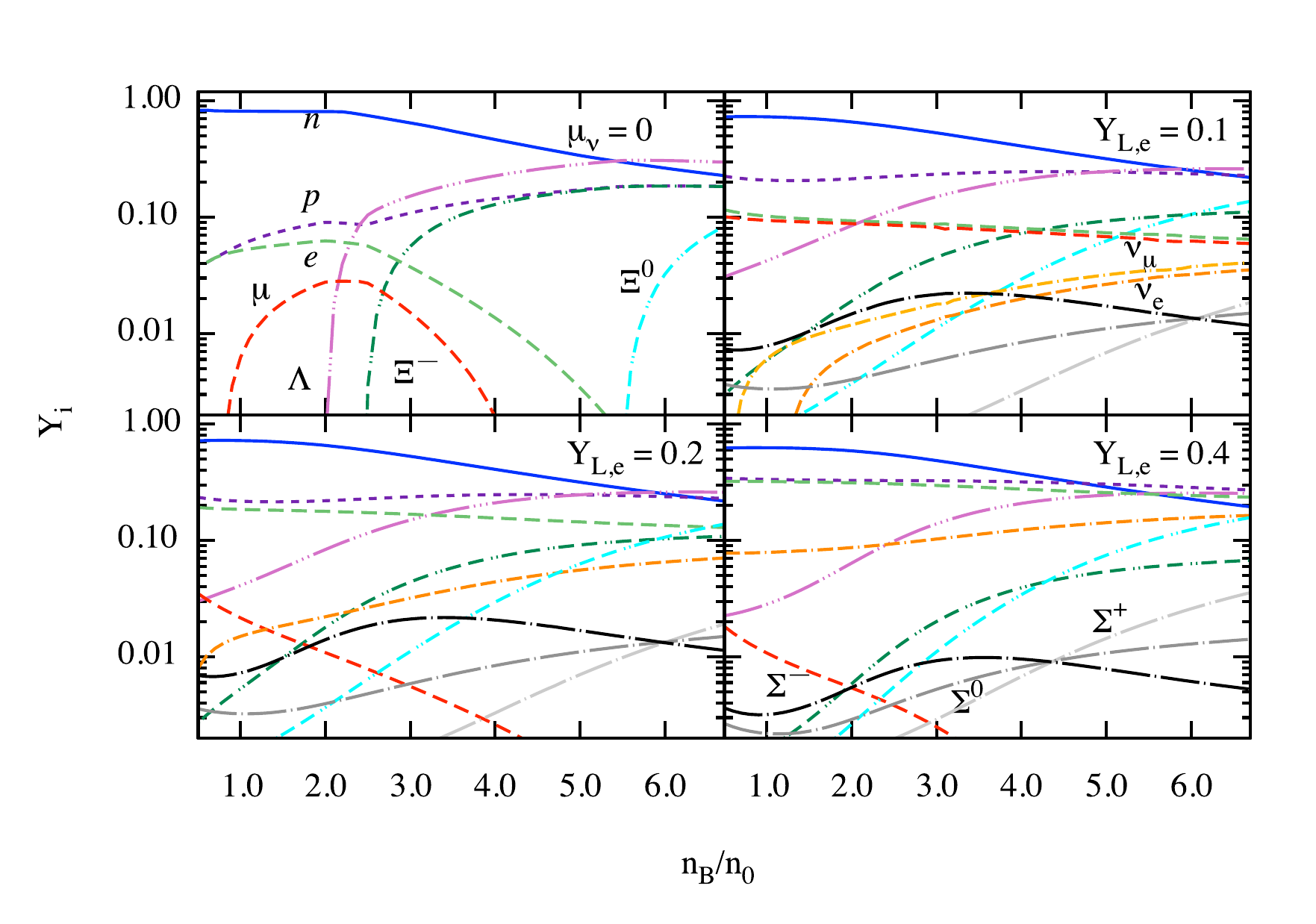}
\caption{
  Particle fractions $Y_i \equiv n_i/n_B$ as a function of baryon density $n_B$
  normalized to the saturation density $n_{0}$~\citep{Sedrakian2021}.
The upper-left panel corresponds to neutrino-free matter in $\beta$-equilibrium
  without (solid) and with (dashed) hyperons at $T=0.1$~MeV. The remaining three panels show the EoS of neutrino-trapped matter for $Y_{L,e}= 0.1, \, 0.2, \, 0.4$ and  temperature 
  $T= 50$~MeV. The muon fractions are adjusted to the conditions of supernova and BNS merger cases as follows: $Y_{L,\mu} = Y_{L,e} = 0.1$ (upper-right panel) and $Y_{L,\mu} = 0$
  and $Y_{L,e}=0.2$ and 0.4 (bottom panels).  At low temperatures (upper left panel), hyperons appear sharply at their thresholds, while at higher temperatures these thresholds are smeared and shift to lower densities, with all $\Sigma$ multiplets being populated.}
\label{fig:fig5}
\end{center}
\end{figure}
%-----------------------------------------

If hyperons are included, the composition becomes more complex, as illustrated in Fig.~\ref{fig:fig5}. At low temperatures, hyperons appear at relatively well-defined threshold densities. The $\Lambda$ hyperon often appears first, around a few times saturation density, followed by negatively charged species such as $\Xi^-$ or $\Sigma^-$ at higher densities, depending on the interaction model. Finite temperature smears these thresholds and produces non-negligible thermal populations below the corresponding zero-temperature onset densities. Thus at $T\sim 50$--$80\,\mathrm{MeV}$ and low $Y_e$, nucleons may still dominate, but $\Lambda$ hyperons can contribute at the percent level, while other strange baryons are thermally populated at lower densities than in cold matter. The charge chemical potential also matters: at higher $Y_e$, neutral hyperons are favored relative to negatively charged species.

These changes in composition feed back directly into the thermodynamics. Additional baryonic species redistribute baryon number among more degrees of freedom and tend to soften the EoS at fixed density. Thermal pressure partly compensates for this softening, so the net effect on the pressure and sound speed depends on the detailed balance between temperature, density, and composition. This is one reason why finite-temperature EoS tables can show nontrivial behavior even when the underlying zero-temperature EoS appears relatively simple, see CompOSE database~\citep{Typel2022} and,
for the case of hypernuclear matter \cite{Tsiopelas2024}.

At subnuclear densities, matter is further complicated by nuclear clustering. In this regime it is described by nuclear statistical equilibrium, consisting of heavy nuclei, light clusters such as deuterons and alpha particles, free nucleons, electrons, positrons, and photons~\citep{Typel2010}. As the temperature increases, nuclei dissolve into nucleons and the system approaches uniform matter. Hyperons are usually negligible at such low densities, but become relevant once the density approaches and exceeds saturation.
%------------------------------------- Fig. 6
\begin{figure}[t] 
\begin{center}
\includegraphics[width=0.5\linewidth,keepaspectratio]{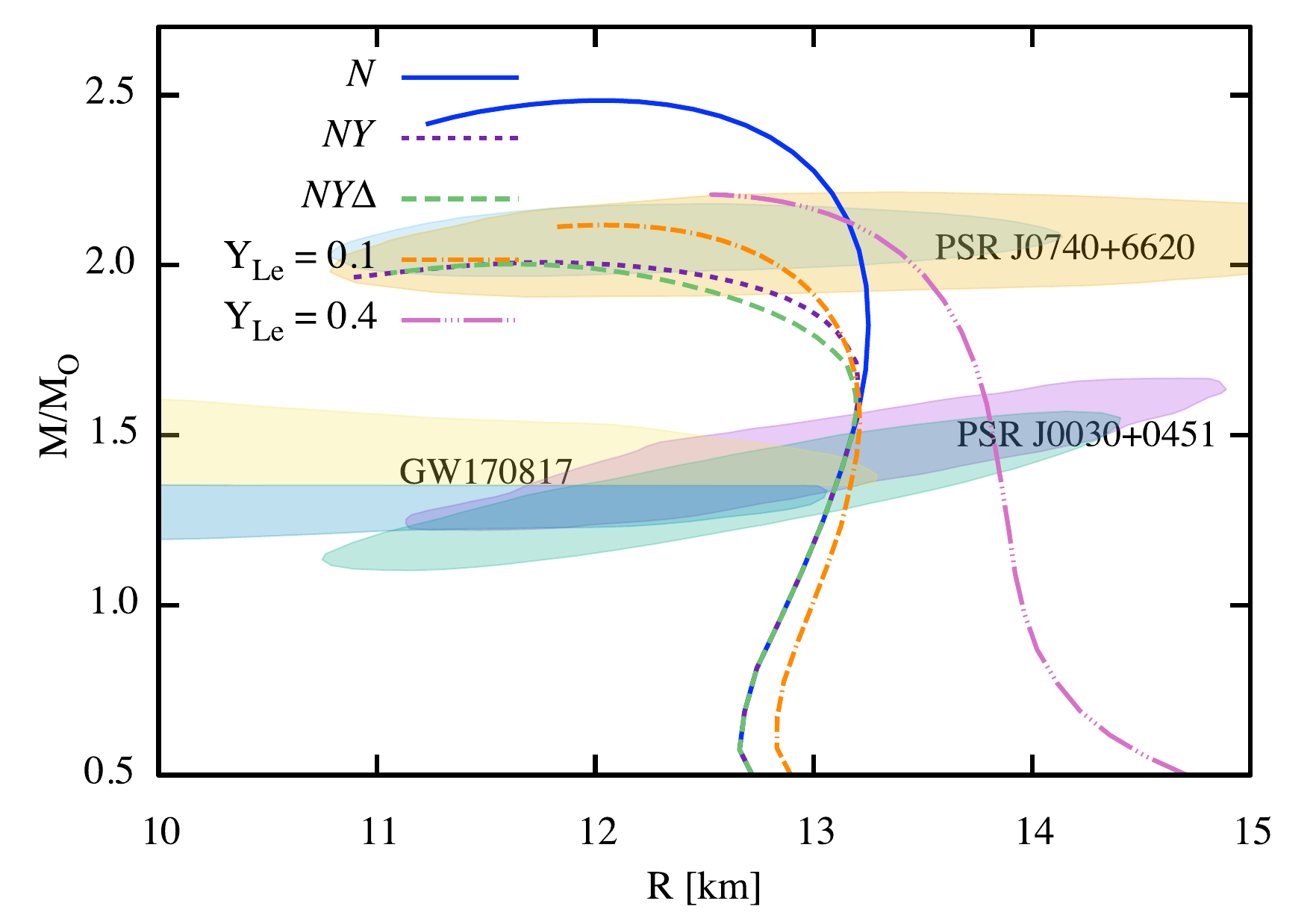}
\caption{Mass–radius relations for non-rotating, spherically symmetric compact stars. Shown are sequences for cold, neutrino-transparent matter in $\beta$-equilibrium with nucleonic ($N$), hypernuclear ($NY$), and $\Delta$-admixed ($NY\Delta$) compositions at $T=0.1$ MeV~\citep{Sedrakian2022}. In addition, isentropic sequences with entropy per baryon $S/A=1$ are displayed for neutrino-trapped matter composed of $NY\Delta$ matter, with fixed lepton fractions $Y_{L,e}=Y_{L,\mu}=0.1$ and $Y_{L,e}=0.4$, $Y_{L,\mu}=0$. The shaded ellipses indicate 90\% confidence regions from NICER observations of PSR~J0030+0451 and PSR~J0740+6620, and from the event GW170817.}
\label{fig:fig6} 
\end{center}
\end{figure}
%-----------------------------------------------------
Another important feature of merger matter is that it is generally not in beta equilibrium. Weak-interaction rates are finite, and the equilibration timescale may be comparable to or longer than the hydrodynamical timescale. The electron fraction $Y_e$ therefore evolves dynamically through electron and positron capture, neutrino emission, and neutrino absorption. As a result, the composition at a given $(n_B,T)$ is not determined solely by equilibrium thermodynamics but also by the history of the fluid element.

Finite-temperature EoS tables must therefore provide not only pressure and energy density, but also detailed composition information, including particle fractions for the relevant species. These tables are typically constructed by solving finite-temperature CDF or related mean-field equations for baryons and mesons, treating leptons and photons as ideal gases, and matching to low-density nuclear statistical equilibrium models. The resulting multidimensional tables encode the thermodynamic response of dense matter, including the effects of hyperons or other additional degrees of freedom when they are included.

The composition of matter, especially the relative abundances of nucleons, hyperons, and leptons, plays a central role in determining neutrino emissivities, opacities, and the evolution of $Y_e$. It also influences the entropy and neutron richness of the ejecta, which are key inputs for $r$-process nucleosynthesis. Understanding the finite-temperature composition of dense matter is therefore essential for connecting microscopic EoS models to the observable signatures of neutron star mergers.

Figure~\ref{fig:fig6} compares the mass-radius relations of non-rotating compact stars computed with both cold and finite-temperature isentropic EoSs for composition that includes hyperons and $\Delta$ baryons.  For illustration, we also show the effect of composition on the mass-radius relations for cold, neutrino-transparent matter in $\beta$-equilibrium, by comparing purely nucleonic ($N$), hypernuclear ($NY$), and $\Delta$-admixed hypernuclear matter ($NY\Delta$), as discussed in Fig.~\ref{fig:fig4}.

The isentropic, neutrino-trapped configurations, characterized by fixed entropy per baryon and fixed lepton fractions, exhibit systematically larger radii and slightly higher maximum masses than their cold counterparts, reflecting the combined effects of thermal pressure and neutrino-induced changes in composition. The dependence on the lepton fraction further elucidates the differences between supernova and BNS merger matter in determining the structure of hot compact stars. The nucleonic model yields a maximum mass $M_{\rm max} = 2.48\,M_\odot$ with a corresponding radius $R \simeq 12.1$ km. The inclusion of hyperons and $\Delta$-resonances softens the EoS, reducing the maximum mass to $M_{\rm max} \simeq 2.0\,M_\odot$ and the radius to $R \simeq 11.6$ km in the $NY\Delta$ case. For canonical masses ($M \sim 1.4\,M_\odot$), the radii of all models are similar, since the onset of hyperons and $\Delta$-resonances occurs at densities exceeding the central densities of such stars. Differences between compositions become significant only at higher masses, with departures from the purely nucleonic sequence emerging at $M \gtrsim 1.5\,M_\odot$.
%-----------------------------------------------------

%\newpage
\section{Initial Conditions and Binary Inspiral}
\label{sec:Chapter3}

The inspiral phase of a BNS system is characterized by a gradual orbital decay driven by GW emission. As the binary loses energy and angular momentum to gravitational radiation, the orbital separation decreases and the GW frequency sweeps upward in the characteristic ``chirp'' pattern, with both amplitude and frequency increasing as the system approaches merger. Although the inspiral may last millions to billions of years from the time of binary formation, the part accessible to ground-based detectors, from $\sim 10\,\mathrm{Hz}$ to merger, spans only the final tens of seconds. It is during this final stage that finite-size effects become important and imprint information about the neutron star interior on the waveform.

To leading post-Newtonian (PN) order, the GW frequency evolution of a quasi-circular binary is~\citep{Blanchet2014}
%--------------------------------
\bea\label{eq:fdot}
\dot{f} = \frac{96}{5} \pi^{8/3} \mathcal{M}^{5/3} f^{11/3},
\eea
%--------------------------------
where $\mathcal{M} = (M_1 M_2)^{3/5} (M_1 + M_2)^{-1/5}$ is the chirp mass. This relation follows from energy balance between the orbital binding energy and the quadrupole GW luminosity, and forms the basis for parameter estimation in GW observations. The chirp mass is the best measured mass combination because it is determined primarily by the rate at which the frequency evolves. The individual masses and the mass ratio require higher-order corrections, spin effects, and finite-size terms to be disentangled.

Higher-order PN corrections introduce dependence on the symmetric mass ratio $\nu = M_1 M_2/(M_1+M_2)^2$, on the individual spins, and on tidal interactions. Over most of the inspiral, the motion is adiabatic: the radiation-reaction timescale is much longer than the orbital period. This separation of timescales allows accurate semi-analytic modeling using PN and effective-one-body  methods~\citep{Buonanno1999}. Close to merger, however, the expansion parameter becomes large, the PN series loses convergence, and numerical relativity is required.

\subsection{Formation Channels of BNSs}

BNSs form through the co-evolution of two massive stars in a gravitationally bound system. In the standard isolated binary channel, the system passes through a sequence of mass-transfer episodes, supernova explosions, and orbital-shrinking phases. Each of these stages introduces uncertainties that are ultimately reflected in the observed distributions of binary masses, spins, eccentricities, and merger times.

The canonical scenario begins with two main-sequence stars of masses $\gtrsim 8\,M_\odot$ in a sufficiently compact orbit. When the more massive star exhausts hydrogen in its core and expands, it may overflow its Roche lobe and transfer mass to its companion. If this transfer is unstable, the system can enter a common-envelope phase, in which the companion spirals through the envelope of the giant~\citep{Ivanova2013}. This stage is difficult to model from first principles, but it is crucial because it can reduce the orbital separation by orders of magnitude. Whether the binary survives depends on the binding energy of the envelope and on the efficiency with which orbital energy is deposited into it.

After envelope ejection, the stripped primary undergoes core collapse and forms the first neutron star. The natal kick imparted during the supernova, arising from asymmetric mass ejection or anisotropic neutrino emission, may unbind the binary or change its eccentricity and orbital inclination. If the system remains bound, it can undergo a second episode of mass transfer before the companion also collapses to form the second neutron star. The newly formed BNS may initially have significant eccentricity, but GW emission circularizes the orbit efficiently before the binary reaches the detector band for typical isolated-field systems.

Population synthesis calculations describe these evolutionary paths statistically~\citep{VignaGomez2018}. They treat uncertain quantities, such as mass-transfer efficiency, common-envelope efficiency, and natal-kick distributions, as model inputs and predict merger rates and binary parameter distributions. These predictions are tested against the observed Galactic BNS population and the merger rates inferred from LIGO/Virgo/KAGRA. The comparison is still limited by small-number statistics, but it already provides important constraints on binary evolution.

There is also a nuclear-physics connection in this formation problem. The relation between progenitor structure, core collapse, and the final neutron star mass depends on the nuclear EoS and on the physics of dense matter in the proto-neutron star. The observed Galactic BNS masses are clustered near $1.35\,M_\odot$, with a relatively narrow spread, whereas the formation of significantly more massive neutron stars requires particular progenitor structures or accretion histories~\citep{Oezel2016}. Dynamically formed binaries, produced for example in dense stellar environments through gravitational encounters, may have different masses and retain measurable eccentricity, providing a possible observational distinction from the isolated-field channel.

\subsection{Orbital Evolution and Tidal Interactions}

The long-term orbital evolution is governed by gravitational radiation reaction. To leading quadrupole order, the binary loses energy at the rate
\bea
\dot{E}_{\rm GW} =
-\frac{32}{5}\frac{G^4}{c^5}
\frac{(M_1M_2)^2(M_1+M_2)}{a^5},
\eea
where $a$ is the semi-major axis. Equating this loss to the time derivative of the orbital binding energy gives the characteristic merger time,
\bea
t_{\rm merge} \simeq
\frac{5}{256}
\frac{c^5 a^4}{G^3 M_1M_2(M_1+M_2)}.
\eea
The strong $a^4$ dependence means that modest differences in the post-supernova separation lead to large differences in the merger time. The Hulse--Taylor pulsar PSR~B1913+16~\citep{Hulse1975} provides the classic example: its observed orbital decay agrees with the prediction of general relativity and implies a merger time of order several hundred
Myr~\citep{Taylor1982}.

As the binary enters the frequency band of ground-based detectors, higher PN corrections become necessary. The expansion is organized in powers of $(v/c)^2 \sim GM/(ac^2)$, with corrections from conservative relativistic dynamics, spin-orbit and spin-spin couplings, and higher-order radiation reaction. For BNS systems in the LIGO/Virgo band, $v/c$ remains below unity but becomes large enough near contact that resummation methods and, ultimately, numerical-relativity calibration are required.

For quasi-circular binaries, the dominant GW frequency is related to the
orbital frequency by $f_{\rm GW}\simeq 2f_{\rm orb}=\Omega/\pi$, where
$\Omega$ is the orbital angular frequency. A rough Newtonian estimate of
the contact frequency is obtained by setting the orbital separation equal
to $R_1+R_2$:
%--------------------------------
\bea
f_{\rm contact} \simeq
\frac{1}{\pi}
\sqrt{\frac{G(M_1+M_2)}{(R_1+R_2)^3}} .
\eea
%---------------------------------
This estimate shows explicitly where the EoS enters the late inspiral:
larger radii, as predicted by stiffer EoS models, imply lower contact
frequencies, so that finite-size effects become important at lower GW
frequencies. Conversely, more compact stars remain well separated until
higher frequencies. In realistic waveforms the EoS dependence appears
already before contact through tidal interactions, but the contact
frequency provides a simple illustration of how stellar radii affect the
late-inspiral signal.

Finite-size effects enter the binary dynamics through tidal interactions, which become increasingly important as the orbital separation approaches the stellar radii. Each neutron star is deformed by the gravitational field of its companion, developing a quadrupole moment whose magnitude depends on the internal density profile and therefore on the EoS.

In the linear adiabatic regime, the induced quadrupole moment $Q_{ij}$
is related to the external tidal tensor $\mathcal{E}_{ij}$ by
\bea
Q_{ij} = -\lambda\,\mathcal{E}_{ij},
\eea
where $\lambda$ is the dimensional tidal deformability defined by 
\bea
\lambda = \frac{2}{3}\frac{k_2 R^5}{G},
\eea
%----------------------------
so the tidal response scales approximately as $R^5$, apart from the additional dependence of the Love number $k_2$ on the internal stellar structure. Consequently, even relatively small differences in radius can lead to large differences in the tidal response. Note that $\lambda$ can be related to the dimensionless deformability defined in Sec.~\ref{sec:Chapter2}, Eq.~\eqref{eq:Lambda}, by $\Lambda = {\lambda c^{10}}/{G^4 M^5}$.

In the gravitational waveform, the leading tidal contribution enters formally at 5PN order relative to the point-particle term. Its dominant contribution to the phase can be written schematically as
\bea
\Psi_{\rm tidal}(f) \propto
-\tilde{\Lambda}\,(\pi\mathcal{M}f)^{5/3},
\eea
where $\tilde{\Lambda}$ is the effective tidal deformability defined in Eq.~\eqref{eq:Lambda_tilde}, and is the tidal parameter most directly constrained in equal- or near-equal-mass binaries.

The value of $\tilde{\Lambda}$ differs substantially among EoS models. Stiff nucleonic EoS models, which produce larger radii, predict larger tidal deformabilities. Softer EoS models, including those with hyperons or quark matter that reduce the stellar radius, lead to smaller values, as discussed in Sec.~\ref{sec:Chapter2}. Hybrid stars with quark cores can have particularly small deformabilities if the transition softens the EoS at intermediate densities. Thus a precise measurement of $\tilde{\Lambda}$ can in principle distinguish between different classes of dense matter models, although individual current-generation events are generally not sufficient to identify the microscopic composition uniquely.

Beyond the adiabatic quadrupolar response, dynamical tides may become relevant close to merger. These arise when the tidal driving frequency approaches the frequencies of stellar oscillation modes, especially the fundamental $f$-mode and composition- or entropy-driven $g$-modes. For typical neutron stars, $f$-mode frequencies lie in the kHz range and therefore affect the late inspiral and merger. The $g$-modes are sensitive to stratification and composition gradients, and may provide additional information on the thermal and chemical structure of the star. These effects are small for current detectors but are important targets for third-generation observatories.

\subsection{Gravitational Wave Signal in Inspiral}

The inspiral waveform is modeled using PN theory, effective one-body (EOB) methods, and numerical-relativity input. The PN expansion provides a systematic perturbative description in powers of $v/c$. The EOB framework reorganizes the two-body problem into the motion of an effective particle in a resummed metric. This resummation captures some strong-field behavior more efficiently than the PN series alone and provides a natural framework for matching numerical-relativity simulations.

Tidal effects are incorporated into EOB models through additional tidal potentials in the effective Hamiltonian. Waveform families such as \texttt{TEOBResumS}~\citep{Bernuzzi2015} and \texttt{SEOBNRv4T}~\citep{Hinderer2016} are used in BNS data analysis. They describe point-particle dynamics, spin effects, and tidal contributions consistently over the frequency range accessible to detectors, while remaining calibrated to numerical-relativity simulations near merger.

In the frequency domain, using the stationary-phase approximation, the GW phase may be decomposed as
\bea
\Psi(f) =
\Psi_{\rm PP}(f)
+\Psi_{\rm spin}(f)
+\Psi_{\rm tidal}(f),
\eea
where $\Psi_{\rm PP}$ denotes the point-particle contribution, $\Psi_{\rm spin}$ contains spin-orbit and spin-spin terms, and $\Psi_{\rm tidal}$ describes finite-size effects. These terms enter at different PN orders and have different frequency dependence, which allows them to be partially separated in parameter estimation.

%------------------------------------- Fig. 7
\begin{figure}[tb] \centering
  \includegraphics[width=0.9\linewidth,keepaspectratio]{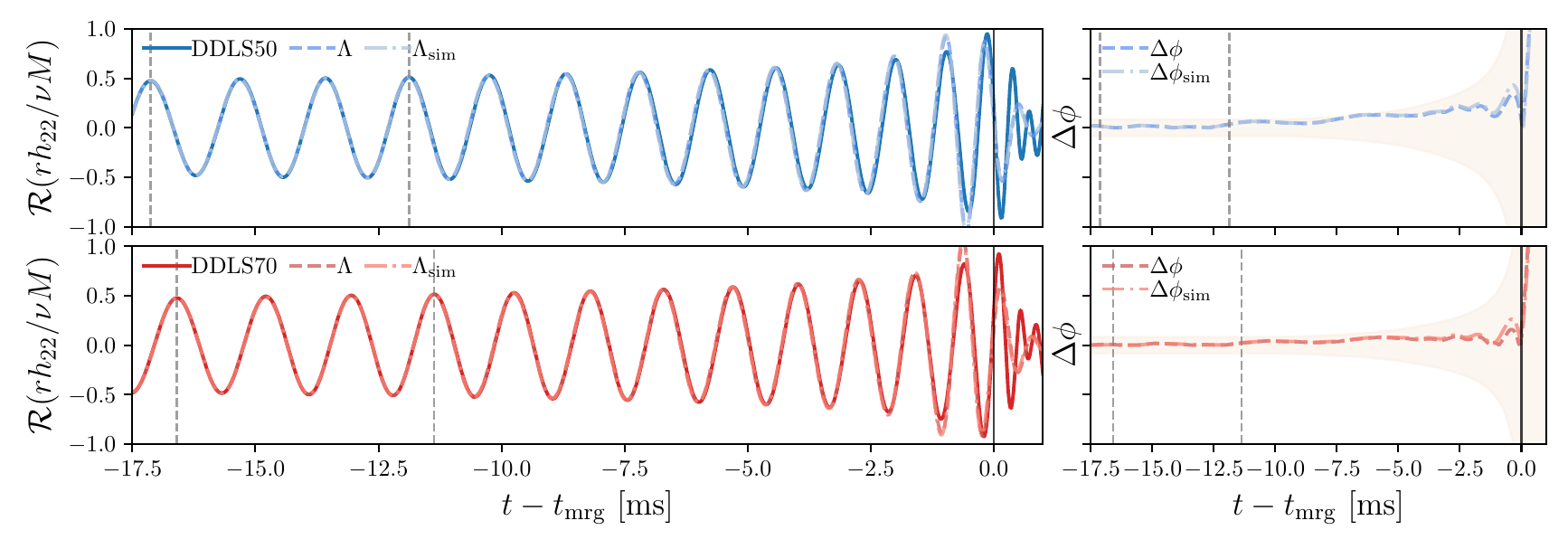} \caption{ GW strain from a BNS inspiral showing the characteristic chirp signal, where amplitude and frequency increase as the orbital separation decreases, for nucleonic EoS with $L_{\rm sym} = 50$~MeV (upper panel) and $70$~MeV (lower panel) at fixed $Q_{\rm sat} = 400$~MeV~\citep{Gieg2025}. The x-axis shows $t-t_{\rm mrg}$, where $t_{\rm mrg}$ is the time of merger. The y-axis shows the dimensionless strain $\mathcal{R} h_{22}/\nu M$, where $\mathcal{R}$ is the GW extraction radius, $h_{22}$ is the dominant $(l=2,m=2)$ mode, $\nu$ is the symmetric mass ratio, and $M = M_1+M_2$ is the total mass. Left panels show waveforms from numerical simulations (thick lines) alongside the {\tt IMRPhenomXAS\_NRTidalv3} approximant using either the static tidal deformability $\Lambda$ (dashed) or the dynamically estimated $\Lambda_{\rm sim}$ (dash-dotted). Right panels show the corresponding dephasings, with shaded bands indicating uncertainty. Finite-size effects such as tidal deformability introduce phase corrections encoding information about the neutron star EoS.}
\label{fig:fig7}
\end{figure}
% ----------------------------------------
The EoS enters the inspiral waveform primarily through finite-size effects encoded in the tidal response of the stars, most commonly characterized by the tidal deformability $\tilde{\Lambda}$, which depends on the stellar masses, radii, and Love numbers. Because these quantities are determined by the pressure of dense matter, gravitational-wave observations are also sensitive, albeit indirectly, to underlying nuclear-matter parameters such as the density dependence of the symmetry energy. The GW170817 measurement favored relatively small tidal deformabilities and therefore disfavored the stiffest nucleonic EoS models with large radii~\citep{Abbott2019}. At the same time, the measurement remains consistent with a broad range of moderately soft hadronic models and does not yet uniquely determine whether exotic phases are present, as discussed in Sec.~\ref{sec:Chapter2}. A larger sample of events, combined with improved waveform modeling, will be needed to sharpen these constraints.

Figure~\ref{fig:fig7} illustrates the accumulated tidal dephasing during the late inspiral for two nucleonic EoS models differing in the slope of the symmetry energy $L_{\rm sym}$. The figure demonstrates that variations in the underlying nuclear interaction, even at fixed overall EoS class, can leave measurable imprints on the inspiral waveform through their effect on the stellar tidal response. The agreement between numerical-relativity waveforms and the {\tt IMRPhenomXAS-NRTidalv3} approximant confirms that semi-analytic models capture the dominant tidal phase evolution, while residual differences at late times reflect higher-order finite-size effects and the onset of non-linear dynamics approaching merger.

\subsection{Effects of Eccentricity  and Spin}

Gravitational radiation efficiently circularizes binaries formed through
the isolated channel; to leading order, the eccentricity decreases as
$e\propto a^{19/12}$ as the orbit shrinks~\citep{Peters1964}.
Consequently, field BNS systems are expected to be nearly circular by
the time they enter the LIGO/Virgo/KAGRA band. In contrast, binaries
formed dynamically in dense stellar environments may retain measurable
eccentricity at $\sim 10\,\mathrm{Hz}$. Detection of such eccentricity
would provide evidence for a non-standard formation channel.

Spin affects the waveform through spin-orbit and spin-spin couplings.
The leading spin-orbit term enters at 1.5PN order and depends on the
projection of the component spins along the orbital angular momentum.
Spin-spin effects enter at higher order. If the spins are misaligned with
the orbital angular momentum, the orbital plane precesses, producing
amplitude and phase modulations that can help break parameter
degeneracies.

In BNS systems formed through the isolated binary channel, the second-born
neutron star is expected to rotate slowly, while the first-born neutron
star may have been recycled by accretion during mass transfer. Observed
Galactic BNS systems typically have spin periods of tens of milliseconds,
corresponding to dimensionless spins
$\chi=cJ/GM^2\lesssim 0.05$. Thus spin effects are usually smaller than
in binary black-hole systems, but they are not entirely negligible. If
ignored, they can bias estimates of the masses and tidal deformabilities.

The connection with the EoS enters through rotational finite-size effects.
A spinning neutron star is not described only by its mass and angular
momentum: rotation changes the stellar shape and generates a spin-induced
quadrupole moment, schematically
$Q_{\rm spin}=-\kappa_{\rm spin} \chi^2 M^3$ in geometrized units, where the
dimensionless coefficient $\kappa_{\rm spin}$ depends on the stellar compactness and
internal density profile, and therefore on the EoS. Similarly, the moment
of inertia $I$ is controlled by the stellar mass distribution. These
quantities are small corrections for the slowly rotating stars expected
in most BNS inspirals, but they provide the rotational analogue of tidal
deformability as probes of dense-matter structure.

Rotation also modifies the tidal response through spin-tidal couplings
and rotational tidal Love numbers, which depend on the internal structure
of the star~\citep{Pani2015}. For the small spins typical of Galactic BNS
systems, these effects are expected to be subdominant for current
detectors, but they may become relevant for high signal-to-noise events
or third-generation observatories.

A useful simplification is provided by nearly universal relations. The
moment of inertia, tidal deformability, and spin-induced quadrupole moment
obey the so-called $I$-Love-$Q$ relations~\citep{Yagi2013}. Although each
quantity separately depends on the EoS, their mutual relations are only
weakly EoS-dependent. These relations are useful in parameter estimation
because they reduce the number of independent matter-dependent parameters
and provide consistency checks on neutron-star structure.

Rapid rotation is not expected to be a dominant property of the
pre-merger stars in standard BNS inspirals. Its main role in the present
context is through slow-rotation corrections to the inspiral waveform.
The qualitatively different problem of a rapidly and differentially
rotating merger remnant is discussed in the following section.

\section{Merger Dynamics and Remnant Structure}
\label{sec:Chapter4}

The merger phase of a BNS system marks the transition from the quasi-adiabatic inspiral to a strongly nonlinear regime governed by strong-field gravity, relativistic hydrodynamics, finite-temperature microphysics, and generation of strong magnetic fields. Once the stars reach merger, the system undergoes rapid compression, shock formation, and angular-momentum redistribution, leading -- depending on the total mass and the EoS -- to either prompt collapse to a black hole or the formation of a massive neutron-star remnant.
The studies of these phase lie at the interface of GW astronomy, nuclear physics at supranuclear densities, and the heavy element production models based on nucleosynthesis.

The observational confirmation of this picture came with GW170817 and its electromagnetic counterpart
AT2017gfo \citep{Cowperthwaite2017,Pian2017,Abbott2019}. This event provided observational support for a sequence of phenomena that had been predicted theoretically and modeled numerically for several decades. The main stages of this sequence are now broadly supported observationally, although many details of the merger phase remain uncertain. It is precisely in this short, violent stage that much of the dense-matter physics becomes accessible.

A useful schematic decomposition of the remnant energy budget is
%--------------------------
\bea
E_{\rm tot} \simeq E_{\rm grav} + E_{\rm th} + E_{\rm rot},
\eea
%--------------------------
where $E_{\rm grav}$ denotes the magnitude of the gravitational binding energy,
$E_{\rm th}$ the internal energy generated mainly by shocks and compression,
and $E_{\rm rot}$ the rotational kinetic energy associated with rapid
differential rotation. The gravitational binding energy is the dominant
contribution, of order $GM^2/R\sim10^{53}$ erg for a remnant of mass
$\sim2.7\,M_\odot$ and radius $\sim12$ km. The thermal and rotational
components are smaller, but still large, typically
$\sim10^{52}$--$10^{53}$ erg, and can therefore delay collapse for
dynamically important times.

The merger occurs on a millisecond timescale. During this interval, matter
reaches densities above nuclear saturation density,
$\rho_0\simeq2.7\times10^{14}\,\mathrm{g\,cm^{-3}}$, temperatures of tens of
MeV, and fluid velocities that are a substantial fraction of the speed of
light. These conditions make the merger phase especially sensitive to the
finite-temperature, high-density EoS discussed in Sec.~\ref{sec:Chapter2}.
They also probe a regime that is not accessible in cold neutron stars or in
terrestrial experiments, making BNS mergers a unique laboratory for dense
matter.

At the extreme densities and temperatures reached during and after merger, the remnant may undergo a transition from hadronic matter to deconfined quark matter even if the inspiraling stars were initially purely hadronic. Numerical-relativity simulations incorporating first-order hadron--quark phase transitions have shown that the appearance of quark matter can substantially modify the remnant dynamics by softening the equation of state at high density, accelerating contraction, and in some cases triggering delayed collapse to a black hole~\citep{Most2019,Most2020}. The formation of a quark core also affects the post-merger gravitational-wave signal, leading to shifts in the dominant oscillation frequencies and potentially producing characteristic spectral features that differ from purely hadronic models. Because the remnant probes densities and temperatures inaccessible in isolated neutron stars, post-merger observations may provide one of the most promising avenues for detecting signatures of deconfined matter in compact stars.

\medskip

%------------------------------------- Fig. 8
\begin{figure}[!tbp]
\centering
\includegraphics[width=1.\linewidth,height=4.5cm]{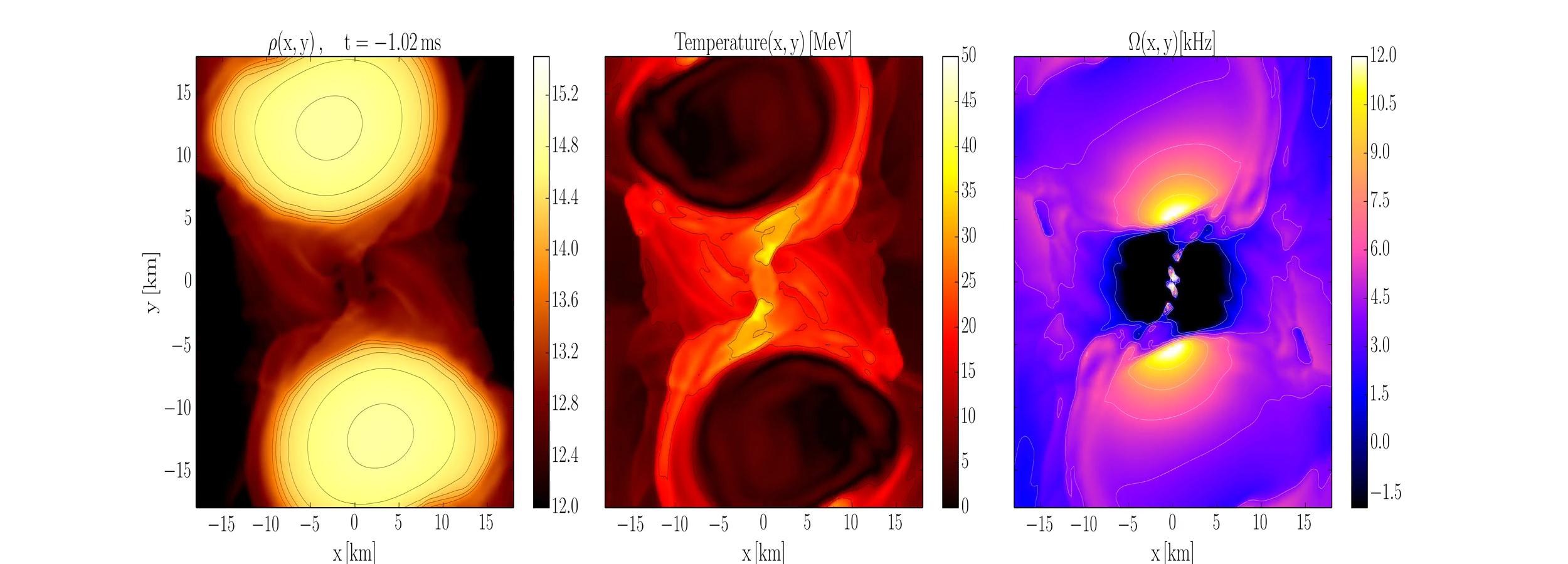}\\
\vskip 0.01cm
\includegraphics[width=1.\linewidth,height=4.5cm]{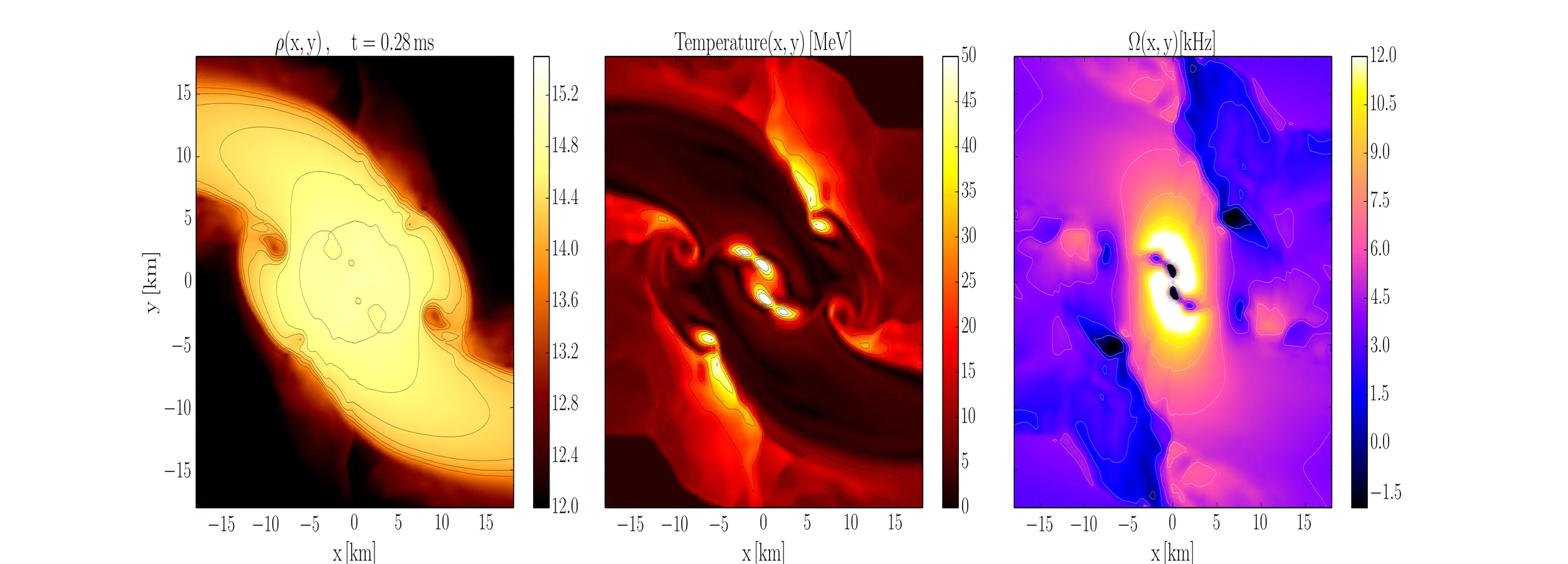}\\
\vskip 0.01cm
\includegraphics[width=1.\linewidth,height=4.5cm]{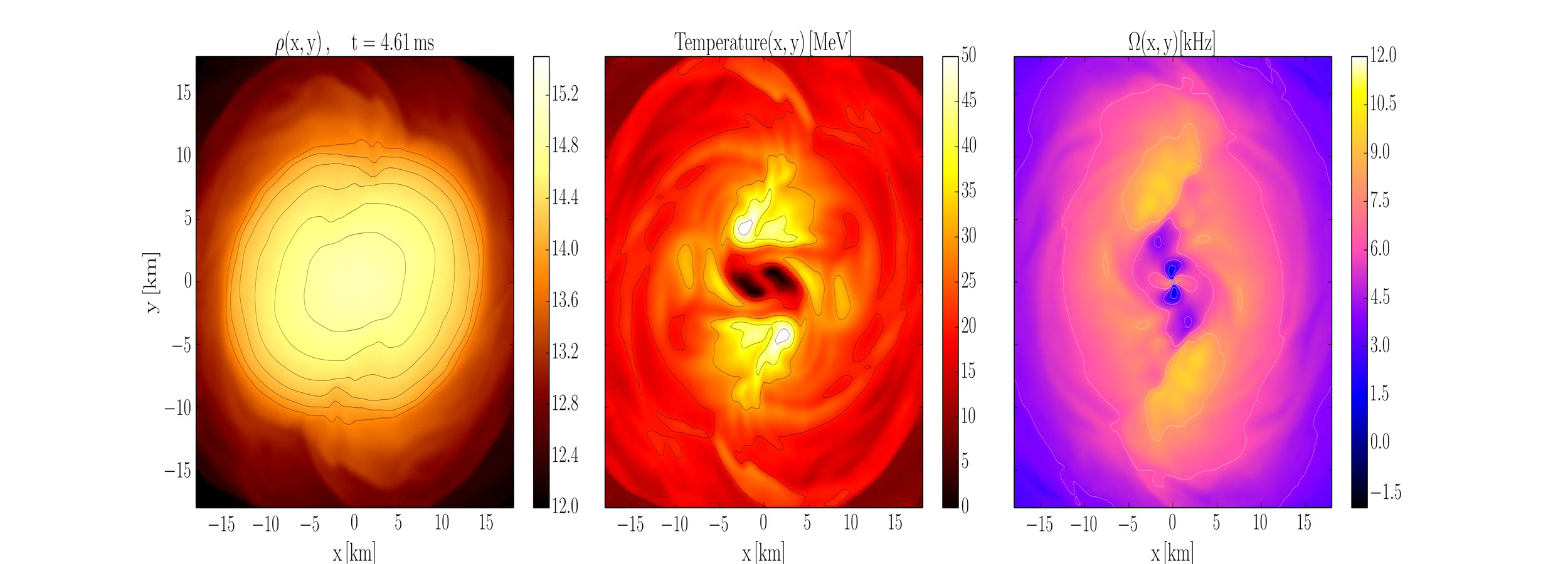}\\
\caption{Snapshots of the merger evolution in the equatorial plane at three
  representative times~\citep{Hanauske2017}.
  From left to right, each row shows the logarithm of the density, temperature (in MeV), and rotational frequency (in kHz). The top row corresponds to the regime close to first contact, where the stars begin to tidally deform and shocks start to develop at the interface. The middle row captures the early post-merger phase, characterized by strong shock heating, the formation of a highly non-axisymmetric remnant, and the emergence of tidal tails. The bottom row shows the later-stage remnant, which has relaxed toward a more axisymmetric configuration with a hot core and differential rotation, prior to its eventual collapse.}
\label{fig:fig8}
\end{figure}
%---------------------------------------

\subsection{Dynamical Phases of Merger}

The merger proceeds through several stages, each characterized by distinct
physical processes, as illustrated in Fig.~\ref{fig:fig8}. During the final
orbits before contact, tidal deformations become increasingly important. The
tidal deformability $\Lambda$ of each star, discussed in
Sec.~\ref{sec:Chapter2}, measures how strongly the star responds to the
gravitational field of its companion. As the separation decreases, the tidal
interaction grows rapidly, scaling roughly as $(R/d)^5$, where $d$ is the
orbital separation. The stellar cores remain approximately intact during this
stage, so the binary can still be viewed as two distinct objects, but the
late-inspiral waveform already carries EoS information through the accumulated
tidal phase.

At contact, two new effects appear almost simultaneously. First, a strong
shear interface forms between the two stellar fluids, which approach each
other with relative velocities of order $0.2$--$0.3\, c$. This shear layer is
unstable to the Kelvin--Helmholtz instability, producing vortices and
small-scale turbulence on sub-millisecond timescales
\citep{Price2006,Baiotti2008}. This instability is especially important for
magnetic-field amplification, as discussed in Sec.~\ref{subsec:BField}.
Second, shocks are launched at the contact interface and in the outer layers
of the merging system. These shocks convert part of the orbital kinetic energy
into heat and raise the temperature locally to tens of MeV.

After contact, the merged core undergoes rapid compression and rebound. The
remnant overshoots its equilibrium configuration and executes large-amplitude,
nonlinear quasi-radial oscillations. These oscillations generate additional
shocks and redistribute entropy throughout the remnant. At the same time, the
non-axisymmetric geometry excites strong non-radial modes, especially the
dominant quadrupolar $m=2$ mode. This mode is responsible for much of the
post-merger GW emission, because it corresponds to a bar-like
deformation radiating at a characteristic frequency discussed in
Sec.~\ref{subsec:PMGW}.

The collision also produces spiral arms and tidal tails extending from the
central remnant. These structures transport angular momentum outward during
the first few milliseconds. Part of the material in the tails becomes unbound
and forms dynamical ejecta. Although the ejected mass is usually only
$\sim10^{-3}$--$10^{-2}\,M_\odot$, its velocity, geometry, and composition
contain important information about the EoS and the thermodynamic history of
the merger. Material that remains bound may settle into the remnant disc and
participate in later secular outflows.

Within a few milliseconds, the system settles into one of several possible
remnant configurations. The subsequent evolution is governed by gravity,
thermal pressure beyond that of the cold EoS and differential rotation, which
provides centrifugal support. The dynamics is further regulated by angular-momentum
redistribution through hydrodynamic flows and magnetic stresses, accompanied
by neutrino emission and mass loss. These processes operate on different
timescales, and their relative importance depends on the EoS, the total mass,
and the mass ratio of the binary.

The mass ratio is especially important for the geometry of the merger. In unequal-mass systems, the less massive star is typically less compact and is more easily tidally disrupted, leading to stronger tidal tails and enhanced equatorial ejecta. In near-equal-mass systems, shock heating at the contact interface is often more prominent, and a transient double-core structure may form before the cores fully merge. This short-lived two-core phase efficiently excites the $m=2$ mode and can produce strong post-merger GW emission. We next discuss these aspects in turn.

The outcome of the merger is classified according to the stability of the remnant, for a schematic view see Fig.~\ref{fig:fig1}. This classification has important implications for the observables: the remnant type determines the form and structure of the GW signal, the level of neutrino irradiation, and the properties of the accretion disc, all of which ultimately affect the observed kilonova emission.

If the total mass exceeds a threshold value $M_{\rm th}$, the remnant
collapses promptly to a black hole, usually within $\lesssim1$--$2$ ms after
contact \citep{Bauswein2013}. In this case, any intermediate neutron-star-like
object is extremely short-lived, the post-merger GW signal is
strongly suppressed, and the amount of shock-heated ejecta is reduced. A black
hole surrounded by an accretion torus may still form, and the subsequent disc
evolution can drive outflows and possibly a relativistic jet. The threshold
mass is typically $M_{\rm th}\sim1.2$--$1.5\,M_{\rm TOV}$. Notably,
stiffer EoS models allow a larger overshoot above $M_{\rm TOV}$ before
collapse.

If the remnant mass exceeds the maximum mass supported by uniform rotation,
but remains below the prompt-collapse threshold, a hypermassive neutron star
(HMNS) forms~\citep{Baumgarte2000}. Such a remnant is supported mainly by
differential rotation and thermal pressure. It is intrinsically transient: as
angular momentum is redistributed and thermal energy is lost, the support is
weakened and collapse to a black hole follows, typically after
$\sim10$--$100$ ms, although the precise lifetime is model dependent. If
angular-momentum redistribution drives the remnant toward uniform rotation
without immediate collapse, the remnant may enter a supramassive phase,
provided its mass is below the maximum mass supportable by uniform rotation.
The HMNS phase is observationally important because it is associated with
strong neutrino emission, continued post-merger GW emission,
and substantial mass ejection.

A supramassive neutron star is supported by uniform rotation and can survive
for much longer timescales, up to seconds or more, before collapsing due to
angular-momentum losses or evolving toward its nonrotating counterpart, if such
a configuration exists. Finally, if the remnant mass lies below $M_{\rm TOV}$,
a stable massive neutron star may form. Such an outcome requires a
sufficiently low total mass and a sufficiently stiff EoS, and is expected to
be less common for typical Galactic BNS masses.

The prompt-collapse threshold has been studied extensively with
numerical-relativity simulations. It is closely correlated with $M_{\rm TOV}$
and with stellar compactness. Empirical fits often express
$M_{\rm th}/M_{\rm TOV}$ in terms of the compactness of the maximum-mass
nonrotating configuration,
%--------------------------
\bea\label{eq:M_th_M_TOV2}
\frac{M_{\rm th}}{M_{\rm TOV}}
\simeq
a_1-a_2\,C_{\rm TOV},
\qquad
C_{\rm TOV}
=
\frac{G M_{\rm TOV}}{R_{\rm TOV}c^2},
\eea
%--------------------------
where the coefficients $a_1$ and $a_2$ are calibrated to simulations~\citep{Hotokezaka2011,Radice2018,Koeppel2019,Tootle2021}.
Equivalent forms use stellar radii such as
$R_{1.6}$ or $R_{1.8}$ as proxies for compactness, where $1.6$ and $1.8$ refer to the mass
of the star in solar units. These relations encapsulate
the key physical point: less compact stars, corresponding to stiffer EoS
models, can support larger remnant masses before collapse. Thus the remnant
outcome provides a direct probe of the high-density EoS.

This classification is also affected by magnetic fields and neutrino cooling.
Magnetic stresses can accelerate the removal of differential rotation, while
neutrino emission removes thermal support. These effects do not change the
basic classification, but they can shift the boundary between a short-lived
HMNS and a longer-lived remnant.

Immediately after merger, the remnant is strongly differentially rotating.
Numerical simulations consistently show a slowly rotating inner core
surrounded by rapidly rotating outer layers, often close to the local
Keplerian rate, as shown in the rightmost column of Fig.~\ref{fig:fig8}. This
profile reflects the merger dynamics: angular momentum from the two stellar
cores is deposited preferentially in the outer remnant and disc. Differential
rotation provides substantial centrifugal support and allows the remnant to
temporarily exceed the maximum mass of a cold nonrotating star.

The rotational energy may be estimated schematically as
%--------------------------
\bea
E_{\rm rot} \simeq
\frac{1}{2}
\int \rho(r)\,\Omega^2(r)\,r^2\,dV,
\eea
%--------------------------
where $\rho$ is the mass density, $\Omega$ the local angular velocity, and the
integration is over the volume of the remnant. For characteristic remnant
densities $\bar{\rho}\sim10^{14}$--$10^{15}\,\mathrm{g\,cm^{-3}}$, radii
$R\sim15$ km, and angular velocities $\Omega\sim10^4\,\mathrm{rad\,s^{-1}}$,
one obtains $E_{\rm rot}\sim10^{52}$--$10^{53}$ erg. This is a significant
fraction of the binding energy and explains why differential rotation can
delay collapse.

Angular momentum is redistributed by several mechanisms. Initially,
hydrodynamic flows and non-axisymmetric pressure gradients exert torques on
fluid elements and transport angular momentum outward. Spiral density waves
transport angular momentum from the inner remnant to the disc and ejecta on
millisecond timescales. GWs also remove angular momentum from
the system, particularly while the remnant remains strongly non-axisymmetric.
This channel is most efficient in the first few tens of milliseconds and
weakens as the remnant becomes more axisymmetric.

Magnetic effects become increasingly important at later times. Differential
rotation winds poloidal magnetic fields into toroidal fields, with a growth
rate schematically given by
$\partial B_\phi/\partial t\sim rB_r\,d\Omega/dr$. The resulting Maxwell
stresses transport angular momentum outward. In regions where
$d\Omega/dr<0$, especially the outer remnant and disc, the magnetorotational
instability (MRI) can generate sustained turbulence and an effective viscosity
\citep{Balbus1991,Siegel2014}. Mass outflows provide an additional
angular-momentum sink, since matter ejected from the outer layers carries
large specific angular momentum.

The combined effect of these processes is to reduce differential rotation. For A hypermassive remnant collapses to a black hole, once differential rotation is sufficiently weakened and thermal support is reduced. The surrounding disc continues to evolve through the same angular-momentum transport mechanisms and provides the reservoir for later winds and possible jet formation.

Shock heating is one of the defining features of the merger and has important
consequences for both remnant stability and electromagnetic emission. The
collision and subsequent quasi-radial oscillations convert orbital kinetic
energy into thermal energy. Temperatures of $30$--$50$~MeV can be reached in
the most strongly shocked regions of the remnant core, while the outer layers
and disc typically reach $\sim5$--$20$ MeV, as illustrated in
Fig.~\ref{fig:fig8}.

It is useful to decompose the pressure into cold and thermal contributions,
%--------------------------
\bea
P = P_{\rm cold}(\rho,Y_i) + P_{\rm th}(\rho,T,Y_i),
\eea
%--------------------------
where $P_{\rm cold}$ is the zero-temperature contribution and $P_{\rm th}$ is
generated by heating during the merger. Here $Y_i$ denotes the relevant
composition variables, including the electron fraction and, in
neutrino-trapped regions, neutrino fractions.

A common approximate description casts the thermal pressure as
%--------------------------
\bea
P_{\rm th} \approx (\Gamma_{\rm th}-1)\rho\,\epsilon_{\rm th},
\eea
%--------------------------
where $\epsilon_{\rm th}$ is the specific thermal energy. For a
non-interacting gas in the nonrelativistic limit $\Gamma_{\rm th}=5/3$, while
in the ultrarelativistic case $\Gamma_{\rm th}=4/3$. Dense nuclear matter
lies between these limits, and the effective thermal index is both density
dependent and EoS dependent. This uncertainty is important because thermal
pressure can delay collapse and modify the post-merger GW
signal.

Thermal support is temporary. Neutrino emission removes energy and entropy
from the remnant on timescales of tens to hundreds of milliseconds, comparable
to or longer than the lifetime of a hypermassive neutron star. In the dense
central regions, neutrino mean free paths are much shorter than the size of
the remnant, so cooling proceeds diffusively rather than by free streaming. A
schematic estimate of the diffusion time is
%--------------------------
\bea
\tau_\nu \sim \frac{R^2}{c\,\ell_\nu},
\eea
%--------------------------
where $\ell_\nu$ is the neutrino mean free path. Neutrino cooling therefore
contributes to delayed collapse, but does not remove thermal support
instantaneously. The underlying opacity and transport physics are discussed in
Sec.~\ref{sec:Chapter6}.

Shock heating also affects the composition of the ejecta. Material heated to
several MeV undergoes rapid weak-interaction processing, and its electron
fraction can increase relative to that of cold neutron-star matter. Thus
shock-heated ejecta are typically less neutron rich than cold tidal ejecta,
which has important consequences for kilonova emission and
$r$-process nucleosynthesis, as discussed in Sec.~\ref{sec:Chapter7}.

\subsection{Remnant Lifetime and Post-Merger GW Emission}
\label{subsec:PMGW}

The remnant lifetime is highly sensitive to the high-density EoS. The EoS
determines the stellar radii, the maximum mass, the pressure profile, and the
resistance of matter to compression. These quantities together control whether
the remnant collapses promptly, survives as a hypermassive neutron star, or
evolves into a longer-lived configuration.

A stiff EoS generally produces larger stellar radii, smaller compactness,
greater pressure support, and a higher prompt-collapse threshold
$M_{\rm th}$. Conversely, a soft EoS leads to more compact stars and a lower
threshold mass. As a result, binaries with total masses similar to GW170817
($M_{\rm tot}\simeq2.7\,M_\odot$) may undergo prompt collapse for sufficiently
soft EoS models, while forming long-lived remnants for stiffer ones. As
discussed above, see Eq.~\eqref{eq:M_th_M_TOV2}, the threshold mass is
tightly correlated with $M_{\rm TOV}$ and stellar compactness.

The remnant lifetime has direct consequences for all major observables. A
prompt-collapse event produces little or no post-merger
GW signal, whereas a long-lived remnant emits in the kilohertz
band for tens of milliseconds or more. A long-lived neutron-star remnant also
sustains neutrino luminosities of order $10^{53}$~erg~s$^{-1}$,
increasing the electron fraction of the surrounding ejecta. In contrast,
prompt collapse rapidly suppresses neutrino irradiation and tends to yield
more neutron-rich outflows. Shock-driven ejecta and neutrino-heated winds are
likewise reduced in prompt-collapse systems.
For GW170817, the presence of both blue and red kilonova components suggests
that the remnant did not collapse immediately, although its precise lifetime
remains uncertain.

If the remnant avoids prompt collapse, it emits a strong post-merger GW signal at frequencies of roughly $1$--$4$ kHz, see Fig.~\ref{fig:fig7}. This signal is generated by the non-axisymmetric oscillations of the remnant and carries information about its size, rotation profile, temperature, and composition. It therefore probes the EoS at densities and temperatures not accessible during the inspiral. The post-merger spectrum is richer than the inspiral spectrum~\citep{Bauswein2012,Hotokezaka2013,Takami2015,Bernuzzi2015a,Rezzolla2016}. Instead of a chirp, it consists of several quasi-monochromatic peaks superposed on a broader background. The dominant feature is the peak frequency $f_{\rm peak}$, also denoted $f_2$, associated with the fundamental quadrupolar oscillation of the remnant. This is the same $l=m=2$ deformation excited during merger and responsible for most of the post-merger GW luminosity.

Schematically,
%-----------------------------
\bea\label{eq:f_peak_schematic}
f_{\rm peak} \simeq f(R,\,M_{\rm tot},\,{\rm EoS}),
\eea
%-----------------------------
and it may be understood roughly as twice a characteristic remnant rotation frequency \citep{Bauswein2012, Hotokezaka2013,Takami2015,Bernuzzi2015a,Rezzolla2016}; note that additional parameters beyond those indicated in Eq.~\eqref{eq:f_peak_schematic} such as mass ratio, spin, and magnetic field may influence $f_{\rm peak}$.

More compact remnants rotate faster and therefore have larger $f_{\rm peak}$. This leads to empirical correlations between $f_{\rm peak}$ and stellar radii such as $R_{1.6}$ or $R_{1.8}$, depending on the binary mass range. These correlations have been established in numerical-relativity simulations using a range of EoS models and form one of the most useful connections between post-merger dynamics and cold neutron-star structure.

A representative fit has the form
%-----------------------------
\bea
f_{\rm peak} \simeq A - B\,R_{1.6},
\eea
%-----------------------------
with $A\sim6$--$7$ kHz and $B\sim0.1$--$0.2$~kHz~km$^{-1}$, depending on the total mass and fit procedure. Thus a remnant with $R_{1.6}\simeq12$ km typically emits near $f_{\rm peak}\sim2.5$--$3.5$ kHz. A measurement of $f_{\rm peak}$ with an accuracy of order $100$ Hz could constrain the stellar radius at the kilometer level.

Subdominant spectral features contain additional information. A
lower-frequency peak, often denoted $f_1$ or $f_-$, can arise from coupling
between the quasi-radial oscillation and the $m=2$ mode. A higher-frequency
feature may appear near the sum of this lower frequency and $f_{\rm peak}$.
Spiral arms in the early post-merger phase also contribute power at lower
frequencies during the first few milliseconds. Unequal-mass systems tend to
produce more asymmetric spectra and stronger secondary features.

The main observational challenge of post-merger GW
observations is detector sensitivity at kilohertz frequencies. Current
LIGO/Virgo/KAGRA sensitivity is generally insufficient to detect post-merger
emission from events at GW170817-like distances. Third-generation detectors
such as the Einstein Telescope \citep{Abac2026} and Cosmic Explorer
\citep{Evans2021} are expected to improve this situation substantially. A
detection of $f_{\rm peak}$ from a well-localized event would provide a direct
probe of supranuclear-density matter at finite temperature.

\subsection{Magnetic Field Amplification}
\label{subsec:BField}

Magnetic fields are usually dynamically subdominant during the inspiral, with typical pre-merger strengths in observed neutron stars of $10^8$--$10^{12}$ G. During and after merger, however, they can be amplified to magnetar-level strengths, locally reaching $\gtrsim10^{15}$ G~\citep{Price2006,Kiuchi2014,Kiuchi2015}. At such strengths, magnetic fields influence angular-momentum transport, mass ejection, and possibly jet
formation~\citep{Duez2006PRL,Duez2006PRD,Ruiz2016}.

The first amplification mechanism is the Kelvin--Helmholtz instability at the shear interface between the merging stars. The growth rate scales as $\gamma_{\rm KHI}\sim\Delta v/\lambda_{\rm KHI}$, where $\Delta v$ is the velocity shear and $\lambda_{\rm KHI}$ the wavelength of the unstable mode. Since $\Delta v\sim0.2$--$0.3\,c$, the instability grows extremely rapidly.  High-resolution MHD simulations show that seed fields can be amplified by several orders of magnitude within the first millisecond.  In principle, the growth saturates when the magnetic energy density approaches the turbulent kinetic-energy density. This gives an equipartition field strength
%-----------------------------
\bea
B_{\rm eq}
\sim
(4\pi\rho v_{\rm turb}^2)^{1/2}
\sim
10^{15}\text{--}10^{16}\ {\rm G},
\eea
%-----------------------------
where $\rho$ is the mass density and $v_{\rm turb}$ is the characteristic turbulent velocity in the remnant. The numerical range corresponds to typical post-merger conditions, with densities below or around nuclear saturation density in the shear layer and outer remnant, and large near-relativistic turbulent velocities.

After the initial turbulent amplification, differential rotation winds
poloidal magnetic fields into toroidal fields on a timescale comparable to the
rotation period. If the toroidal field becomes sufficiently strong, magnetic
buoyancy may bring magnetic flux toward the surface and contribute to a
dynamo-like cycle. In the outer remnant and disc, the MRI can drive sustained turbulence
whenever $d\Omega/dr<0$. Its growth rate is of order $\Omega$, while the
fastest-growing wavelength is
\bea
\lambda_{\rm MRI}\sim \frac{v_A}{\Omega},
\qquad
v_A=\frac{B}{\sqrt{4\pi\rho}},
\eea
where $v_A$ is the Alfvén velocity, $B$ is the magnetic-field strength,
and $\rho$ is the mass density. For $B\sim10^{15}\,{\rm G}$ and typical
remnant densities, this wavelength can be only tens of meters, which
remains difficult to resolve in global simulations.

Magnetic stresses transport angular momentum through the Maxwell stress
tensor. Schematically,
%-----------------------------
\bea
\dot{J}_{\rm mag}
\simeq
\int r\,T^{r\phi}_{\rm EM}\,dA,
\eea
%-----------------------------
where $T^{r\phi}_{\rm EM}$ is the relevant magnetic stress. These stresses
become especially important after the initial hydrodynamic phase, when
turbulence and differential rotation have built up strong magnetic fields.

Magnetic fields also help launch outflows. A spinning black hole surrounded by
a magnetized disc can power a relativistic jet through the Blandford--Znajek
mechanism~\citep{Blandford1977}, which is a leading model
for sGRBs~\citep{Rezzolla2011,Ruiz2018}. If instead a neutron-star remnant survives, a
magnetized wind from the rotating remnant can inject energy into the
surrounding ejecta and may contribute to extended emission or enhanced
kilonova luminosity.

\subsection{Mass Ejection and Connection to Kilonovae}

Matter is ejected during and after the merger through several channels with
different timescales, geometries, and velocities. These ejecta provide the
material for kilonova emission and for subsequent $r$-process nucleosynthesis,
but their basic properties are set by the merger dynamics.

The first component is dynamical ejecta, launched on millisecond timescales.
Tidal ejecta originate from the outer layers of the neutron stars and are
expelled mainly near the orbital plane. This channel is enhanced in
unequal-mass systems, where the less massive star is more easily disrupted.
Shock-driven ejecta originate from the heated interface between the two stars
and are often launched toward higher latitudes. Near-equal-mass systems tend
to produce relatively stronger shock-heated ejecta, while unequal-mass systems
favor tidal tails. The total dynamical ejecta mass is typically
$\sim10^{-4}$--$10^{-2}\,M_\odot$, with substantial dependence on the mass
ratio, total mass, and EoS.

The second major component is the post-merger disc or secular outflow. After
merger, an accretion disc or torus of mass
$M_{\rm disc}\sim10^{-2}$--$10^{-1}\,M_\odot$ may surround the remnant or
black hole. Over tens of milliseconds to seconds, this disc ejects matter
through viscous heating, magnetic stresses, neutrino irradiation, and nuclear
recombination \citep{Fernandez2013,Siegel2017}. These outflows can remove a
substantial fraction of the initial disc mass and provide an important
contribution to the total ejecta budget.

The remnant outcome strongly affects all ejecta channels. Prompt collapse
suppresses shock-driven ejecta and rapidly reduces neutrino irradiation,
leading to more neutron-rich outflows. A longer-lived neutron-star remnant, by
contrast, sustains shocks and neutrino emission, increasing the electron
fraction of polar and disc-wind material. Thus the lifetime of the remnant
controls the relative amount of lanthanide-poor and lanthanide-rich ejecta and
therefore the relative strength of blue and red kilonova components. The
detailed nucleosynthesis and opacity connection are discussed in
Sec.~\ref{sec:Chapter7}.

The kilonova associated with GW170817 showed evidence for both blue and red
components, indicating a multi-component ejecta structure. The blue component
points to relatively high-$Y_e$, lanthanide-poor material, likely produced in
polar outflows exposed to neutrino irradiation, while the red component
requires neutron-rich, lanthanide-bearing ejecta. This suggests that the
remnant survived long enough to irradiate part of the outflow, but not so long
as to eliminate neutron-rich material entirely
\citep{Cowperthwaite2017,Pian2017}. This connection makes the dynamics of the
merger important not only to compact-object astrophysics, but also to the origin
of the heavy elements.

\subsection{Future of Multi-Messenger Modeling of Merger Remnants}

The merger phase is the link between the quasi-adiabatic inspiral and the
observable post-merger aftermath. It determines whether the remnant collapses
promptly or survives temporarily, how much matter is ejected, how angular
momentum and heat are redistributed, and what gravitational-wave and
electromagnetic signals are produced. Each observable depends on a different
combination of microphysical and macrophysical inputs, so the strongest
constraints come from combining messengers.

From the nuclear-physics perspective, the merger probes matter at densities of
a few times $n_0$, temperatures of $10$--$50$ MeV, rapid rotation, and
strong departure from weak equilibrium. This region of the phase diagram is
not probed by cold neutron-star observations or by terrestrial experiments.
The finite-temperature EoS enters through thermal pressure, composition, and
neutrino opacity, and its effects appear in the remnant lifetime,
post-merger gravitational-wave frequency, ejecta composition, and kilonova
color.

A sufficiently well-observed event can in principle constrain the tidal
deformability from the inspiral, the prompt-collapse threshold from the
remnant lifetime, the post-merger peak frequency from the kilohertz
gravitational-wave spectrum, and the ejecta mass and composition from
kilonova light curves and spectra. These quantities together constrain the
EoS over a broad range of densities and temperatures.

The same multimessenger picture also connects the remnant to relativistic jet
formation. If a jet launches successfully, its delay, opening angle, Lorentz
factor, and afterglow structure depend on the disc and on the lifetime of the
central remnant. Thus short gamma-ray bursts and afterglows provide another
indirect route to the EoS and remnant structure.

Looking ahead, third-generation gravitational-wave detectors, improved optical
and X-ray facilities, and more realistic numerical simulations will be
essential for extracting this physics. In particular, full general-relativistic
MHD simulations incorporating spectral neutrino transport and
finite-temperature EoS tables will be needed for quantitative interpretation
\citep{Foucart2016,Radice2018a}. The discussion above underlines that the merger
phase lies at the intersection of nuclear astrophysics, gravitational-wave
astronomy, and transient electromagnetic astronomy, and provides the physical
bridge to the kilonova phenomena discussed in the following chapters.

%\newpage
\section{Transport Properties and Dissipation}
\label{sec:Chapter5}
Transport phenomena in dense matter provide the connection between microscopic interactions and macroscopic dynamics in neutron stars and BNS mergers. In these systems, matter is driven away from equilibrium by rapid compression, strong velocity shear, and weak interaction processes operating on dynamical timescales. Dissipative effects then govern the relaxation toward local thermodynamic equilibrium and, in doing so, influence the evolution of the merger remnant, the damping of oscillations, and the observable multimessenger signals.

Transport in compact star matter has traditionally been studied in the cold (essentially zero-temperature) and dense regime, where the constituents form degenerate quantum liquids. This regime is relevant for mature isolated or accreting neutron stars, as well as for the interiors of white dwarfs. The problem of transport in the BNS merger context differs substantially from this traditional picture. The densities involved extend up to several times the nuclear saturation density $n_0$
and place the system in a strongly interacting and only partially degenerate regime, where perturbative descriptions are of limited applicability and must be supplemented by nuclear many-body theory. At the same time, temperatures of order $T \sim 10-80$~MeV are sufficiently high to significantly modify the transport coefficients relative to the cold neutron-star limit. The relevant matter is therefore neither a cold, fully degenerate Fermi liquid in strict equilibrium nor a weakly coupled thermal gas, but rather a hot, dense, strongly interacting fluid.

Relativistic hydrodynamics and MHD form the basis for the large-scale description of dense matter in BNS mergers~\citep{RezzollaZanotti2013}. In the dissipative case, these frameworks require transport coefficients as essential inputs. These coefficients encode the underlying microphysics of the matter--such as scattering rates and cross sections, correlations, and background composition--and determine, for example, the dissipation of currents and the decay of magnetic fields, the dispersion of plasma waves, etc. Consequently, understanding dissipative MHD is essential to the interpretation of a broad range of observational signatures.

Relativistic MHD involves several key dissipative transport coefficients: shear (first) viscosity $\eta$, bulk (second) viscosity $\zeta$, thermal conductivity $\kappa$, and diffusion coefficients associated with conserved charges in multicomponent systems, as well as electrical conductivity in charged matter. These coefficients characterize the response of matter to gradients in velocity, density and composition, temperature, chemical potentials, and electromagnetic fields. In the strong magnetic fields expected in BNS mergers, scalar transport coefficients become tensorial, giving rise to effects such as
electrical and thermal Hall effects. Furthermore, the simultaneous presence of electric fields and temperature gradients leads to cross-coupling phenomena, such as thermoelectric effects (e.g., thermopower).

The transport coefficients depend sensitively on the composition of matter--whether nucleonic, hyperonic, or quark--and on the dominant strong- and weak-interaction processes at a given density and temperature. Their determination is part of the broader program of studying the properties of strongly interacting matter, including its composition and EoS, with the ultimate goal of connecting microphysics to observable merger dynamics. Weak interactions, particularly those involving neutrinos, play a crucial role in both composition and transport in neutron star mergers. These interactions are fast enough to modify the composition on relevant timescales, but not sufficiently fast to maintain overall exact beta equilibrium. As a result, the remnant can sustain chemical imbalances generated by compression, expansion, and fluid motion. This competition between macroscopic driving and microscopic equilibration is the basic origin of bulk viscosity, discussed in detail below, and is one of the clearest examples of how dense-matter microphysics enters the GW signal.

\subsection{Bulk viscosity}

Bulk viscosity has been studied extensively in the BNS context.  Consider first the simplest case of composition, i.e., nuclear matter composed of neutrons, protons, electrons, and muons~\citep{Alford2019,Alford2021b,Most2024}. The leptonic equilibration processes are too slow to be included in the dynamics; the contribution of muons to bulk viscosity is discussed by \cite{Alford2022,Alford2023}. In any event they always contribute to the static thermodynamic quantities such as susceptibilities.

For temperatures above the trapping temperature,
$T_{\rm tr}\simeq 5$ MeV~\citep{Alford2018a}, neutrinos are trapped and beta equilibrium is
established through Urca  (neutron $\beta$-decay and electron capture) processes,
% -----------------------
\bea\label{eq:Urca_def}
n\rightleftarrows p+e^-+\bar\nu_e,\qquad
p+e^-\rightleftarrows n+\nu_e .
\eea
%----------------
The corresponding equilibrium condition is
$
\mu_p+\mu_e=\mu_n+\mu_\nu 
$
where $\mu_n$, $\mu_p$, $\mu_e$, and $\mu_{\nu_e}$ are the chemical
potentials of neutrons, protons, electrons, and electron neutrinos.
At given $T$, baryon density $n_B=n_n+n_p$, and lepton densities
$n_{L_l}=n_l+n_{\nu_l}=Y_{L_l}n_B$, the particle fractions are fixed by
this condition, its muonic analogue
$\mu_p+\mu_\mu=\mu_n+\mu_{\nu_\mu}$, and charge neutrality
$n_p=n_e+n_\mu$. If  muon reactions are neglected, the lepton fractions
must be specified separately for each flavor.

In cold $npe$ matter direct Urca is kinematically allowed only above the proton-fraction threshold $Y_p \simeq 1 / 9$; the threshold is modified in $npe\mu$ matter.
In the low-temperature regime and for low proton fractions (typically $Y_p\le 1/9$ for $npe$ matter) the direct Urca is suppressed. The modified Urca reactions provide the next available channel,
%-------------------------------------
\bea
n+N \to p+N+e^-+\bar{\nu}_e,
\qquad
p+N+e^- \to n+N+\nu_e,
\eea
%------------------------------------
where $N$ is a spectator nucleon that absorbs the excess momentum. These reactions are slower because of the additional particle in the initial or final state and the corresponding reduction of phase space. In cold neutron stars this distinction is crucial for cooling~\citep{Page2006}. In merger remnants, where the temperatures are much higher and the matter is only partially degenerate in some regions, both direct and modified processes may contribute.

For $T\lesssim T_{\rm tr}\sim 5$~MeV the matter is neutrino-transparent:
neutrinos escape, cannot appear in initial states, and the above Urca
processes proceed only in the left-to-right direction~\citep{Alford2019a,Alford2023}. The composition is
then usually determined from the zero-temperature equilibrium conditions
$\mu_n=\mu_p+\mu_e$ and $\mu_\mu=\mu_e$. Finite-temperature corrections
become important for $T\gtrsim 1$ MeV, but have only
a minor impact on the bulk viscosity in the transparent regime.

Compression and rarefaction drive matter out of beta equilibrium. The
chemical imbalance in neutrino trapped matter is
\bea
\mu_\Delta=\mu_n+\mu_\nu-\mu_p-\mu_e ,
\eea
which vanishes in equilibrium. For small-amplitude density oscillations
with frequency $\omega$, baryon and lepton conservation imply
$
\delta n_j(t)=-({\theta}/{i\omega}) n_{j0}, \quad j=n,p,e,\nu, 
$
where $\theta$ is the velocity divergence. Particle-density perturbations
are decomposed as
$n_j(t)=n_{j0}+\delta n_j(t),$ and
$\delta n_j=\delta n_j^{\rm eq}+\delta n'_j ,$ 
where $\delta n_j^{\rm eq}$ is the instantaneous beta-equilibrium shift,
and $\delta n'_j$ is the non-equilibrium part.
Because $\mu_\Delta$ is nonzero, the continuity equations acquire source terms from beta reactions. These terms enter with a plus sign for $p$ and $e$ and with a minus sign for $n$ and $\nu$; for example,
%-------------------------
\bea
\frac{\partial}{\partial t}\delta n_n
=-\theta n_{n0}-\lambda_{\rm Urca}\mu_\Delta .
\eea
%-------------------------
Here $\lambda_{\rm Urca}>0$ is a measure of disquilibrium between the rates of neutron decay
and electron capture
$\Gamma_\Delta\equiv\Gamma_{n\to pe\bar\nu}-\Gamma_{pe\to n\nu }=\lambda_{\rm Urca}\mu_\Delta$.

The bulk viscous pressure follows from
%-------------------------
\bea
\Pi=\sum_j\frac{\partial p}{\partial n_j}\delta n'_j
=\sum_{lj}n_{l0}\frac{\partial\mu_l}{\partial n_j}\delta n'_j ,
\eea
where we used Gibbs-Duhem relation
$dp=sdT+\sum_l n_l d \mu_l$. With $\Pi=-\zeta\theta$,
one obtains
\bea\label{eq:zeta}
\zeta=\frac{C^2}{A}\frac{\gamma}{\omega^2+\gamma^2}.
\eea
% -----------------------------------
The prefactor $C^2/A$ is purely thermodynamic and determined by the EoS, whereas $\gamma=\lambda_{\rm Urca} A$ contains the microscopic weak-interaction rates; for definitions of $A$ and $C$, which, respectively, are the susceptibility associated with changing the composition and a measure of the coupling between compression and chemical imbalance, see e.g.,~\cite{Alford2023} Eqs.~(37)-(38) and (44)-(45). These equations account also for muonic counterparts of the processes
on electrons and the coupling between the leptonic sectors. 

The physical meaning of Eq.~\eqref{eq:zeta} is transparent: if $\omega\ll\gamma$, reactions are fast and the matter remains close to equilibrium, so little dissipation is produced. If $\omega\gg\gamma$, the composition is effectively frozen during an oscillation cycle, and dissipation is again suppressed. The largest dissipation occurs when the microscopic relaxation time $\gamma^{-1}$ is comparable to the macroscopic oscillation time $\omega^{-1}$.

The coefficient $\lambda_{\rm Urca}$ can be obtained from the microscopic rates of Urca processes. For neutron decay one has
%------------------------------------------------------
\bea \label{eq:Gamma1_def}
\Gamma_{n\to pe\bar\nu} &=& \int\!\! \frac{d^3p_n}{(2\pi)^32E_n} \int\!\!
\frac{d^3p_p}{(2\pi)^32E_p} \int\!\! \frac{d^3k_e}{(2\pi)^32E_e}
\int\!\! \frac{d^3k_\nu}{(2\pi)^32E_\nu}\sum \vert 
{\cal M}_{\rm Urca}\vert^2 \nonumber\\
& \times & \bar{f}(k_e)\bar{f}(p_p) \bar{f}(k_\nu) f(p_n) (2\pi)^4\delta^{(4)}(k_e+k_\nu+p_p-p_n),
\eea
% ------------------------------------------------------
where $f(p_j)=\{\exp\left[(E_j(p)-\mu)/T\right]+1\}^{-1}$ etc. are the Fermi distribution functions of particles, with $E_j(p)$ being the single-particle spectrum for momentum $p_j$, and $\bar{f}(p_j)=1-f(p_j)$. In the neutrino-transparent regime the final state neutrino/anti-neutrino Pauli-blocking can be set to 1.  The spin-averaged relativistic matrix element of the Urca processes reads
%-----------------------------------------------------------
\be\label{eq:matrix_el_full}
\sum \vert {\cal M}_{\rm Urca}\vert^2 =32 G_F^2\cos^2
\theta_c \Big[(1+g_A )^2(k\cdot p) (k'\cdot p')
+(1-g_A)^2(k\cdot p') (k'\cdot p)
+(g_A^2-1)m^*_n m^*_p(k\cdot k')\Big],
\ee 
%----------------------------------------------------------- 
where $G_F=1.166\cdot 10 ^{-5}$ GeV$^{-2}$ is the Fermi coupling constant, $\theta_c$ is the Cabibbo angle with $\cos\theta_c=0.974$, $g_A=1.26$ is the axial-vector coupling constant, and $m_n^*/m_p^*$ is the effective neutron/proton mass.  The inverse rate is obtained by interchanging $f_j\leftrightarrow\bar f_j$.  While $\lambda_{\rm Urca}$ generally must be evaluated numerically, analytic expressions in the degenerate regime are available that  make the main scaling relations explicit not far from this limit.  For neutrino-trapped matter (assuming equal neutron and proton effective masses $m^*$) one finds
%--------------------------------
\bea
\lambda_{\rm Urca} = \frac{1}{12\pi^3} m^{*2}\tilde G^2T^2 p_{Fe}p_{F\nu} (p_{Fp}+p_{Fe}+p_{F\nu}-p_{Fn}) ,
\eea
%--------------------------------
where $\tilde G = G_F^2 \cos ^2 \theta_c\left(1+3 g_A^2\right)$,  whereas in neutrino-transparent matter
%--------------------------------
\bea
\lambda_{\rm Urca} = \frac{17}{240\pi} m^{*2}\tilde G^2T^4p_{Fe}\, \theta(p_{Fp}+p_{Fe}-p_{Fn}) .
\eea
%--------------------------------
Thus, close to the degenerate limit the trapped-matter rate scales as
$\lambda_{\rm Urca}\propto T^2$, while the transparent-matter rate scales as
$\lambda_{\rm Urca}\propto T^4$. The step function in the latter expression
encodes the direct-Urca threshold: in neutrino-transparent matter the process is
blocked unless proton and electron Fermi momenta are sufficiently large
to satisfy momentum conservation. In neutrino-trapped matter, by contrast, the
additional neutrino Fermi momentum relaxes this constraint, and the rate
remains finite.

%------------------------------------- Fig. 9
\begin{figure}[!tbp] \begin{center} 
\includegraphics[width=0.45\linewidth,keepaspectratio]{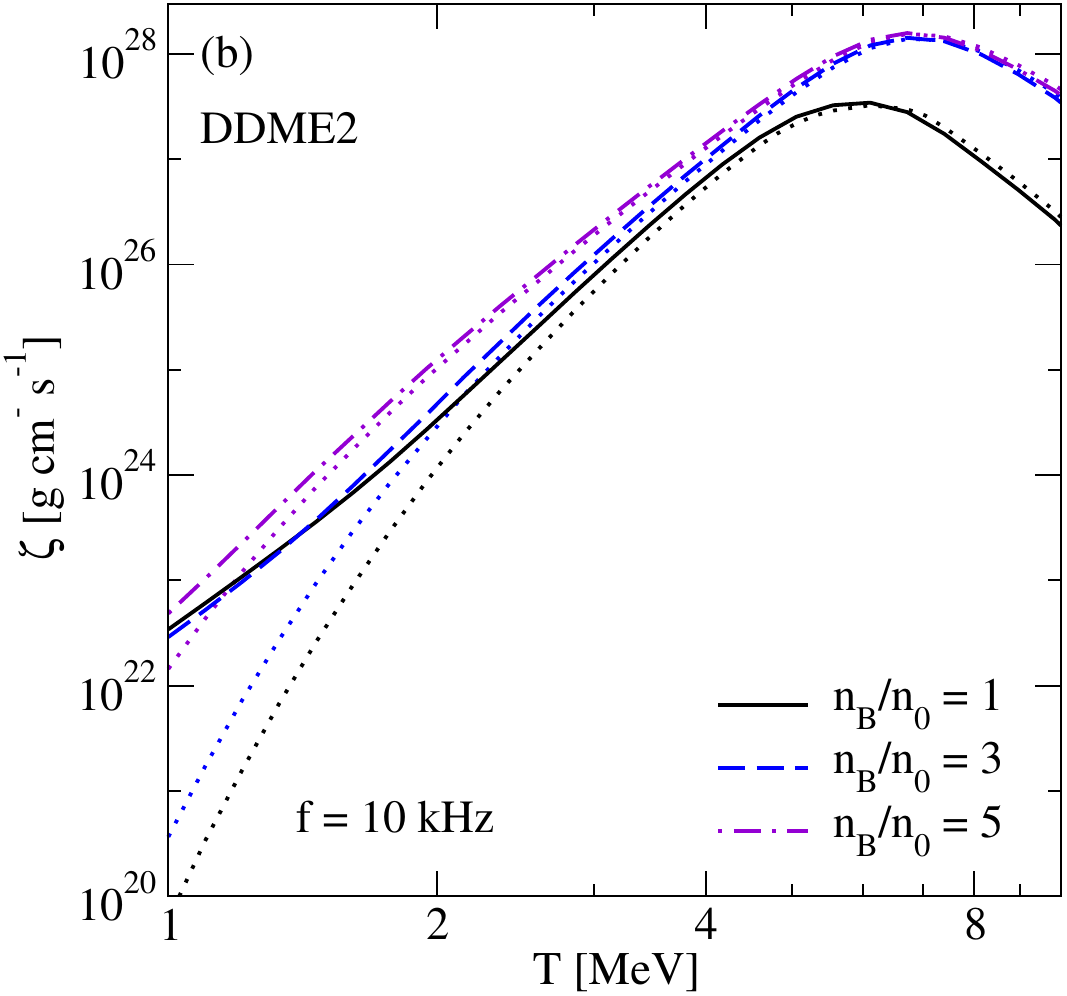}
\hspace{0.2cm} \includegraphics[width=0.45\linewidth,keepaspectratio]{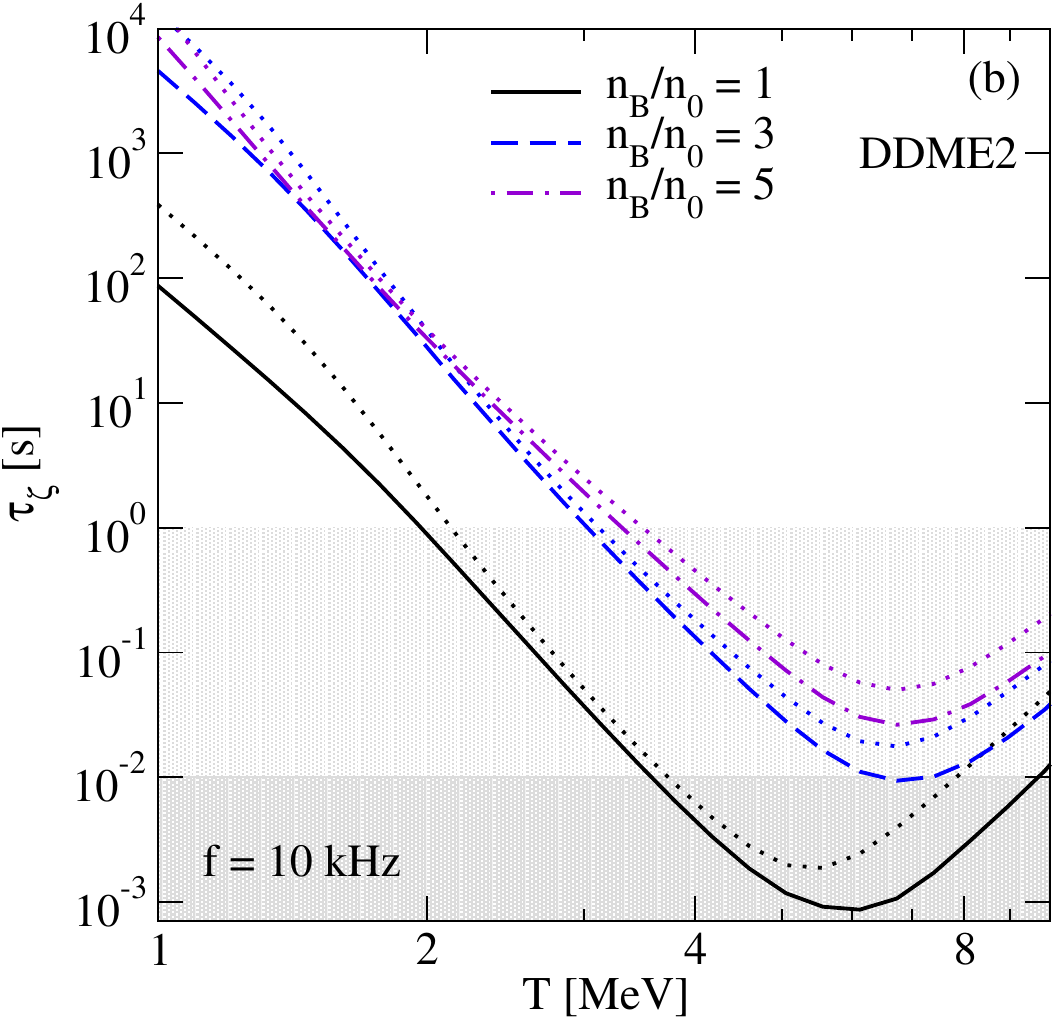} \caption{ Dependence of the bulk viscosity (left panel) and damping timescale (right panel) of relativistic $npe\mu$ matter on temperature, for several densities indicated in the plots, using the DDME2 CDF model and a frequency $f=10$ kHz~\citep{Alford2023}.  The dotted lines correspond to results obtained when modified Urca processes are neglected.
}
\label{fig:fig9}
\end{center}
\end{figure}
% ---------------------------------------
The bulk viscous damping timescale is given by~\citep{Alford2023}
%--------------------------------------------------
\bea\label{eq:damping_time}
\tau_{\zeta} =\frac{1}{9}\frac{Kn_B}{\omega^2\zeta},
\eea 
%--------------------------------------------------
where the incompressibility of nuclear matter is $K=9n_B(\partial^2\epsilon/\partial n_B^2)$ and $\epsilon$ is the energy density. As seen from Fig.~\ref{fig:fig9} left panel, the bulk viscosity exhibits a resonant peak at the temperature where the equilibration rate matches the oscillation frequency $(\gamma=\omega)$, with the peak shifting to higher temperatures and decreasing in magnitude as the frequency increases. Below the peak $(\gamma\ll\omega$) the viscosity decreases as $\zeta\propto \omega^{-2}$, while above it $(\gamma\gg\omega)$ it becomes frequency-independent and controlled solely by the equilibration rate. Fig.~\ref{fig:fig9} shows that the dominant contribution comes from the direct Urca process in the relevant temperature range, the modified Urca process being important only in the low-temperature regime $T\le 2$~MeV.  The shape of the damping timescale shown in right panel of Fig.~\ref{fig:fig9} reflects its scaling with $\zeta$. The shaded regions indicate where the damping timescale becomes shorter than the characteristic evolution timescales of a BNS merger remnant: short-term ($\simeq10\,$ms dark shading) and long-term ($\simeq 1\,$s light shading). For the low oscillation frequency $f=10$ kHz considered here, the bulk viscous damping is generally inefficient on short timescales, with $\tau_\zeta$ typically exceeding the dynamical timescale across the relevant density and temperature range. However, for long-lived remnants the damping can become relevant, particularly at temperatures
$T\sim 2-10$~MeV. The minimum of the damping timescale occurs at sub- to near-saturation densities, $n_B \lesssim n_0$, and temperatures $T \sim 5$ MeV, where $\tau_\zeta$ approaches the long-term ($\sim$1 s) evolution timescale.

The dominant equilibration mechanism depends strongly on the composition of dense matter. In nucleonic matter, beta equilibration proceeds through direct or modified Urca processes, with rates that have steep temperature dependence because of the restrictions on the phase space once matter is the degenerate fermion regime. If hyperons appear~\citep{Alford2021b}, non-leptonic weak reactions among baryons are the fastest processes that equilibrate independent imbalances $\mu_n - \mu_\Lambda,$ and $\mu_n + \mu_n - \mu_p - \mu_{\Sigma^-}$ via the reactions
%-------------------------
\bea
 n + n \rightleftarrows p + \Sigma^-, \quad 
 n + p \rightleftarrows p + \Lambda ,
\eea
% --------------------------
for matter containing $n,p,\Lambda,\Sigma^-$,  neglecting double-strangeness hyperons. All other reactions can be generated from these combinations. \cite{Alford2021b}  finds that bulk viscosity from strangeness-changing processes is negligible for MeV temperatures relevant to BNS mergers, but becomes significant at keV temperatures in accord with the earlier studies restricted to cold neutron stars.

Bulk viscosity in quark matter is also relevant for BNS mergers, as the densities and temperatures can place matter in the deconfined region of the QCD phase diagram. The study of bulk viscosity in quark matter has been carried out for both unpaired matter~\citep{CruzRojas2024,Hernandez2024} and paired matter in the 2SC phase~\citep{Alford2024a}, where one color (blue) remains unpaired while the other two (red and green) form a BCS condensate. Since the 2SC phase coexists with an unpaired sector of blue quarks, the transport properties are largely governed by excitations of these unpaired quarks. Background properties of quark matter are typically modeled using approaches such as perturbative QCD, holography, the MIT bag model, or effective models such as the Nambu–Jona-Lasinio (NJL) model, see~\cite{CruzRojas2024,Hernandez2024, Alford2024a}. These frameworks are then used to evaluate the contributions of non-leptonic and semi-leptonic weak processes that control bulk viscosity.

Consider for illustration neutrino-transparent matter, where bulk viscosity is most dominant --
typically in the range $T\lesssim 10$ MeV. The main equilibration processes are the 
semi-leptonic quark Urca processes
%----------------------------------
\bea
d\to u+e^-+\bar\nu_e,\qquad
u+e^-\to d+\nu_e,
\eea
%----------------------------------
and analogous reactions involving the strange quark. In $uds$ matter there is also the non-leptonic process $u+d \leftrightarrow u+s$, which is typically much faster than the semi-leptonic channels and therefore enforces $\mu_d\simeq\mu_s$. The relevant slow imbalance is $ \mu_\Delta=\mu_d-\mu_u-\mu_e , $ analogous to the beta-equilibration imbalance in nuclear matter.

The microscopic rates in quark matter nevertheless differ from the nucleonic case for kinematical reasons. For ultrarelativistic light quarks and electrons, Urca kinematics is close to the threshold for momentum conservation, so thermal smearing of the Fermi surfaces supplies the available phase space.  As a result, the low-temperature scaling of the light-quark Urca equilibration coefficient and that for  massive strange quark~\citep{Alford2024a}
%------------------------------
\bea
\lambda_d\simeq 0.2~ G_F^2 p_{Fd}^2 T^5 , \quad \lambda_s \simeq 0.03 G_F^2 \sin ^2 \theta_c \mu_s^* m_s^{* 2} T^4,
\eea
% ------------------------------
where $p_{Fd}$ is the Fermi momentum of $d$-quarks, $\mu_s^*$ and $m_s^*$ the (effective) chemical potentials of the strange quarks. 
The non-leptonic rate is faster still,
%------------------------------
$
\lambda_{\rm non\text{-}lep}\propto
G_F^2\mu_d^5T^2 ,
$
which justifies treating it as effectively instantaneous in the MeV
temperature range~\citep{Hernandez2024}.

Recent calculations of 2SC quark matter in a vector-enhanced NJL model show that, despite these microscopic differences, the resulting bulk viscosity in the range $1\leq T\leq 10$~MeV and $4n_0\leq n_B\leq 7n_0$ is broadly similar to that of nucleonic matter \citep{Alford2023,Alford2024a,Alford2025}. Varying the vector coupling changes quark chemical potentials and therefore both the static susceptibilities and weak-interaction phase space; quantitatively, a factor-of-two variation in the vector coupling changes $\zeta$ and the damping time by factors of about $3$--$20$, mainly through the susceptibility $C$. Changes in the diquark coupling and, thus, pairing gap, have a smaller effect, primarily through the suppression of paired red-green quark contributions, although charge neutrality and beta equilibrium keep the paired and unpaired sectors thermodynamically coupled~\citep{Alford2024a,Alford2025}.

Thus, in the MeV temperature range most relevant for merger remnants,
bulk viscous damping in 2SC quark matter is not qualitatively distinct
from that in nuclear matter and may be difficult to use as a clean
diagnostic of deconfinement. A possible exception occurs at much lower
temperatures, $T\lesssim 0.1$ MeV, where the finite rate of the
non-leptonic process produces an additional low-temperature peak in
$\zeta$ that is absent in nuclear matter
\citep{Alford2006,CruzRojas2024,Hernandez2024}. The semi-leptonic MeV
peak discussed here is therefore a separate feature and, by
extrapolation, may also make unpaired quark matter difficult to
distinguish from nuclear matter through bulk viscosity alone.

The discussion above was limited to the linear response, which requires $\delta\mu\ll T$. In merger remnants this condition may fail. Large-amplitude density oscillations can generate chemical potential imbalances comparable to, or larger than, the temperature. In that regime, the reaction rates can no longer be linearized in $\delta\mu$, and the effective bulk viscosity becomes amplitude dependent.
The physical reason is simple: large departures from equilibrium open additional phase space for the reactions that restore equilibrium, thereby increasing the dissipated energy per oscillation cycle. The damping is then strongest in the regions where the oscillation amplitude is largest, so the dissipation becomes spatially inhomogeneous. This is particularly relevant for the post-merger remnant, where the density oscillations are highly nonlinear during the first several milliseconds.

Nonlinear bulk viscosity may also generate higher harmonics and broaden spectral features in the gravitational-wave signal. These effects are still being developed theoretically and are not yet standard in merger simulations~\citep{Chabanov2025}. Nevertheless, they represent an important correction to the linear picture, especially when one attempts to connect post-merger damping quantitatively to weak-interaction rates.

\subsection{Electrical and Thermal Conductivities}

Transport coefficients can be derived either from kinetic theory or from the evaluation of equilibrium correlation functions within linear response theory, although modern formulations of relativistic hydrodynamics generally require the inclusion of terms beyond first order in gradients. In kinetic theory, dissipation arises because microscopic collisions and reactions require a finite time to restore local equilibrium after a perturbation. In linear response theory, the same coefficients are expressed in terms of equilibrium correlation functions, providing a more formal framework that is particularly useful when quasiparticle descriptions become unreliable, as is often the case in strongly correlated systems.

In dense baryonic matter, the relevant processes include strong interactions among nucleons, electromagnetic scattering of leptons, and weak interactions such as Urca reactions that change the flavor composition of the matter. In the traditionally studied cold regime, neutron star matter is highly degenerate and only quasiparticles within a thermal shell of width $\sim T$ around their Fermi surfaces participate efficiently. This phase-space restriction gives rise to the characteristic temperature dependence of transport coefficients, each degenerate fermion contributing a factor $T/E_F$, where $E_F$ is the relevant Fermi energy. Note, however, that the matrix elements are modified by the dense medium through the Pauli blocking of intermediate propagation states in the effective scattering $T$-matrix, mean-field self-energies and screening. However, in BNS mergers the temperatures are sufficiently high that the matter is often only partially or semi-degenerate, with $T/E_F$ no longer being a small parameter for all particle species. As a result, the sharp Fermi-surface restriction is relaxed, thermal excitations involve a much broader region of momentum space, and the standard low-temperature scaling laws of transport coefficients and reaction rates do not apply.

{\it Electrical conductivity} $\sigma$ in BNS mergers determines the Ohmic diffusion timescale for  length scale $L$ 
%------------------------
\bea
\tau_{\rm Ohm}\propto L^2\sigma,
\eea
%------------------------
and therefore controls the decay, reconnection, and dissipation of magnetic-field structures in resistive MHD. High conductivity drives the system toward the ideal-MHD limit, whereas regions of reduced conductivity enhance magnetic diffusion and reconnection. At nuclear densities and above, the matter remains highly degenerate even at the elevated temperatures reached in BNS mergers. Consequently, the electrical conductivity is extremely large, and the dense regions of the merger are well approximated by ideal MHD.

\begin{figure}[tb] 
\centering
% --- Left Minipage ---
  \begin{minipage}{0.48\textwidth}
    \centering
    \includegraphics[height=5.8cm, width=\textwidth, keepaspectratio]{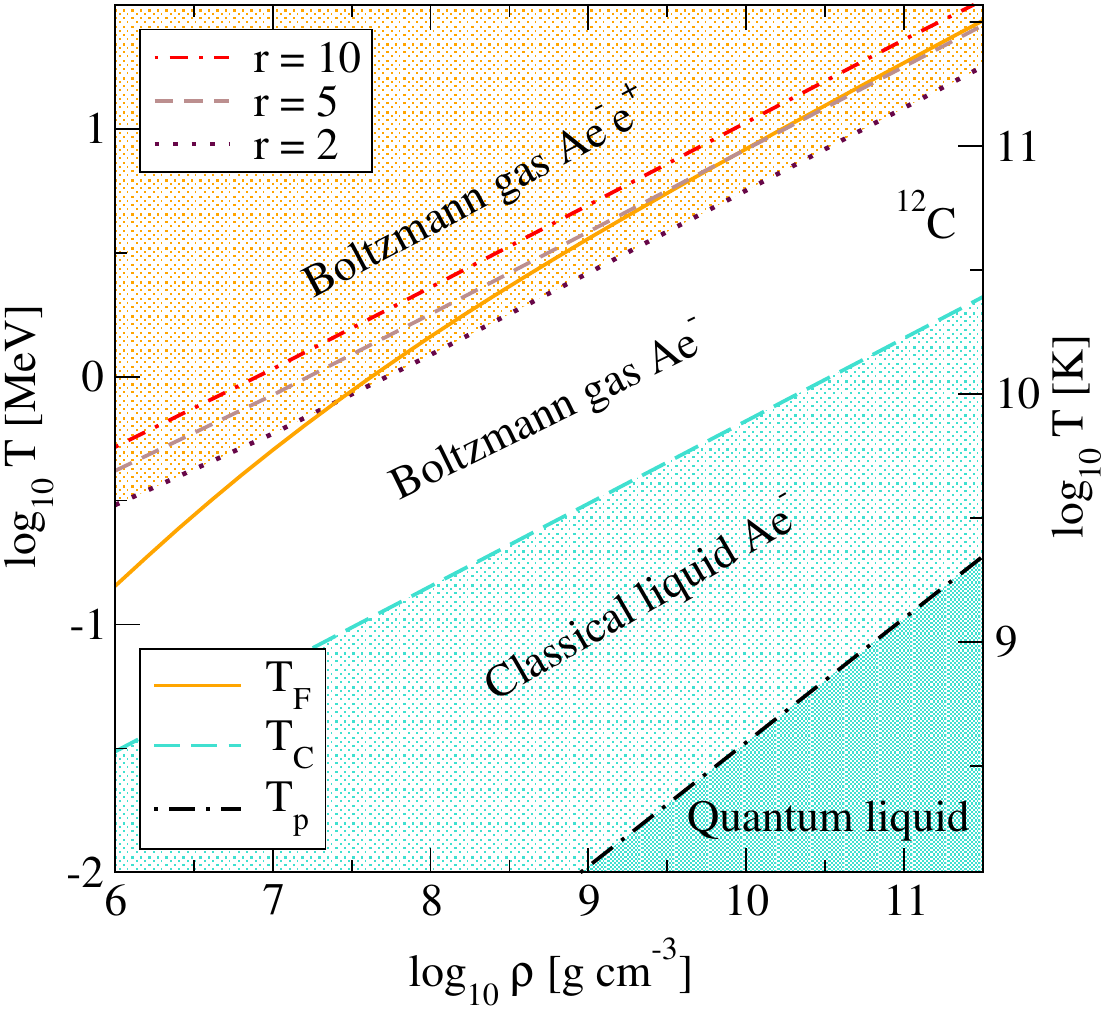}
  \end{minipage}
  % --- Right Minipage ---
  \begin{minipage}{0.48\textwidth}
    \centering
    \includegraphics[height=5.5cm, width=\textwidth, keepaspectratio]{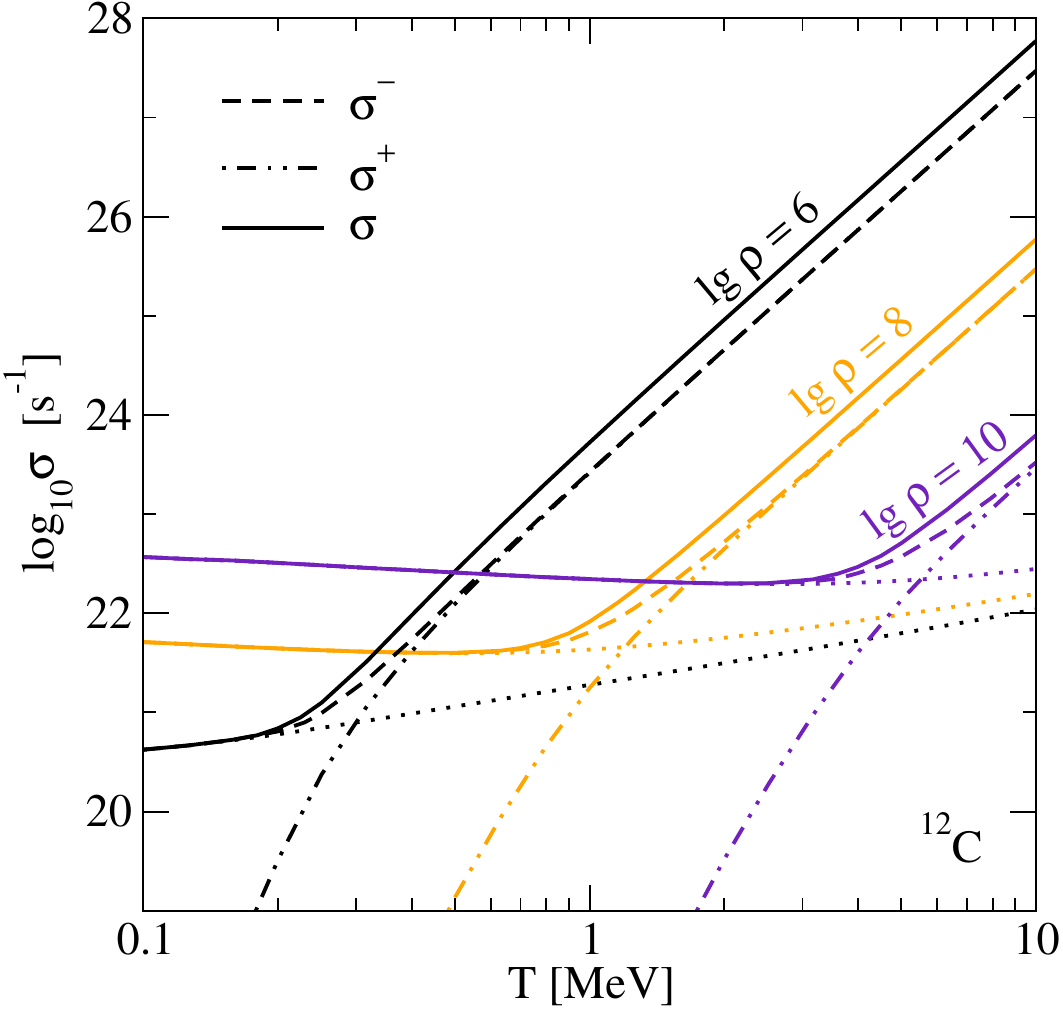}
  \end{minipage}

\caption{Left: Temperature-density phase diagram for a crustal plasma of stellar matter assuming a single nucleus of $^{12}$C~\citep{Petrosyan2026}. The three curves near $T_F$ mark the loci where the ratio of total to net electron density, $n^-/n_e$, equals $r$. Right: Temperature dependence of the electronic (dashed lines) and positronic (dash–double-dotted lines) partial conductivities, together with their total contribution (solid lines), for the three densities indicated in the figure. The conductivities shown include only electron–ion ($ei$) and positron–ion ($pi$) scattering processes. For comparison, the dotted lines display the conductivity obtained by \cite{Harutyunyan2016}, where the positron contribution was neglected.
}
\label{fig:fig10}
\end{figure}

In the outer layers of merger remnants, as well as in proto-neutron stars and hot white dwarfs, the plasma consists of ions embedded in a relativistic electron gas. Charge neutrality implies $n_e = Z n_i$, where $Z$ is the ion charge.
The state of the ionic component is characterized by the Coulomb coupling parameter
%-----------------------------------
\bea
\Gamma=\frac{Z^2e^2}{a_iT},\qquad
a_i=\left(\frac{4\pi n_i}{3}\right)^{-1/3}.
\eea
%-----------------------------------
For $\Gamma \ll 1$, the ions form a weakly coupled Boltzmann gas, whereas for $\Gamma \gtrsim 1$ they form a strongly correlated liquid and eventually crystallize at $\Gamma_m \simeq 160$. The phase diagram of matter is shown in Fig.~\ref{fig:fig10}.  Depending on temperature, the ionic component exists as a Boltzmann gas (above $T_C$), a classical liquid ($T_p\leq T\leq T_C$), or a quantum liquid (below $T_p$), where $T_C=Z^2e^2/a_i$ is the Coulomb temperature, defined so that $\Gamma=T_C/T$ and $T_p=\hbar\omega_p$, with $\omega_p=(4\pi Z^2e^2n_i/m_i)^{1/2}$ is the ion plasma temperature below which quantum effects in the ion motion become important.
Electrons become degenerate below the Fermi temperature $T_F=\epsilon_F$, with $\epsilon_F$ being the electron Fermi energy.  When at low temperatures the system solidifies, the relevant degrees of freedom become the lattice phonons, with electrons occupying Bloch states. In this case one- and two-phonon scattering processes dominate the conductivity.  In warm crustal matter that is in the liquid state, ion--ion correlations play an important role and enter the scattering rates through the static structure factor $S(q)$.
%----------------------------------------------------
%----------------------------------------------------

The kinetic description is based on the Boltzmann equation for electrons
in electric and magnetic fields~\citep{Harutyunyan2016}
%-----------------------------------------------------------
\bea
\frac{\partial f}{\partial t}
+\bm v\cdot\bm \nabla f
-e(\bm E+\bm v\times\bm B)\cdot
\frac{\partial f}{\partial\bm p}
=I[f],
\eea
%-----------------------------------------------------------
where $I[f]$ is the collision integral. The dominant scattering channel
in the crust is electron--ion scattering, with the Coulomb interaction
screened by the electron plasma and modified by ion correlations and
finite nuclear size.
For electron-ion 
%-----------------------------------------------------------
\bea\label{eq:collision_ei}
I_{ei} &=& -(2\pi)^4\sum\limits_{234}|{\cal M}_{12\to 34}^{ei}|^2
\delta^{(4)}(p_1+p_2-p_3-p_4)\nonumber\\
&&\times \Big[f_1g_2(1-f_3)(1\pm g_4)-f^-_3g_4^i(1-f_1)(1\pm g_2^i)\Big],
\eea
%-----------------------------------------------------------
where the indices $1$ and $3$ label the incoming and outgoing
electrons or positrons, respectively, $g_{2,4} $ are
the ion distribution functions before and after the collision, and
${\cal M}_{12\to 34}^{ei}$ is the electron–ion scattering
matrix element, and we  used 
the shorthand notation $\sum\limits_i = \int
d\vecp_i/(2\pi)^3$. Assuming that ions remain in thermal equilibrium,
and are heavy enough one can use Boltzmann distribution
\bea
g(p) = n_i\left(\frac{2 \pi\beta}{M }\right)^{3 / 2}e^{-\beta(\varepsilon_p - \mu)}
\eea
where $\varepsilon_p = p^2/2M$, and $M$ and $\mu$ are the mass
and chemical potential of the ion, while $\beta = T^{-1}$ is the
inverse temperature. 
At sufficiently high temperatures, electron--positron
pairs can become abundant and modify the electrical conductivity,
especially in hot dilute regions, see Fig.~\ref{fig:fig10}.  Their effect is usually to enhance
charge transport relative to calculations that neglect pair production~\citep{Petrosyan2026}.
Using the  linearization procedure around a local equilibrium Fermi distribution $f^0$,
one obtains the {\it energy-dependent} relaxation-time~\citep{Harutyunyan2016}
%------------------------------------------
\bea
\tau^{-1}(\varepsilon)=\frac{\pi Z^2 e^4 n_i}{\varepsilon p^3} \int_0^{2 p} d q q^3 S(q) F^2(q) \frac{4 \varepsilon^2-q^2}{\left|q^2+\Pi_L^{\prime}\right|^2} ,
\eea
% ------------------------------------------
where the form-factor $F(q) $ accounts for the finite size of nuclei, $S(q)$ is the ion-ion correlation function already mentioned above, and the real and imaginary parts of the longitudinal polarization are given by $\Pi_L(q, \omega)=q_D^2 \chi_l$ where, using the hard-thermal-loop approximation of quantum electrodynamics~\citep{Braaten1990},
%------------------------------------------
\bea
\operatorname{Re} \chi_l(q, \omega) \simeq 1-\left(\frac{\omega}{q}\right)^2, \quad \operatorname{Im} \chi_l(q, \omega) \simeq-\frac{\pi \omega}{2 q},
\eea
% ------------------------------------------
$q_D$ is the Debye wave-number, $\Pi_L^{\prime}$ is the polarization tensor of plasma accounting for the screening of the longitudinal plasma exchange between electrons.  The integral in this expression defines the Coulomb logarithm (CL), which has been widely adopted in the literature to account for correlation effects on transport in electron-ion plasma. Written in this form, it remains valid at arbitrary temperatures provided finite-temperature corrections to the ionic component and recoil effects are negligible. Many-body physics enters through two quantities: the ion dynamical structure factor $S(q, \omega)$, which encodes correlations in the ionic component, and the longitudinal polarization tensor of the electrons $\Pi_L(\omega, q)$. State-of-the-art evaluations of $S(q, \omega)$, typically performed in the static limit $\omega=0$, employ hypernetted-chain or Monte Carlo methods for the classical one-component plasma.

In a magnetic field, the relevant control parameter is the Hall parameter  $\omega_c\tau$
%------------------------------------------
\bea
\omega_c=\frac{eB}{\varepsilon},
\eea
%------------------------------------------
where $\tau$ is the microscopic relaxation time. For $\omega_c\tau\ll1$ the conductivity is nearly isotropic, whereas for $\omega_c\tau\gtrsim1$ it becomes tensorial and includes transverse and Hall components. The conductivity tensor has a simple form when the magnetic field is along the $z$-direction
%------------------------------------------
\bea
\hat{\sigma}=\left(\begin{array}{ccc}
\sigma_0 & -\sigma_1 & 0 \\
\sigma_1 & \sigma_0 & 0 \\
0 & 0 & \sigma
                   \end{array}\right) .
                 \eea
%------------------------------------------
Its elements reduce to a simple form in the limiting cases of highly degenerate matter,
%------------------------------------------                 
\bea\label{eq:sigma_deg}
\sigma=\frac{n_e e^2 \tau_F}{\varepsilon_F}, \quad \sigma_0=\frac{\sigma}{1+\left(\omega_{c F} \tau_F\right)^2},
\quad \sigma_1=\left(\omega_{c F} \tau_F\right) \sigma_0,
\eea
%------------------------------------------
where $\tau_F$ and $\omega_{c F}  $ are evaluated at the Fermi energy.  

Finite-temperature calculations show that the conductivity can behave non-monotonically as a function of temperature near the transition between non-degenerate and degenerate electrons~\citep{Harutyunyan2016,Petrosyan2026}, see Fig.~\ref{fig:fig10}.  In semidegenerate matter, increasing temperature opens phase space and can strongly increase the number of charge carriers, while at higher temperatures lepton--lepton scattering becomes increasingly important. A realistic treatment of $\sigma$ in BNS environments must therefore include finite-temperature screening, ion correlations, possible pair production, and the full dependence on density, temperature, composition, and magnetic field.  For non-degenerate matter, the conductivity tensor components can be written as \bea\label{eq:sigma_nondeg}
%-----------------
\sigma_n = \frac{n_e e^2}{3T} \left\langle v^2 \, \frac{\tau(\omega_c\tau)^n}{1+(\omega_c\tau)^2} \right\rangle, \qquad n=0,1,2 ,
\eea
%-----------------
where the angular brackets denote a thermal average over the Boltzmann distribution. The quantities inside the brackets may be evaluated, to logarithmic accuracy, at a representative thermal energy $\bar{\varepsilon}\sim T$. In the ultrarelativistic Boltzmann limit, one may use $v^2/(3T)\simeq 1/\bar{\varepsilon}$, so that Eq.~\eqref{eq:sigma_nondeg} takes the same form as Eq.~\eqref{eq:sigma_deg}, with the Fermi energy replaced by the average thermal energy $\bar{\varepsilon}$.

With conductivity and other microphysical ingredients in place, a natural question arises: how do the resulting conductivities shape the large-scale electromagnetic evolution of the merger? Specifically, do the finite-temperature corrections and the non-monotonic behaviour of $\sigma$ near the degenerate--non-degenerate transition translate into significant resistive effects on dynamically relevant length- and timescales?
Neglecting positron contributions, which further enhance the conductivity in the high-temperature and low-density regime, \cite{Harutyunyan2018} showed that across the entire density-temperature range relevant for BNS mergers, the magnetic-field decay time greatly exceeds all dynamically relevant timescales, establishing the validity of the ideal MHD approximation throughout the merger process. This conclusion extends down to length scales of order one meter, at least an order of magnitude below the resolution of current numerical grids. These results call for a reassessment of resistive MHD treatments of BNS mergers, which have hitherto relied on conductivity models that vanish in the low-density limit~\citep{Dionysopoulou2015}. Indeed, the breakdown of ideal MHD is not triggered by the onset of dissipation, but rather by the failure of the hydrodynamic description of matter itself.

Next, in addition to the response to electric and magnetic fields discussed above, we consider the response to a thermal gradient, characterized by the thermal conductivity~\citep{Harutyunyan2024}.  In the electron-ion liquid heat is carried mainly by electrons moving through an ionic background. Again, the specifics of BNS mergers and related systems such as proto-neutron stars, is that the temperatures can be large enough that electrons are only partially degenerate or even non-degenerate, requiring a finite-temperature kinetic treatment.

Electric fields and temperature gradients enter the linearized
Boltzmann equation through the combination
%--------------------
\bea
\bm F=\bm E'
+\frac{\varepsilon-\mu}{eT}\nabla T,
\qquad
e\bm E'=e\bm E+\nabla\mu .
\eea
%--------------------
Consequently, charge and heat transport are coupled. The linear response
relations can be written as
%--------------------
\bea
\bm E'=\hat\rho\,\bm j-\hat Q\nabla T,
\qquad
\bm q=-\hat\kappa\nabla T-T\hat Q\,\bm j ,
\eea
%--------------------
where $\hat\rho$ is the resistivity tensor, $\hat Q$ is the thermopower
tensor, and $\hat\kappa$ is the thermal conductivity tensor.

For a magnetic field along the $z$ axis, the thermal conductivity tensor
takes the form
\bea
\hat\kappa=
\begin{pmatrix}
\kappa_0 & -\kappa_1 & 0\\
\kappa_1 & \kappa_0 & 0\\
0 & 0 & \kappa
\end{pmatrix}.
\eea
Here $\kappa$ is the longitudinal conductivity along the magnetic field,
$\kappa_0$ is the transverse conductivity, and $\kappa_1$ is the Hall
component. The latter gives the thermal Hall, or Righi--Leduc, effect,
with coefficient
\bea
L=-\frac{\kappa_1}{B}.
\eea
Thus, in a magnetized plasma, heat does not flow only along
$-\bm\nabla T$, but can also be redirected perpendicular to both
$\bm\nabla T$ and $\bm B$.

For warm crustal matter composed of $^{56}{\rm Fe}$ or
$^{12}{\rm C}$, the longitudinal thermal conductivity grows with
density approximately as
$
\kappa\propto\rho^\alpha,
$
with $\alpha\simeq0.4$ in the degenerate regime and a much weaker
$\alpha\simeq0.08$ in the non-degenerate regime~\citep{Harutyunyan2024}. Its temperature
dependence may be written as
$
\kappa\propto T^\gamma ,
$
with $\gamma\simeq0.9$ for degenerate electrons and
$\gamma\simeq1.8$ for non-degenerate electrons. The ratio $\kappa/T$
develops a minimum near $T\simeq0.1$--$0.15\,T_F$, analogous to the minimum
in the electrical conductivity, reflecting the
crossover between degenerate and non-degenerate transport. Since
$\tau^{-1}\sim Z^2n_i$, lighter nuclei generally give larger
conductivities, approximately $\kappa\sim Z^{-1}$.

Magnetic fields suppress heat transport transverse to the field once
$\omega_c\tau\gtrsim1$. In the strongly magnetized limit,
\bea
\kappa_0\simeq
\frac{\kappa}{(\omega_c\tau)^2}\ll\kappa ,
\eea
so heat flows much more efficiently along magnetic field lines than
across them. The Hall component behaves as
\bea
\kappa_1\simeq
\begin{cases}
(\omega_c\tau)\kappa, & \omega_c\tau\ll1,\\[2mm]
\kappa/(\omega_c\tau), & \omega_c\tau\gg1 ,
\end{cases}
\eea
and therefore reaches a maximum near $\omega_c\tau\simeq1$. Since the
Hall parameter decreases with density, low-density crustal regions are
the most anisotropic. For $B\sim10^{12}$ G anisotropy can already appear
in the outer crust, while for $B\gtrsim10^{13}$--$10^{14}$ G much of the
outer crust may be magnetized.

These effects are potentially important in BNS merger remnants and proto-neutron stars, where strong magnetic fields coexist with large thermal gradients. Thermal conduction smooths temperature inhomogeneities, while the Hall and Righi–Leduc terms couple heat flow to the magnetic-field geometry. As discussed above, thermal conduction, together with electrical conductivity and thermopower, constitutes an essential part of the coupled magneto-thermal transport problem relevant for resistive MHD simulations of hot compact-star remnants. This is especially important in low-density, subnuclear-density plasma regions, and potentially also in the dilute accretion disks formed in the aftermath of the merger.

\subsection{Shear Viscosity}

Shear viscosity describes the transport of transverse momentum between neighboring fluid elements. Its magnitude is controlled by the mean free path of the particles that carry momentum: a longer mean free path implies a larger shear viscosity, because momentum is transported over greater distances before collisions restore local equilibrium. In dense baryonic matter, neutron--neutron and neutron--proton scattering dominate in some regimes, while leptonic contributions, especially from electrons and muons, become important when their mean free paths are long. In the case of nuclear collisions the  relevant scattering amplitudes are strongly modified by the medium (Pauli-blocking effects, self-energy, etc). In Coulomb system screening of the plasmon exchange modifies the interaction, and collective modes alter the long-range part of the force~\citep{Schmitt2018}. 

The role of shear viscosity in the pre-merger inspiral was assessed by \cite{Bildsten1992}, who showed that viscous tidal spin-up cannot bring the stars to corotation before contact for any physically reasonable viscosity. The viscous timescale for momentum transport would need to reach the light-crossing time of the star to produce tidal locking, requiring values of shear or bulk viscosity that are unphysically large. However, \cite{Ripley2023} showed that viscosity can modify the GW phase at 4PN order and can be constrained by observations, offering a novel way to probe the internal physics of neutron stars.
A complementary proposal is that the large bulk viscosity generated by non-leptonic weak reactions involving hyperons may produce substantial tidal heating during the late inspiral and an observable gravitational-wave phase shift, thereby offering a potential probe of strangeness in neutron-star cores \citep{Ghosh2024}.

For the post-merger remnant, the situation is different in principle. Simple estimates of dissipation timescales indicate that thermal transport and shear viscosity are unlikely to play a significant role in the remnant unless neutrino trapping occurs, which requires temperatures above roughly 10 MeV and gradients on scales of 0.1 km or less, see \cite{Alford2018b}. This is in contrast to bulk viscosity driven by Urca processes, discussed above. 

Dedicated calculations of shear viscosity at the finite temperatures characteristic of BNS merger remnants -- ranging from a few to tens of MeV -- are still absent from the literature. Calculations motivated by heavy-ion collisions cover the relevant temperature range but not the density regime of merger remnants and the high isospin asymmetries expected there.

\subsection{Astrophysical Implications}

Transport processes influence the post-merger remnant in several interconnected ways. It is now well-established that bulk viscosity damps compressional oscillations and can thereby reduce the amplitude of the dominant post-merger gravitational-wave modes~\citep{Most2024,Chabanov2025}. Since the relevant equilibration rates depend sensitively on composition, the damping timescale may in principle carry information about whether the remnant is purely nucleonic or contains hyperons or quark matter. However, quantitative estimates show that nucleonic matter and quark matter -- including the two-flavor color-superconducting phase -- yield analogous damping timescales, which complicates any straightforward compositional inference from the gravitational-wave signal alone.

The electrical conductivity in BNS merger remnants is high enough to validate the ideal MHD approximation in all regimes where the hydrodynamic description of matter remains valid. An exception arises in the dilute matter of tidal tails and post-merger discs, where the low density can render resistive effects non-negligible.

The roles of microscopic shear viscosity and thermal conductivity in the post-merger remnant remain comparatively unexplored~\citep{Alford2018b,Harutyunyan2024}. Current estimates suggest that microscopic shear viscosity does not dominate momentum transport in the remnant, where turbulence and magnetic stresses provide effective angular-momentum redistribution on shorter timescales. Nevertheless, it sets the baseline dissipation rate and enters the interpretation of any hydrodynamic damping not attributable to bulk processes or magnetic effects. However, these estimates are largely based on cold-matter results or heavy-ion extrapolations. On larger scales, the dominant effective viscosity is of turbulent rather than microscopic origin. The MRI drives magneto-turbulence in the remnant that facilitates angular-momentum transport and heats the matter. 
The effective turbulent stress is commonly parameterized through a Shakura–Sunyaev $\alpha$-viscosity prescription. However, resolving MHD turbulence self-consistently in global merger simulations remains computationally prohibitive. Consequently, most current simulations either neglect turbulent transport altogether or incorporate it through sub-grid effective-viscosity models with phenomenological choices of the parameter $\alpha$~\citep{Radice2017,Fujibayashi2018,Duez2020,Radice2024}.

When neutrinos are trapped in the dense hot core of the remnant, they contribute an additional momentum-transport and non-zero neutrino shear viscosity~\citep{Alford2018b}. In this regime, neutrinos form a nearly diffusive gas whose mean free path is short enough for them to carry transverse momentum between fluid layers before escaping. Neutrino viscosity has been identified as one of the mechanisms that can enforce solid-body rotation in the remnant core, where the MRI cannot operate because the angular velocity increases inward. Although neutrino viscosity is subdominant on most length scales once the MRI develops in the outer remnant, it may be the primary transport channel in the innermost core during the neutrino-trapped phase.

It should be noted that electrical and thermal conductivities are coupled in the BNS context, giving rise to an interesting feedback loop between thermal gradients, magnetic fields, and electric currents. Beyond this coupling, thermal conduction smooths temperature gradients and regulates neutrino emission, which in turn modifies the electron fraction of the outflows and their nucleosynthetic yield.

All these processes are coupled. Neutrino emission and absorption alter the composition; the composition determines weak reaction rates; those rates set the bulk viscosity and lepton-number transport; and the resulting dissipation feeds back on the remnant lifetime and mass ejection. A complete treatment therefore requires transport coefficients to be evaluated self-consistently with the thermodynamic state and composition of the matter. Transport coefficients thus represent one of the main remaining links between dense-matter microphysics and multimessenger observables.

%\newpage

\section{Neutrino Microphysics and Transport}
\label{sec:Chapter6}

We now address weak interactions and neutrino transport in BNS mergers,
focusing on the microphysical processes that determine neutrino emission,
absorption, scattering, and the evolution of the electron fraction. We have
already seen that neutrinos determine the composition of dense matter, control
the loss of energy and lepton number, and set the conditions under which
$r$-process nucleosynthesis proceeds. Once the two neutron stars come into
contact, the system evolves under the combined action of gravity, strong
interactions, hydrodynamics, and weak processes. While gravity and the strong
interaction determine the bulk dynamics and pressure support of matter, weak
interactions regulate the neutron-to-proton ratio, cool the remnant, and
determine the electron fraction $Y_e$ of the ejecta. Since $Y_e$ is one of the
key quantities controlling the $r$-process outcome, weak interaction physics is
essential for connecting merger simulations to nucleosynthesis and kilonova
observations.

At the temperatures and densities reached in BNS merger remnants, the strong
temperature dependence of the leading neutrino processes, such as direct Urca
reactions, can generate intense neutrino emission, with luminosities that may
reach $10^{53}$--$10^{54}\,\mathrm{erg\,s^{-1}}$, comparable to those in
core-collapse supernovae but released over a much shorter time. Neutrinos then
play several roles: they remove energy from the hot remnant, drive the matter
toward weak equilibrium, and change the composition of the ejecta through
absorption on free nucleons. In this way, neutrinos provide a coupling between
the dense interior and the observable outflows. The color of the kilonova, the
velocity and composition of the ejecta, and the final $r$-process abundance
pattern all depend on this coupling. 

\subsection{Weak Interaction Processes}

The weak processes relevant for mergers fall into three broad categories:
charged-current reactions, neutral-current pair processes, and scattering
\citep{Burrows2006}. Charged-current processes, mediated by $W^\pm$ exchange,
change neutrons into protons and vice versa and therefore directly determine
$Y_e$. Neutral-current processes, mediated by $Z^0$ exchange, produce or
scatter neutrino--antineutrino pairs without changing the composition. At the
level of elementary weak matrix elements, scattering and emission processes are
related by crossing symmetry, although their in-medium rates differ because of
phase-space blocking and many-body response effects.

The direct and modified Urca reactions have been discussed in the previous
section. The classification into direct and modified processes is well defined
within a quasiparticle description of neutron-star matter, where nucleons are
treated as sharp excitations near the Fermi surface. An alternative and more
general framework collects these processes into a single response function by
treating nucleons through their spectral functions, which naturally accommodate
a finite quasiparticle width and capture physics beyond the quasiparticle
approximation \citep{Alford2024c,Sedrakian2024}.

Turning to neutrino pair processes, electron--positron annihilation,
%-------------------------------------
\bea
e^-+e^+ \to \nu+\bar{\nu},
\eea
%-------------------------------------
is efficient in hot regions where positrons are thermally abundant. It is
especially important for producing heavy-flavor neutrinos, collectively
denoted $\nu_x$, since muon and tau neutrinos are not produced by ordinary
charged-current reactions at merger temperatures. Nucleon--nucleon
bremsstrahlung,
%-------------------------------------
\bea
N+N\to N+N+\nu+\bar{\nu},
\eea
%-------------------------------------
is another significant source of heavy-flavor neutrinos and contributes to the
energy loss of the nuclear fluid. Within the quasiparticle picture, the
single-nucleon process $N\to N+\nu+\bar{\nu}$ is forbidden by
energy-momentum conservation; it becomes allowed in the spectral-function
framework, where higher-order processes are absorbed into the finite width of
the quasiparticle. Plasmon decay,
%-------------------------------------
\bea
\tilde{\gamma}\to \nu+\bar{\nu},
\eea
%-------------------------------------
can occur in a degenerate electron plasma, where collective electromagnetic
excitations acquire an effective mass. In merger remnants it is generally
subdominant relative to pair annihilation and bremsstrahlung, but it can
contribute in specific regions of the density--temperature plane.

Heavy-flavor neutrinos interact only through neutral currents and therefore do
not affect $Y_e$ directly. They nevertheless carry away energy and contribute
to the cooling of the remnant, so their luminosities and mean energies remain
important inputs for the thermal evolution.

Neutral-current scattering on nucleons,
%-------------------------------------
\bea
\nu+N\to \nu+N,
\eea
%-------------------------------------
is the dominant opacity source for all neutrino flavors in the dense interior~\citep{Reddy1999,Burrows2006}.
Charged-current absorption,
%-------------------------------------
\bea
\nu_e+n\to p+e^-,
\qquad
\bar{\nu}_e+p\to n+e^+,
\eea
%-------------------------------------
provides the dominant opacity for electron-type neutrinos and directly
modifies the electron fraction of matter exposed to the neutrino flux.
Neutrino-electron/positron scattering,
%-------------------------------------
\bea
\nu+e^\pm\to \nu+e^\pm,
\eea
%-------------------------------------
is often subdominant in total opacity at high density, but it is important for
energy exchange and for determining the neutrino spectrum. The relative
importance of these processes varies across the remnant and depends on the
composition of matter. In the dense core, nucleon scattering dominates the
opacity. In the intermediate-density decoupling region, charged-current
absorption determines where $\nu_e$ and $\bar{\nu}_e$ decouple. In the disc
and ejecta, absorption on free nucleons controls neutrino heating and the
evolution of $Y_e$.

\subsection{Neutrino Opacities and Mean Free Paths}

The macroscopic behavior of neutrinos is determined by their opacity and mean
free path. Both depend on neutrino energy, flavor, density, temperature, and
composition. A useful leading-order estimate for a charged-current weak cross
section is~\citep{Burrows2006}
%------------------------
\bea
\sigma_{\rm cc}
\simeq
\sigma_0 (1+3g_A^2)
\left(\frac{E_\nu}{m_e c^2}\right)^2
\sim
10^{-41}
\left(\frac{E_\nu}{10\,\mathrm{MeV}}\right)^2
\mathrm{cm}^2 ,
\eea
%------------------------
where $g_A$ is the axial coupling and
$\sigma_0=4G_F^2(m_e c^2)^2/[\pi(\hbar c)^4]\simeq
1.76\times10^{-44}\,{\rm cm^2}$ is the reference weak-interaction cross
section. The numerical estimate is meant only as an order-of-magnitude
charged-current opacity scale; detailed rates include recoil, weak magnetism,
phase-space blocking, final-state lepton masses, and in-medium corrections.
The quadratic dependence on energy means that high-energy neutrinos have a
larger cross section and therefore decouple farther out, at lower densities
and usually lower temperatures.

The mean free path can be estimated from the kinetic formula
%------------------------
\bea
\ell_\nu^{-1}=n_{\rm scatt.}\sigma_{\rm eff},
\eea
%------------------------
where $n_{\rm scatt.}$ is the number density of scattering centers and $\sigma_{\rm eff}$
is an effective cross section averaged over the local density and temperature.
In the dense, hot remnant core, $\ell_\nu$ can be much smaller than the system
size, so neutrinos are trapped and propagate diffusively. In the outer disc
and ejecta, the mean free path can exceed the local scale height, and
neutrinos stream freely. The merger remnant therefore contains optically thick
and optically thin regions simultaneously.

Neutrino propagation in matter is intimately related to the nuclear many-body
problem. Pauli blocking suppresses reactions when the final-state nucleon or
lepton levels are already occupied, and can substantially reduce emission and
absorption rates in semidegenerate matter even at the finite temperatures
characteristic of BNS mergers. Many-body correlations introduce further
in-medium modifications of the weak response functions, including collective
excitations of nuclear matter that alter the opacities~\citep{Reddy1999}. These effects are
quantitatively important because neutrino luminosities and mean energies
determine the irradiation of the ejecta and hence the final electron fraction.

The transition from diffusion to free streaming occurs across an extended
decoupling region rather than at a sharp surface, commonly referred to as the
neutrinosphere. The decoupling density and temperature are energy dependent,
so strictly speaking a separate neutrinosphere exists for each neutrino
energy. Furthermore, electron neutrinos ($\nu_e$), electron antineutrinos
($\bar{\nu}_e$), and heavy-flavor neutrinos ($\nu_x$) decouple at different
densities because their interactions differ. Electron neutrinos have
charged-current opacity on neutrons, electron antineutrinos have
charged-current opacity on protons, and heavy-flavor neutrinos interact only
through neutral currents. Since the remnant is neutron rich, the $\nu_e$
opacity is typically larger than the $\bar{\nu}_e$ opacity, while $\nu_x$
decouple deeper and consequently at higher temperatures. This leads to
flavor-dependent luminosities and mean energies, which are central to
determining the equilibrium value of $Y_e$ in the irradiated ejecta.

\subsection{Neutrino Transport Methods}

A complete description of neutrino transport requires solving the Boltzmann
equation for the neutrino distribution function
$f_\nu(\mathbf{x},\mathbf{p},t)$. In the special-relativistic limit, neglecting
gravitational redshift and spacetime curvature, this takes the form
%-------------------------------------
\bea
\frac{\partial f_\nu}{\partial t}
+
{\bm v}\cdot{\bm \nabla} f_\nu
=
\mathcal{C}[f_\nu],
\eea
%-------------------------------------
where the left-hand side describes free streaming and $\mathcal{C}[f_\nu]$ is
the collision integral encoding emission, absorption, and scattering. In
general-relativistic merger simulations the equation must be formulated on a
curved spacetime background, which introduces additional terms coupling the
distribution function to the metric. The full problem is extremely demanding
because $f_\nu$ depends on position, momentum direction, energy, time, and
flavor. For this reason, a hierarchy of approximations is used in practice
\citep{Foucart2023}.

The simplest approach is a leakage scheme. In this method, local emission
rates are estimated from the thermodynamic state. These rates are then reduced
in optically thick regions based on an approximate diffusion timescale.
Leakage schemes are computationally efficient and capture the overall cooling
of the remnant. However, they do not evolve the neutrino radiation field. This
limits their ability to describe non-local absorption, angular effects, or
neutrino heating of the ejecta. As a result, they are often less reliable for
predicting $Y_e$ in disc winds and polar outflows.

Moment methods improve on leakage by evolving angular moments of the
distribution function, usually including the energy density and flux. They
also require a closure relation for the pressure tensor. The M1 scheme is
popular because it offers a balance between diffusion and free-streaming
limits at a reasonable computational cost for three-dimensional simulations.
It models neutrino irradiation and absorption more realistically than leakage.
Its main drawback is the approximate closure, which may fail when the
angular distribution is complex, for example, in regions with crossing beams or
near neutrino decoupling surfaces where the angular structure is important for
flavor conversion.

Full Boltzmann transport solves for the angular and energy dependence of
$f_\nu$ without using a closure approximation. It represents the most complete
classical treatment. However, it is still too expensive for fully dynamical
three-dimensional merger simulations. Monte Carlo methods offer a different
approach by replacing closure errors with statistical noise, which can be
reduced by increasing the number of particle packets. Currently, Monte Carlo
transport is mainly used in post-processing or in limited dynamical
calculations. It provides important benchmarks for leakage and moment schemes.

The choice of transport method significantly affects nucleosynthesis
predictions. More accurate methods typically show stronger neutrino
irradiation of the disc and ejecta compared to leakage schemes. This leads to
higher $Y_e$ in some outflow components. This difference can change the
balance of light and heavy $r$-process nuclei production, impacting the
kilonova color and peak brightness.

\subsection{Neutrino Flavor Oscillations}

Neutrinos are produced and interact in flavor eigenstates, but propagate as
superpositions of mass eigenstates. This mismatch gives rise to flavor
oscillations. In ordinary environments the main effects are vacuum
oscillations and matter-enhanced Mikheyev--Smirnov--Wolfenstein (MSW)
conversion. In merger remnants the situation is more complicated because both
the matter density and the neutrino number density are large, and
neutrino--neutrino forward scattering renders the flavor evolution nonlinear
\citep{Wu2017,Capozzi2022,Froustey2024}.

The flavor evolution is described by a density matrix $\varrho$ obeying
%--------------------
\bea
i\hbar\,\partial_t\varrho=[H_{\rm eff},\varrho],
\qquad
H_{\rm eff}=H_{\rm vac}+H_{\rm MSW}+H_{\nu\nu},
\eea
%--------------------
where $H_{\rm vac}$, $H_{\rm MSW}$, and $H_{\nu\nu}$ denote the vacuum,
matter, and neutrino--neutrino contributions to the effective Hamiltonian,
respectively. The vacuum term is
%--------------------
\bea
H_{\rm vac}
=
\frac{1}{2E_\nu}
U\,{\rm diag}(0,\Delta m_{21}^2,\Delta m_{31}^2)\,U^\dagger,
\eea
%--------------------
where $U$ is the Pontecorvo--Maki--Nakagawa--Sakata (PMNS) mixing matrix and
$\Delta m_{ij}^2\equiv m_i^2-m_j^2$ are the mass-squared splittings. For
typical merger neutrino energies, the vacuum oscillation scale is
macroscopic, but in the dense remnant the matter and neutrino
self-interaction potentials dominate the flavor Hamiltonian. Vacuum mixing is
therefore strongly suppressed in the interior and becomes relevant mainly at
larger radii where these potentials decrease.

Matter effects arise from coherent forward scattering on electrons,
%--------------------
\bea
H_{\rm MSW}
=
\sqrt{2}G_F n_e\,{\rm diag}(1,0,0),
\eea
%--------------------
with the opposite sign for antineutrinos. When this matter potential matches
the vacuum oscillation scale, the effective mixing angle in matter is enhanced
and an MSW resonance occurs. In mergers, the corresponding densities are
typically reached outside the remnant, that is in the disc or ejecta. Whether
this conversion affects nucleosynthesis depends on where the resonance occurs
relative to the region in which neutrino absorption sets $Y_e$.

Near the neutrino decoupling region, neutrino--neutrino forward scattering can
dominate the flavor Hamiltonian. The corresponding term is
%--------------------
\bea
H_{\nu\nu}
=
\sqrt{2}G_F
\int \frac{d^3p'}{(2\pi)^3}
(\varrho_{{\bf p}'}-\bar{\varrho}_{{\bf p}'})
(1-\cos\theta_{{\bf p}{\bf p}'}),
\eea
%--------------------
where the angular factor $(1-\cos\theta_{\mathbf{p}\mathbf{p}'})$ reflects
the current--current structure of the weak interaction. This term couples the
flavor evolution of different neutrino trajectories and makes the problem
nonlinear. It can give rise to collective oscillations, spectral swaps, and
other phenomena that have no analog in vacuum or MSW physics.

A particularly important instability, also prominent in the supernova context,
is fast flavor conversion \citep{Richers2022,Froustey2024}. This instability is driven by
crossings in the angular distribution of the electron lepton number, that is,
by configurations in which the angular distributions of $\nu_e$ and
$\bar{\nu}_e$ cross as a function of propagation direction. The associated
length scales can be as short as centimeters to meters, far below ordinary
vacuum oscillation scales, and the growth rates can be correspondingly rapid.
Such electron lepton number crossings may occur near the neutrinosphere of
merger remnants, in particular where $\nu_e$ and $\bar{\nu}_e$ decouple from
different surfaces and develop different angular distributions.

The quantitative impact of flavor conversion on merger nucleosynthesis remains
uncertain. If fast flavor conversion modifies the relative spectra of
electron-flavor and heavy-flavor neutrinos ($\nu_x$), it can shift the
equilibrium electron fraction $Y_e$ in irradiated ejecta. Because small
changes in $Y_e$ can produce large changes in lanthanide production, this
remains an important open problem at the intersection of quantum many-body
physics, neutrino transport, and astrophysics. A fully consistent treatment
ultimately requires coupling quantum flavor evolution to neutrino transport
and hydrodynamics, which remains beyond the reach of current global merger
simulations.

\section[r-process Nucleosynthesis in Neutron Star Mergers]{$r$-process Nucleosynthesis in Neutron Star Mergers}
\label{sec:Chapter7}

Binary neutron-star mergers provide conditions that are especially favorable for rapid neutron-capture ($r$-process) nucleosynthesis: neutron-rich matter, rapid expansion, and, in some ejecta components, high entropy. They are therefore widely regarded as a major source of the heaviest elements in the Universe, including lanthanides and actinides \citep{Goriely2011,Metzger2010,Radice2020,Cowan2021}. The multimessenger observation of GW170817 and its kilonova counterpart AT2017gfo provided strong empirical support for the coalescence of two neutron stars in the gravitational wave event GW170817, followed by its kilonova, providing strong observational evidence that neutron-star mergers synthesize $r$-process material~\citep{Abbott2017,Kasen2017,Cowperthwaite2017,Metzger2017}.

The nucleosynthetic outcome is controlled primarily by the electron fraction
$Y_e$, entropy, and expansion timescale. These quantities differ substantially
among tidal ejecta, shock-heated ejecta, neutrino-driven winds, and viscous or
magnetically driven disc outflows. The following subsections first discuss how
weak interactions determine $Y_e$, and then describe how the different ejecta
components map onto distinct $r$-process abundance patterns.

\subsection{Electron Fraction Evolution}

The electron fraction (equal to the proton fraction by charge neutrality),
%--------------------
\bea
Y_e=\frac{n_p}{n_b}
=
\frac{n_{e^-}-n_{e^+}}{n_b},
\eea
%--------------------
where $n_p$, $n_b$, $n_{e^-}$, and $n_{e^+}$ are the proton, baryon,
electron, and positron number densities, respectively, and the second equality
follows from charge neutrality, is the primary compositional variable
controlling $r$-process nucleosynthesis
\citep{Qian1996,Lippuner2015}. Its evolution is governed by the competition
between electron and positron captures and neutrino and antineutrino
absorption,
%--------------------
\bea
\frac{dY_e}{dt}
=
-\Gamma_{e^-p\to n\nu_e}
+
\Gamma_{e^+n\to p\bar{\nu}_e}
+
\Gamma_{\nu_e n\to pe^-}
-
\Gamma_{\bar{\nu}_e p\to n+e^+}
+\cdots ,
\eea
%--------------------
where the $\Gamma$'s from left to right correspond to the rates per baryon for electron capture on protons, positron capture on neutrons, neutrino absorption on neutrons, and antineutrino absorption on protons~\citep{Qian1996,Foucart2023}.  The ellipsis denotes additional charged-current contributions, including muonic weak processes when muons are present.  Simulations employing five-species neutrino transport show that the inclusion of muons and muonic $\beta$-processes enhances post-merger cooling and produces more neutron-rich ejecta, thereby increasing lanthanide production and suppressing the overproduction of light $r$-process elements, particularly for softer EoSs~\citep{Ng2025}.

In the dense interior of the neutron star merger, where matter and neutrinos are close to thermal and chemical equilibrium, the net rate vanishes and the composition is determined by the $\beta$-equilibrium condition
%--------------------
$
\mu_n+\mu_{\nu_e}=\mu_p+\mu_e,
$
% --------------------
see also Eq. \eqref{eq:Urca_def}.
 In this regime $Y_e$ is small, reflecting the neutron-rich
character of dense matter.

In the disc and ejecta, the neutrino field is not in equilibrium with the
matter. The electron fraction is then driven toward an irradiation equilibrium
value,
%--------------------
\bea
Y_{e,\rm eq}
\simeq
\left(
1+\frac{\Gamma_{\bar{\nu}_ep\to ne^+}}
{\Gamma_{\nu_e n\to p e^-}}
\right)^{-1}.
\eea
%--------------------
The corresponding rates scale approximately $\propto L_\nu\langle E_\nu^2\rangle/r^2$, where $L_\nu$ is the neutrino luminosity and $\langle E_\nu^2\rangle$ is the mean squared neutrino energy \citep{Qian1996,Foucart2023}.  Because the $\nu_e$ and $\bar\nu_e$ absorption rates depend on both their luminosities and spectral energies, their competition can drive the irradiated matter toward substantially larger electron fractions than those of cold neutron-star matter.  The resulting equilibrium value can lie near $Y_{e,\rm eq}\sim0.4$--$0.5$, far above the value in cold neutron-star matter \citep{Qian1996,Metzger2014,Perego2014}.

Whether an ejecta element reaches this equilibrium value depends on the
competition between the weak interaction timescale
$\tau_{\rm weak}\sim(\Gamma_{\nu_e n}+\Gamma_{\bar{\nu}_e p})^{-1}$ and the
expansion timescale $\tau_{\rm exp}\sim r/v$, where $v$ is the outflow
velocity. Rapidly expanding tidal ejecta preserve their initial neutron-rich
composition and can have $Y_e\lesssim0.1$, whereas slower disc winds, exposed
for longer times to neutrino irradiation, can reach $Y_e\sim0.3$--$0.5$
\citep{Wanajo2014,Sekiguchi2015,Just2015,Perego2014}. This broad distribution
of electron fractions is one of the main reasons why merger ejecta can produce
both heavy lanthanide-rich material and lighter $r$-process nuclei
\citep{Wanajo2014,Just2015,Lippuner2015}.

The geometry of the neutrino radiation field is also important. Polar ejecta
receive a stronger neutrino flux from the remnant and disc, while equatorial
tidal ejecta are more shielded. This naturally produces a latitude dependence
in $Y_e$: higher-$Y_e$, lanthanide-poor material is preferentially polar,
whereas low-$Y_e$, lanthanide-rich material is concentrated near the orbital
plane \citep{Perego2014,Sekiguchi2015,Foucart2023}. This angular structure is
directly relevant for the viewing-angle dependence of kilonova emission
\citep{Kasen2013,Metzger2014,Lippuner2015}.

\subsection{Ejecta Mechanisms and Thermodynamic Conditions}

The merger ejecta consist of several components with distinct physical origins, thermodynamic histories, and nucleosynthetic signatures. Dynamical ejecta are launched within milliseconds of merger and arise from tidal torques and shock heating. Tidal ejecta originate predominantly from the cold outer layers of the neutron stars and are typically extremely neutron rich, with $Y_e\lesssim0.1$, and potentially substantially lower values in weakly irradiated material.~\citep{Bauswein2013a}. Their very low $Y_e$, and hence large neutron-to-seed ratio, favors a robust heavy $r$-process extending to the actinides. Shock-driven ejecta are produced near the collision interface, where heating and weak interactions can raise the electron fraction to $Y_e\sim0.1$--$0.4$ \citep{Wanajo2014}. This component can therefore produce a broader distribution of nuclei, including lighter $r$-process material. The total mass of dynamical ejecta typically ranges from $10^{-4}$ to $10^{-2}\,M_\odot$, with a strong dependence on the binary mass ratio and the stiffness of the EoS. Full general-relativistic simulations with several finite-temperature EoSs found dynamical ejecta masses of order $10^{-3}\,M_\odot$, while the resulting heavy $r$-process abundance pattern was comparatively robust against variations in the EoS and binary masses~\citep{Bovard2017}.

On longer timescales, a massive accretion disc forms around the remnant.
Viscous heating, magnetic stresses, nuclear recombination, and neutrino
irradiation drive disc outflows that can eject a significant fraction of the
disc mass. These outflows generally have higher electron fractions,
$Y_e\sim0.2$--$0.4$, and contribute preferentially to the lighter
$r$-process. Their properties depend strongly on the remnant lifetime
\citep{Fernandez2013,Siegel2017}: long-lived neutron-star remnants produce
stronger neutrino irradiation and higher $Y_e$, whereas prompt collapse leads
to more neutron-rich outflows. Neutrino-driven winds from the remnant surface
and magnetically driven outflows can also contribute. These components usually
exhibit higher entropy and intermediate to high $Y_e$, and may play a role in
setting the lighter $r$-process abundances.

The nucleosynthetic outcome is controlled mainly by entropy $s$, expansion timescale $\tau_{\rm exp}$, and electron fraction $Y_e$. Entropy regulates the balance between free nucleons and seed nuclei. Low-entropy ejecta form seeds efficiently, while high-entropy ejecta can maintain a larger neutron-to-seed ratio.  In mergers, entropies vary from a few $k_B$ per baryon in tidal ejecta to $\sim100\,k_B$ in shock-heated regions~\citep{Oechslin2007}.  The expansion timescale determines when nuclear reactions freeze out. Fast expansion leads to early freeze-out, whereas slower expansion allows more complete processing.  Among these variables, $Y_e$ is usually the dominant control parameter: low-$Y_e$ ejecta produce a heavy $r$-process, while higher-$Y_e$ ejecta produce lighter nuclei. Neutrino irradiation can modify $Y_e$, driving material toward the irradiation equilibrium discussed above. This effect is strongest in slowly expanding polar ejecta and disc winds, whereas rapidly expanding tidal ejecta preserve a more neutron-rich composition.

\subsection{Neutrino-Driven Winds}

Neutrino-driven winds are thermally driven outflows from the remnant surface
and inner disc, analogous to the proto-neutron-star (PNS) wind in
core-collapse supernovae (CCSNe) \citep{Qian1996,Dessart2008,
Perego2014,Foucart2023}. Neutrinos streaming out of the dense remnant are
absorbed in the outer layers, deposit energy, and generate a pressure gradient
that can lift material out of the gravitational potential well
\citep{Dessart2008,Perego2014}.

The net heating rate per unit mass from charged-current $\nu_e$ and
$\bar{\nu}_e$ absorption can be written schematically as
%--------------------
\bea
\dot{q}_{\rm abs}
\simeq
\frac{\sigma_0(1+3g_A^2)}{4m_u(m_e c^2)^2}
\left(
Y_n
\frac{L_{\nu_e}}{4\pi r^2}
\frac{\langle E_{\nu_e}^3\rangle}{\langle E_{\nu_e}\rangle}
+
Y_p
\frac{L_{\bar{\nu}_e}}{4\pi r^2}
\frac{\langle E_{\bar{\nu}_e}^3\rangle}{\langle E_{\bar{\nu}_e}\rangle}
\right),
\eea
%--------------------
where $Y_n$ and $Y_p$ are the neutron and proton fractions, $L_{\nu_e}$ and
$L_{\bar{\nu}_e}$ are the electron-neutrino and electron-antineutrino
luminosities, and $\langle E_\nu\rangle$ denotes an average over the
corresponding neutrino spectrum. Here $m_u$ is the atomic mass unit, $m_e$ is
the electron mass, $g_A$ is the axial coupling, and
$\sigma_0=4G_F^2(m_e c^2)^2/[\pi(\hbar c)^4]\simeq
1.76\times10^{-44}\,{\rm cm^2}$ is the reference weak-interaction cross
section. For order-of-magnitude estimates, the spectral factor is often
written in terms of an effective squared energy,
$\langle E_\nu^3\rangle/\langle E_\nu\rangle\sim\langle E_\nu^2\rangle$,
with weak-magnetism, recoil, threshold, and angular corrections treated
separately \citep{Qian1996,Foucart2023}. This heating competes with neutrino
cooling and gravity. The gain region is defined as the region where the net
heating rate is positive and matter can be accelerated outward.

The wind mass-loss rate $\dot{M}_{\rm wind}$ depends on the neutrino
luminosities, mean energies, remnant lifetime, and disc geometry
\citep{Qian1996,Perego2014,Martin2015}. In the first tens to hundreds of
milliseconds after merger, the luminosities are largest and the wind can
contribute significantly to the total ejecta mass. Neutrino-driven winds are
typically less neutron rich than cold tidal ejecta, with electron fractions
that can reach $Y_e\sim0.3$--$0.5$ in strongly irradiated regions, although
the precise range depends on the remnant lifetime, latitude, and neutrino
transport treatment \citep{Perego2014,Martin2015,Foucart2023}. As a result,
these winds tend to produce light $r$-process nuclei, and in the most
proton-rich cases iron-group or trans-iron nuclei, rather than a robust
lanthanide-rich component \citep{Martin2015,Lippuner2015}.

Merger winds differ from supernova winds in several important respects. The remnant is rapidly and differentially rotating, the geometry is strongly non-spherical, the disc collimates and partly shields the outgoing neutrino flux, and the neutrino luminosities evolve on dynamical timescales \citep{Dessart2008,Perego2014,Foucart2023}. Supernova wind formulae, therefore, provide useful intuition, but they cannot be applied quantitatively without merger-specific simulations.

\subsection{$r$-process Path and Nuclear Physics Inputs}

\begin{figure}[!]
\begin{center}
\includegraphics[width=0.8\hsize]{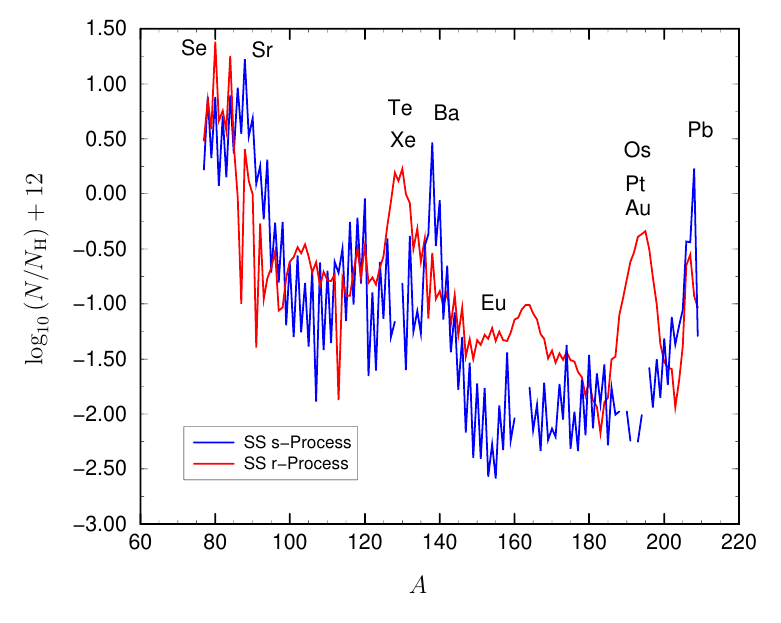}
\caption{
  Solar-system (SS) abundances decomposed into contributions from slow ($s$-process) and rapid ($r$-process) neutron capture,
  plotted against mass number $A$,  where $N_i$ and $N_{\mathrm{H}}$ are the number abundances of isotope $i$ and hydrogen,
  respectively. The $r$-process peaks occur near $A\simeq80,\ 130,$ and $195$, reflecting the influence of the neutron shell
  closures at $N=50,\ 82,$ and $126$. The $s$-process peaks lie near $A\simeq88,\, 138,$ and $208$.
  The solar $r$-process distribution provides a reference for comparison with neutron-star merger
  nucleosynthesis calculations~\citep{Goriely2011,Metzger2017}. Reproduced from \cite{Truran2002}, Fig. 1.  }
\label{fig:fig11}
\end{center}
\end{figure}

The key feature of the $r$-process is the extreme neutron flux. Neutron number
densities can reach $n_n\sim10^{20}$--$10^{28}$~cm$^{-3}$, far above
those characteristic of the slow neutron-capture process
\citep{Cowan2021}. In these conditions, neutron-capture timescales are much
shorter than $\beta$-decay timescales, and the nuclear flow proceeds far from
stability. When neutron captures and photodissociations are in equilibrium 
%-------------------------------
\bea\label{eq:n_gamma}
(Z,A)+n \rightleftarrows (Z,A+1)+\gamma ,
\eea
% -------------------------------
the abundance ratio within an isotopic chain is approximately described by
the schematic Saha form~\citep{Mumpower2016}
%-------------------------------
\bea
\frac{Y(Z,A+1)}{Y(Z,A)} \propto n_n T^{-3/2} \exp\left(\frac{S_n}{k_B T}\right),
\eea
% -------------------------------
where $S_n$ is the one-neutron separation energy. In the hot $r$-process
regime, rapid neutron captures and photodisintegrations establish an equilibrium
abundance distribution within each isotopic chain. Its maximum typically lies
at separation energies of a few MeV, with the precise values determined by the
temperature and neutron density. The dominant nuclei in each chain constitute the waiting points:
further flow to higher $Z$ proceeds through slower $\beta^-$ decays. Waiting points are therefore not restricted to magic neutron numbers. At the shell closures $N=50,82,126$, abrupt changes in neutron separation energies and relatively long $\beta$-decay lifetimes produce pronounced bottlenecks. The resulting accumulation of material gives rise, after decay toward stability, to the characteristic abundance peaks near $A\simeq80,130,$ and $195$, illustrated by the solar $r$-process distribution in Fig.~\ref{fig:fig11}~\citep{Truran2002}.

As the ejecta expand and cool, neutron captures and photodisintegrations can no longer maintain
$(n,\gamma)\rightleftarrows(\gamma,n)$  equilibrium given by Eq.~\eqref{eq:n_gamma}. Subsequent evolution toward stability involves $\beta$-decays, $\beta$-delayed neutron emission, and residual neutron captures, which reshape the final abundance distribution. In sufficiently neutron-rich ejecta, the reaction flow reaches nuclei that undergo fission. If enough free neutrons remain, the fragments capture neutrons and return to the heavy-nucleus region, producing fission recycling. This can reduce sensitivity to the initial composition, although the resulting abundance pattern remains dependent on the fission-fragment distributions, other nuclear inputs, and the ejecta conditions~\citep{Goriely2011}.

Reliable $r$-process modeling requires nuclear masses, neutron-capture rates, $\beta$-decay rates, and fission properties for nuclei often far from stability. These inputs are mostly theoretical and therefore introduce significant uncertainties.  Nuclear masses determine the $r$-process path and the locations of the abundance peaks~\citep{Arnould2007}. Neutron-capture rates become crucial during freeze-out, $\beta$-decay rates affect the speed of the process, and fission properties influence recycling and fragment distributions. Each of these inputs introduces uncertainties that can shift the predicted abundance peaks, radioactive heating rates, and kilonova observables.

%------------------------------------- Fig.
%-------------------------------------

\subsection{Observational Constraints}

Neutron-star merger nucleosynthesis calculations can reproduce the main features of the solar $r$-process abundance distribution, particularly the second and third peaks, for sufficiently neutron-rich ejecta. The first peak is more sensitive to $Y_e$ and requires contributions from higher-$Y_e$ ejecta.  Metal-poor stars provide an independent constraint: many show a nearly universal heavy $r$-process pattern, while lighter elements exhibit larger scatter, suggesting multiple sources or multiple ejecta components \citep{Roederer2014}. Galactic chemical-evolution models indicate that mergers can account for much of the observed $r$-process enrichment, although this conclusion depends on merger rates, delay-time distributions, and early enrichment histories \citep{Cote2018}.

Kilonova emission provides a more direct probe of merger ejecta. It is powered by the radioactive decay of freshly synthesized $r$-process nuclei and is strongly affected by the ejecta composition. Lanthanide-rich material has high opacity and produces redder, longer-lived emission, whereas lanthanide-poor material produces bluer and earlier emission \citep{Tanaka2013}. The kilonova AT2017gfo associated with GW170817 displayed evidence for both blue and red emission components, consistent with a multi-component ejecta structure~\citep{Villar2017}. Interpreting such observations requires detailed nuclear reaction networks, opacity calculations, and radiation-transport simulations. Future observations, together with improved nuclear data and increasingly sophisticated merger simulations, will make $r$-process nucleosynthesis an important diagnostic tool for the nuclear physics of dense matter and neutron-rich nuclei.

\section{Summary and conclusions}
\label{sec:Chapter8}

In this review we examined the nuclear physics of BNS mergers, linking the theory of dense matter to the multimessenger signals accessible to current and future observations. The EoS of dense matter is the central microphysical input governing the inspiral, merger, and post-merger evolution. We reviewed the theoretical frameworks used to construct the EoS, ranging from chiral effective field theory and many-body approaches at low densities to covariant density functionals and phenomenological models at higher densities, where additional degrees of freedom such as hyperons, $\Delta$ resonances, or deconfined quarks may appear. Observational constraints from massive pulsars, X-ray measurements, and the tidal deformability inferred from GW170817 already place important limits on viable EoS models. In particular, GW170817 favored relatively small tidal deformabilities and therefore moderately soft matter at intermediate densities, whereas the existence of neutron stars with masses near $2\,M_\odot$ requires sufficient stiffness at higher densities. Reconciling these constraints within a unified EoS remains a major challenge.

The dynamics of mergers depend closely on the EoS due to tidal interactions during inspiral, shock formation at contact, and the structure and stability of the remnant. If prompt collapse does not occur, the remnant produces a detailed post-merger gravitational-wave signal. Its oscillation spectrum carries information about the finite-temperature EoS, rotation, composition, and potential phase transitions. Extracting this information will call for precise theoretical templates that include finite-temperature effects, neutrino physics, magnetic fields, and dissipative processes.

We also discussed transport phenomena in merger remnants. The electrical conductivity of dense merger matter is typically high enough to justify the ideal-MHD approximation in the body of the remnant. However, resistive effects become significant in low-density tidal tails and accretion flows. Bulk viscosity from out-of-equilibrium weak interactions can damp post-merger oscillations on millisecond timescales and may reveal details about the composition of dense matter. Turbulent angular-momentum transport, often described by an effective Shakura–Sunyaev $\alpha$, is likely to surpass microscopic viscosity on larger scales, although its precise measurement remains uncertain.

Weak interactions and neutrino transport shape the electron fraction $Y_e$ of the ejecta and thus affect the conditions necessary for $r$-process nucleosynthesis. We discussed charged-current and neutral-current neutrino processes and the shift between neutrino-trapped and transparent states. We also addressed briefly the flavor-dependent decoupling surfaces of $\nu_e$, $\bar{\nu}_e$, and $\nu_x$. The final value of $Y_e$ relies on the balance between weak equilibration and expansion timescales. Tidal ejecta stay neutron-rich, while neutrino-irradiated winds trend towards $Y_{e,\rm eq}\sim0.4$--$0.5$. Additionally, we reviewed neutrino flavor oscillations, including fast flavor conversions near the neutrinosphere, which could significantly modify the neutrino spectra and introduce an additional source of uncertainty in nucleosynthesis predictions.

Several critical open questions remain. The composition of matter above roughly $2\,n_0$, the role of exotic phases, or the possibility of deconfinement transitions in merger remnants is still not fully understood. Binary neutron-star mergers uniquely probe matter at high density, large isospin asymmetry, and finite temperature. This includes situations where neutrinos are trapped and matter is far from equilibrium. At the same time, many aspects of the essential microphysics—finite-temperature EoSs, neutrino transport, dissipative processes, magnetic turbulence, and nuclear reaction networks—still need significant improvement. While it is well-known that BNS mergers are key sites for producing heavy elements, explaining the specific abundance patterns and the chemical evolution history of the Galaxy remains to be explored.

Future advances are expected from various directions. These include better nuclear theory and new experimental constraints on nuclei both near and far from stability. More sophisticated merger simulations that incorporate consistent microphysics, transport processes, and weak interactions will improve the predictive capability of theoretical models. Technological developments and next-generation gravitational-wave detectors should allow for direct observation of the post-merger phase, which is a crucial missing element in our understanding of BNS mergers. Electromagnetic multimessenger observations of future merger events will offer complementary constraints. We conclude that the nuclear physics of BNS mergers plays a key role in these developments, making this field one of the most promising areas of nuclear astrophysics for the next decade.

\section*{Acknowledgment}

The author gratefully acknowledges discussions and collaborations with many colleagues who shared their insights and advice on various topics. In particular, thanks are due to M.~Alford, T. Dietrich, A.~Harutyunyan, J.~J.~Li, M. Oertel,
L.~Rezzolla, and F.~Weber.  This work has been supported by the Polish NCN Grant No. 2023/51/B/ST9/02798 and, in part, by DFG grant No. SE 1836/6-1.

\bibliographystyle{Harvard}
\begin{thebibliography*}{182}
\providecommand{\bibtype}[1]{}
\providecommand{\natexlab}[1]{#1}
{\catcode`\|=0\catcode`\#=12\catcode`\@=11\catcode`\\=12
|immediate|write|@auxout{\expandafter\ifx\csname
  natexlab\endcsname\relax\gdef\natexlab#1{#1}\fi}}
\renewcommand{\url}[1]{{\tt #1}}
\providecommand{\urlprefix}{URL }
\expandafter\ifx\csname urlstyle\endcsname\relax
  \providecommand{\doi}[1]{doi:\discretionary{}{}{}#1}\else
  \providecommand{\doi}{doi:\discretionary{}{}{}\begingroup
  \urlstyle{rm}\Url}\fi
\providecommand{\bibinfo}[2]{#2}
\providecommand{\eprint}[2][]{\url{#2}}

\bibtype{Article}%
\bibitem[Abac et al.(2026)]{Abac2026}
\bibinfo{author}{Abac A}, \bibinfo{author}{Abramo R}, \bibinfo{author}{Albanesi
  S}, \bibinfo{author}{Albertini A}, \bibinfo{author}{Agapito A},
  \bibinfo{author}{Agathos M} and  et al. (\bibinfo{year}{2026}),
  \bibinfo{month}{Mar.}
\bibinfo{title}{The science of the einstein telescope}.
\bibinfo{journal}{{\em Journal of Cosmology and Astroparticle Physics}}
  \bibinfo{volume}{2026} (\bibinfo{number}{03}): \bibinfo{pages}{081}.
ISSN \bibinfo{issn}{1475-7516}.
  \bibinfo{doi}{\doi{10.1088/1475-7516/2026/03/081}}.

\bibtype{Article}%
\bibitem[{Abbott} et al.(2017)]{Abbott2017}
\bibinfo{author}{{Abbott} BP}, \bibinfo{author}{{Abbott} R},
  \bibinfo{author}{{Abbott} TD}, \bibinfo{author}{{Acernese} F},
  \bibinfo{author}{{Ackley} K}, \bibinfo{author}{{Adams} C} and  et al.
  (\bibinfo{year}{2017}), \bibinfo{month}{Oct.}
\bibinfo{title}{{{GW170817}: Observation of Gravitational Waves from a Binary
  Neutron Star Inspiral}}.
\bibinfo{journal}{{\em \prl}} \bibinfo{volume}{119} (\bibinfo{number}{16}),
  \bibinfo{eid}{161101}. \bibinfo{doi}{\doi{10.1103/PhysRevLett.119.161101}}.
\eprint{1710.05832}.

\bibtype{Article}%
\bibitem[Abbott et al.(2019)]{Abbott2019}
\bibinfo{author}{Abbott B}, \bibinfo{author}{Abbott R}, \bibinfo{author}{Abbott
  T}, \bibinfo{author}{Acernese F}, \bibinfo{author}{Ackley K},
  \bibinfo{author}{Adams C} and  et al. (\bibinfo{year}{2019}),
  \bibinfo{month}{Jan.}
\bibinfo{title}{Properties of the binary neutron star merger {GW170817}}.
\bibinfo{journal}{{\em Physical Review X}} \bibinfo{volume}{9}
  (\bibinfo{number}{1}): \bibinfo{pages}{011001}.
ISSN \bibinfo{issn}{2160-3308}. \bibinfo{doi}{\doi{10.1103/physrevx.9.011001}}.

\bibtype{Article}%
\bibitem[{Abbott} et al.(2020)]{LVC_GW190814}
\bibinfo{author}{{Abbott} R}, \bibinfo{author}{{Abbott} TD},
  \bibinfo{author}{{Abraham} S}, \bibinfo{author}{{Acernese} F},
  \bibinfo{author}{{Ackley} K}, \bibinfo{author}{{Adams} C} and  et al.
  (\bibinfo{year}{2020}), \bibinfo{month}{Jun.}
\bibinfo{title}{{{GW190814}: Gravitational Waves from the Coalescence of a 23
  Solar Mass Black Hole with a 2.6 Solar Mass Compact Object}}.
\bibinfo{journal}{{\em \apjl}} \bibinfo{volume}{896} (\bibinfo{number}{2}),
  \bibinfo{eid}{L44}. \bibinfo{doi}{\doi{10.3847/2041-8213/ab960f}}.
\eprint{2006.12611}.

\bibtype{Article}%
\bibitem[{Alford} and {Haber}(2021)]{Alford2021b}
\bibinfo{author}{{Alford} MG} and  \bibinfo{author}{{Haber} A}
  (\bibinfo{year}{2021}), \bibinfo{month}{Apr.}
\bibinfo{title}{{Strangeness-changing rates and hyperonic bulk viscosity in
  neutron star mergers}}.
\bibinfo{journal}{{\em \prc}} \bibinfo{volume}{103} (\bibinfo{number}{4}),
  \bibinfo{eid}{045810}. \bibinfo{doi}{\doi{10.1103/PhysRevC.103.045810}}.
\eprint{2009.05181}.

\bibtype{Article}%
\bibitem[{Alford} and {Harris}(2018)]{Alford2018a}
\bibinfo{author}{{Alford} MG} and  \bibinfo{author}{{Harris} SP}
  (\bibinfo{year}{2018}), \bibinfo{month}{Dec.}
\bibinfo{title}{{{\ensuremath{\beta}} equilibrium in neutron-star mergers}}.
\bibinfo{journal}{{\em \prc}} \bibinfo{volume}{98} (\bibinfo{number}{6}),
  \bibinfo{eid}{065806}. \bibinfo{doi}{\doi{10.1103/PhysRevC.98.065806}}.
\eprint{1803.00662}.

\bibtype{Article}%
\bibitem[Alford and Harris(2019)]{Alford2019a}
\bibinfo{author}{Alford MG} and  \bibinfo{author}{Harris SP}
  (\bibinfo{year}{2019}).
\bibinfo{title}{Damping of density oscillations in neutrino-transparent nuclear
  matter}.
\bibinfo{journal}{{\em Physical Review C}} \bibinfo{volume}{100}
  (\bibinfo{number}{3}): \bibinfo{pages}{035803}.
ISSN \bibinfo{issn}{2469-9993}.
  \bibinfo{doi}{\doi{10.1103/physrevc.100.035803}}.

\bibtype{Article}%
\bibitem[Alford and Schmitt(2006)]{Alford2006}
\bibinfo{author}{Alford MG} and  \bibinfo{author}{Schmitt A}
  (\bibinfo{year}{2006}), \bibinfo{month}{Nov.}
\bibinfo{title}{Bulk viscosity in 2sc quark matter}.
\bibinfo{journal}{{\em Journal of Physics G: Nuclear and Particle Physics}}
  \bibinfo{volume}{34} (\bibinfo{number}{1}): \bibinfo{pages}{67--101}.
ISSN \bibinfo{issn}{1361-6471}.
  \bibinfo{doi}{\doi{10.1088/0954-3899/34/1/005}}.

\bibtype{Article}%
\bibitem[{Alford} and {Sedrakian}(2017)]{Alford2017}
\bibinfo{author}{{Alford} M} and  \bibinfo{author}{{Sedrakian} A}
  (\bibinfo{year}{2017}), \bibinfo{month}{Oct.}
\bibinfo{title}{{Compact Stars with Sequential QCD Phase Transitions}}.
\bibinfo{journal}{{\em \prl}} \bibinfo{volume}{119} (\bibinfo{number}{16}),
  \bibinfo{eid}{161104}.
ISSN \bibinfo{issn}{1079-7114}.
  \bibinfo{doi}{\doi{10.1103/PhysRevLett.119.161104}}.
\eprint{1706.01592}.

\bibtype{Article}%
\bibitem[Alford et al.(2005)]{Alford2005}
\bibinfo{author}{Alford M}, \bibinfo{author}{Braby M}, \bibinfo{author}{Paris
  M} and  \bibinfo{author}{Reddy S} (\bibinfo{year}{2005}),
  \bibinfo{month}{Aug.}
\bibinfo{title}{Hybrid stars that masquerade as neutron stars}.
\bibinfo{journal}{{\em The Astrophysical Journal}} \bibinfo{volume}{629}
  (\bibinfo{number}{2}): \bibinfo{pages}{969--978}.
ISSN \bibinfo{issn}{1538-4357}. \bibinfo{doi}{\doi{10.1086/430902}}.

\bibtype{Article}%
\bibitem[{Alford} et al.(2008)]{Alford2008}
\bibinfo{author}{{Alford} MG}, \bibinfo{author}{{Schmitt} A},
  \bibinfo{author}{{Rajagopal} K} and  \bibinfo{author}{{Sch{\"a}fer} T}
  (\bibinfo{year}{2008}), \bibinfo{month}{Oct.}
\bibinfo{title}{{Color superconductivity in dense quark matter}}.
\bibinfo{journal}{{\em RvMP}} \bibinfo{volume}{80} (\bibinfo{number}{4}):
  \bibinfo{pages}{1455--1515}. \bibinfo{doi}{\doi{10.1103/RevModPhys.80.1455}}.
\eprint{0709.4635}.

\bibtype{Article}%
\bibitem[Alford et al.(2013)]{Alford2013}
\bibinfo{author}{Alford MG}, \bibinfo{author}{Han S} and
  \bibinfo{author}{Prakash M} (\bibinfo{year}{2013}), \bibinfo{month}{Oct.}
\bibinfo{title}{Generic conditions for stable hybrid stars}.
\bibinfo{journal}{{\em Physical Review D}} \bibinfo{volume}{88}
  (\bibinfo{number}{8}): \bibinfo{pages}{083013}.
ISSN \bibinfo{issn}{1550-2368}.
  \bibinfo{doi}{\doi{10.1103/physrevd.88.083013}}.

\bibtype{Article}%
\bibitem[Alford et al.(2018)]{Alford2018b}
\bibinfo{author}{Alford MG}, \bibinfo{author}{Bovard L},
  \bibinfo{author}{Hanauske M}, \bibinfo{author}{Rezzolla L} and
  \bibinfo{author}{Schwenzer K} (\bibinfo{year}{2018}), \bibinfo{month}{Jan.}
\bibinfo{title}{Viscous dissipation and heat conduction in binary neutron-star
  mergers}.
\bibinfo{journal}{{\em Physical Review Letters}} \bibinfo{volume}{120}
  (\bibinfo{number}{4}): \bibinfo{pages}{041101}.
ISSN \bibinfo{issn}{1079-7114}.
  \bibinfo{doi}{\doi{10.1103/physrevlett.120.041101}}.

\bibtype{Article}%
\bibitem[Alford et al.(2019)]{Alford2019}
\bibinfo{author}{Alford M}, \bibinfo{author}{Harutyunyan A} and
  \bibinfo{author}{Sedrakian A} (\bibinfo{year}{2019}), \bibinfo{month}{Nov.}
\bibinfo{title}{Bulk viscosity of baryonic matter with trapped neutrinos}.
\bibinfo{journal}{{\em Physical Review D}} \bibinfo{volume}{100}
  (\bibinfo{number}{10}): \bibinfo{pages}{103021}.
ISSN \bibinfo{issn}{2470-0029}.
  \bibinfo{doi}{\doi{10.1103/physrevd.100.103021}}.

\bibtype{Article}%
\bibitem[{Alford} et al.(2022)]{Alford2022}
\bibinfo{author}{{Alford} M}, \bibinfo{author}{{Harutyunyan} A} and
  \bibinfo{author}{{Sedrakian} A} (\bibinfo{year}{2022}), \bibinfo{month}{Sep.}
\bibinfo{title}{{Bulk Viscosity of Relativistic npe{\ensuremath{\mu}} Matter in
  Neutron-Star Mergers}}.
\bibinfo{journal}{{\em Particles}} \bibinfo{volume}{5} (\bibinfo{number}{3}):
  \bibinfo{pages}{361--376}. \bibinfo{doi}{\doi{10.3390/particles5030029}}.
\eprint{2209.04717}.

\bibtype{Article}%
\bibitem[Alford et al.(2023)]{Alford2023}
\bibinfo{author}{Alford M}, \bibinfo{author}{Harutyunyan A} and
  \bibinfo{author}{Sedrakian A} (\bibinfo{year}{2023}), \bibinfo{month}{Oct.}
\bibinfo{title}{Bulk viscosity from urca processes: npe$\mu$ matter in the
  neutrino-transparent regime}.
\bibinfo{journal}{{\em Physical Review D}} \bibinfo{volume}{108}
  (\bibinfo{number}{8}): \bibinfo{pages}{083019}.
ISSN \bibinfo{issn}{2470-0029}.
  \bibinfo{doi}{\doi{10.1103/physrevd.108.083019}}.

\bibtype{Article}%
\bibitem[Alford et al.(2024{\natexlab{a}})]{Alford2024a}
\bibinfo{author}{Alford M}, \bibinfo{author}{Harutyunyan A},
  \bibinfo{author}{Sedrakian A} and  \bibinfo{author}{Tsiopelas S}
  (\bibinfo{year}{2024}{\natexlab{a}}).
\bibinfo{title}{Bulk viscosity of two-color superconducting quark matter in
  neutron star mergers}.
\bibinfo{journal}{{\em Physical Review D}} \bibinfo{volume}{110}
  (\bibinfo{number}{6}): \bibinfo{pages}{l061303}.
ISSN \bibinfo{issn}{2470-0029}.
  \bibinfo{doi}{\doi{10.1103/physrevd.110.l061303}}.

\bibtype{Article}%
\bibitem[Alford et al.(2024{\natexlab{b}})]{Alford2024c}
\bibinfo{author}{Alford MG}, \bibinfo{author}{Haber A} and
  \bibinfo{author}{Zhang Z} (\bibinfo{year}{2024}{\natexlab{b}}),
  \bibinfo{month}{Nov.}
\bibinfo{title}{Beyond modified urca: The nucleon width approximation for
  flavor-changing processes in dense matter}.
\bibinfo{journal}{{\em Physical Review C}} \bibinfo{volume}{110}
  (\bibinfo{number}{5}): \bibinfo{pages}{l052801}.
ISSN \bibinfo{issn}{2469-9993}.
  \bibinfo{doi}{\doi{10.1103/physrevc.110.l052801}}.

\bibtype{Article}%
\bibitem[Alford et al.(2025)]{Alford2025}
\bibinfo{author}{Alford M}, \bibinfo{author}{Harutyunyan A},
  \bibinfo{author}{Sedrakian A} and  \bibinfo{author}{Tsiopelas S}
  (\bibinfo{year}{2025}).
\bibinfo{title}{Bulk viscosity of two-flavor color superconducting quark matter
  in neutron star mergers}.
\bibinfo{journal}{{\em Frontiers in Astronomy and Space Sciences}}
  \bibinfo{volume}{12}.
ISSN \bibinfo{issn}{2296-987X}.
  \bibinfo{doi}{\doi{10.3389/fspas.2025.1648066}}.

\bibtype{Article}%
\bibitem[Arnould et al.(2007)]{Arnould2007}
\bibinfo{author}{Arnould M}, \bibinfo{author}{Goriely S} and
  \bibinfo{author}{Takahashi K} (\bibinfo{year}{2007}).
\bibinfo{title}{The r-process of stellar nucleosynthesis: Astrophysics and
  nuclear physics achievements and mysteries}.
\bibinfo{journal}{{\em Physics Reports}} \bibinfo{volume}{450}
  (\bibinfo{number}{4-6}): \bibinfo{pages}{97--213}.
ISSN \bibinfo{issn}{0370-1573}.
  \bibinfo{doi}{\doi{10.1016/j.physrep.2007.06.002}}.

\bibtype{Article}%
\bibitem[{Baiotti}(2019)]{Baiotti2019}
\bibinfo{author}{{Baiotti} L} (\bibinfo{year}{2019}), \bibinfo{month}{Nov.}
\bibinfo{title}{{Gravitational waves from neutron star mergers and their
  relation to the nuclear equation of state}}.
\bibinfo{journal}{{\em Progress in Particle and Nuclear Physics}}
  \bibinfo{volume}{109}, \bibinfo{eid}{103714}.
  \bibinfo{doi}{\doi{10.1016/j.ppnp.2019.103714}}.
\eprint{1907.08534}.

\bibtype{Article}%
\bibitem[{Baiotti} and {Rezzolla}(2017)]{Baiotti2017}
\bibinfo{author}{{Baiotti} L} and  \bibinfo{author}{{Rezzolla} L}
  (\bibinfo{year}{2017}), \bibinfo{month}{Sep.}
\bibinfo{title}{{Binary neutron star mergers: a review of
  Einstein{\textquoteright}s richest laboratory}}.
\bibinfo{journal}{{\em Reports on Progress in Physics}} \bibinfo{volume}{80}
  (\bibinfo{number}{9}), \bibinfo{eid}{096901}.
  \bibinfo{doi}{\doi{10.1088/1361-6633/aa67bb}}.
\eprint{1607.03540}.

\bibtype{Article}%
\bibitem[Baiotti et al.(2008)]{Baiotti2008}
\bibinfo{author}{Baiotti L}, \bibinfo{author}{Giacomazzo B} and
  \bibinfo{author}{Rezzolla L} (\bibinfo{year}{2008}), \bibinfo{month}{Oct.}
\bibinfo{title}{Accurate evolutions of inspiralling neutron-star binaries:
  Prompt and delayed collapse to a black hole}.
\bibinfo{journal}{{\em Physical Review D}} \bibinfo{volume}{78}
  (\bibinfo{number}{8}): \bibinfo{pages}{084033}.
ISSN \bibinfo{issn}{1550-2368}.
  \bibinfo{doi}{\doi{10.1103/physrevd.78.084033}}.

\bibtype{Article}%
\bibitem[Balbus and Hawley(1991)]{Balbus1991}
\bibinfo{author}{Balbus SA} and  \bibinfo{author}{Hawley JF}
  (\bibinfo{year}{1991}).
\bibinfo{title}{A powerful local shear instability in weakly magnetized disks.
  i - linear analysis. ii - nonlinear evolution}.
\bibinfo{journal}{{\em The Astrophysical Journal}} \bibinfo{volume}{376}:
  \bibinfo{pages}{214}.
ISSN \bibinfo{issn}{1538-4357}. \bibinfo{doi}{\doi{10.1086/170270}}.

\bibtype{Article}%
\bibitem[Baumgarte et al.(2000)]{Baumgarte2000}
\bibinfo{author}{Baumgarte TW}, \bibinfo{author}{Shapiro SL} and
  \bibinfo{author}{Shibata M} (\bibinfo{year}{2000}), \bibinfo{month}{Jan.}
\bibinfo{title}{On the maximum mass of differentially rotating neutron stars}.
\bibinfo{journal}{{\em The Astrophysical Journal}} \bibinfo{volume}{528}
  (\bibinfo{number}{1}): \bibinfo{pages}{L29--L32}.
ISSN \bibinfo{issn}{0004-637X}. \bibinfo{doi}{\doi{10.1086/312425}}.

\bibtype{Article}%
\bibitem[Bauswein and Janka(2012)]{Bauswein2012}
\bibinfo{author}{Bauswein A} and  \bibinfo{author}{Janka HT}
  (\bibinfo{year}{2012}), \bibinfo{month}{Jan.}
\bibinfo{title}{Measuring neutron-star properties via gravitational waves from
  neutron-star mergers}.
\bibinfo{journal}{{\em Physical Review Letters}} \bibinfo{volume}{108}
  (\bibinfo{number}{1}): \bibinfo{pages}{011101}.
ISSN \bibinfo{issn}{1079-7114}.
  \bibinfo{doi}{\doi{10.1103/physrevlett.108.011101}}.

\bibtype{Article}%
\bibitem[Bauswein et al.(2013{\natexlab{a}})]{Bauswein2013}
\bibinfo{author}{Bauswein A}, \bibinfo{author}{Baumgarte TW} and
  \bibinfo{author}{Janka HT} (\bibinfo{year}{2013}{\natexlab{a}}).
\bibinfo{title}{Prompt merger collapse and the maximum mass of neutron stars}.
\bibinfo{journal}{{\em Physical Review Letters}} \bibinfo{volume}{111}
  (\bibinfo{number}{13}): \bibinfo{pages}{131101}.
ISSN \bibinfo{issn}{1079-7114}.
  \bibinfo{doi}{\doi{10.1103/physrevlett.111.131101}}.

\bibtype{Article}%
\bibitem[Bauswein et al.(2013{\natexlab{b}})]{Bauswein2013a}
\bibinfo{author}{Bauswein A}, \bibinfo{author}{Goriely S} and
  \bibinfo{author}{Janka HT} (\bibinfo{year}{2013}{\natexlab{b}}).
\bibinfo{title}{Systematics of dynamical mass ejection, nucleosynthesis, and
  radioactively powered electromagnetic signals from neutron-star mergers}.
\bibinfo{journal}{{\em The Astrophysical Journal}} \bibinfo{volume}{773}
  (\bibinfo{number}{1}): \bibinfo{pages}{78}.
ISSN \bibinfo{issn}{1538-4357}.
  \bibinfo{doi}{\doi{10.1088/0004-637x/773/1/78}}.

\bibtype{Article}%
\bibitem[Baym et al.(2018)]{Baym2018}
\bibinfo{author}{Baym G}, \bibinfo{author}{Hatsuda T}, \bibinfo{author}{Kojo
  T}, \bibinfo{author}{Powell PD}, \bibinfo{author}{Song Y} and
  \bibinfo{author}{Takatsuka T} (\bibinfo{year}{2018}), \bibinfo{month}{Mar.}
\bibinfo{title}{From hadrons to quarks in neutron stars: a review}.
\bibinfo{journal}{{\em Reports on Progress in Physics}} \bibinfo{volume}{81}
  (\bibinfo{number}{5}): \bibinfo{pages}{056902}.
ISSN \bibinfo{issn}{1361-6633}. \bibinfo{doi}{\doi{10.1088/1361-6633/aaae14}}.

\bibtype{Article}%
\bibitem[Bernuzzi et al.(2015{\natexlab{a}})]{Bernuzzi2015a}
\bibinfo{author}{Bernuzzi S}, \bibinfo{author}{Dietrich T} and
  \bibinfo{author}{Nagar A} (\bibinfo{year}{2015}{\natexlab{a}}),
  \bibinfo{month}{Aug.}
\bibinfo{title}{Modeling the complete gravitational wave spectrum of neutron
  star mergers}.
\bibinfo{journal}{{\em Physical Review Letters}} \bibinfo{volume}{115}
  (\bibinfo{number}{9}): \bibinfo{pages}{091101}.
ISSN \bibinfo{issn}{1079-7114}.
  \bibinfo{doi}{\doi{10.1103/physrevlett.115.091101}}.

\bibtype{Article}%
\bibitem[Bernuzzi et al.(2015{\natexlab{b}})]{Bernuzzi2015}
\bibinfo{author}{Bernuzzi S}, \bibinfo{author}{Nagar A},
  \bibinfo{author}{Dietrich T} and  \bibinfo{author}{Damour T}
  (\bibinfo{year}{2015}{\natexlab{b}}), \bibinfo{month}{Apr.}
\bibinfo{title}{Modeling the dynamics of tidally interacting binary neutron
  stars up to the merger}.
\bibinfo{journal}{{\em Physical Review Letters}} \bibinfo{volume}{114}
  (\bibinfo{number}{16}): \bibinfo{pages}{161103}.
ISSN \bibinfo{issn}{1079-7114}.
  \bibinfo{doi}{\doi{10.1103/physrevlett.114.161103}}.

\bibtype{Article}%
\bibitem[Bildsten and Cutler(1992)]{Bildsten1992}
\bibinfo{author}{Bildsten L} and  \bibinfo{author}{Cutler C}
  (\bibinfo{year}{1992}), \bibinfo{month}{Nov.}
\bibinfo{title}{Tidal interactions of inspiraling compact binaries}.
\bibinfo{journal}{{\em The Astrophysical Journal}} \bibinfo{volume}{400}:
  \bibinfo{pages}{175}.
ISSN \bibinfo{issn}{1538-4357}. \bibinfo{doi}{\doi{10.1086/171983}}.

\bibtype{Article}%
\bibitem[Blanchet(2014)]{Blanchet2014}
\bibinfo{author}{Blanchet L} (\bibinfo{year}{2014}), \bibinfo{month}{Feb.}
\bibinfo{title}{Gravitational radiation from post-newtonian sources and
  inspiralling compact binaries}.
\bibinfo{journal}{{\em Living Reviews in Relativity}} \bibinfo{volume}{17}
  (\bibinfo{number}{1}).
ISSN \bibinfo{issn}{1433-8351}. \bibinfo{doi}{\doi{10.12942/lrr-2014-2}}.

\bibtype{Article}%
\bibitem[Blandford and Znajek(1977)]{Blandford1977}
\bibinfo{author}{Blandford RD} and  \bibinfo{author}{Znajek RL}
  (\bibinfo{year}{1977}).
\bibinfo{title}{Electromagnetic extraction of energy from kerr black holes}.
\bibinfo{journal}{{\em Monthly Notices of the Royal Astronomical Society}}
  \bibinfo{volume}{179} (\bibinfo{number}{3}): \bibinfo{pages}{433--456}.
ISSN \bibinfo{issn}{1365-2966}. \bibinfo{doi}{\doi{10.1093/mnras/179.3.433}}.

\bibtype{Article}%
\bibitem[Bombaci and Logoteta(2018)]{Bombaci2018}
\bibinfo{author}{Bombaci I} and  \bibinfo{author}{Logoteta D}
  (\bibinfo{year}{2018}), \bibinfo{month}{Jan.}
\bibinfo{title}{Equation of state of dense nuclear matter and neutron star
  structure from nuclear chiral interactions}.
\bibinfo{journal}{{\em Astronomy and Astrophysics}} \bibinfo{volume}{609}:
  \bibinfo{pages}{A128}.
ISSN \bibinfo{issn}{1432-0746}.
  \bibinfo{doi}{\doi{10.1051/0004-6361/201731604}}.

\bibtype{Article}%
\bibitem[Bovard et al.(2017)]{Bovard2017}
\bibinfo{author}{Bovard L}, \bibinfo{author}{Martin D},
  \bibinfo{author}{Guercilena F}, \bibinfo{author}{Arcones A},
  \bibinfo{author}{Rezzolla L} and  \bibinfo{author}{Korobkin O}
  (\bibinfo{year}{2017}), \bibinfo{month}{Dec.}
\bibinfo{title}{r -process nucleosynthesis from matter ejected in binary
  neutron star mergers}.
\bibinfo{journal}{{\em Physical Review D}} \bibinfo{volume}{96}
  (\bibinfo{number}{12}): \bibinfo{pages}{124005}.
ISSN \bibinfo{issn}{2470-0029}.
  \bibinfo{doi}{\doi{10.1103/physrevd.96.124005}}.

\bibtype{Article}%
\bibitem[Braaten and Pisarski(1990)]{Braaten1990}
\bibinfo{author}{Braaten E} and  \bibinfo{author}{Pisarski RD}
  (\bibinfo{year}{1990}).
\bibinfo{title}{Soft amplitudes in hot gauge theories: A general analysis}.
\bibinfo{journal}{{\em Nuclear Physics B}} \bibinfo{volume}{337}
  (\bibinfo{number}{3}): \bibinfo{pages}{569--634}.
ISSN \bibinfo{issn}{0550-3213}.
  \bibinfo{doi}{\doi{10.1016/0550-3213(90)90508-b}}.

\bibtype{Article}%
\bibitem[Buonanno and Damour(1999)]{Buonanno1999}
\bibinfo{author}{Buonanno A} and  \bibinfo{author}{Damour T}
  (\bibinfo{year}{1999}), \bibinfo{month}{Mar.}
\bibinfo{title}{Effective one-body approach to general relativistic two-body
  dynamics}.
\bibinfo{journal}{{\em Physical Review D}} \bibinfo{volume}{59}
  (\bibinfo{number}{8}): \bibinfo{pages}{084006}.
ISSN \bibinfo{issn}{1089-4918}.
  \bibinfo{doi}{\doi{10.1103/physrevd.59.084006}}.

\bibtype{Article}%
\bibitem[Burrows et al.(2006)]{Burrows2006}
\bibinfo{author}{Burrows A}, \bibinfo{author}{Reddy S} and
  \bibinfo{author}{Thompson TA} (\bibinfo{year}{2006}), \bibinfo{month}{Oct.}
\bibinfo{title}{Neutrino opacities in nuclear matter}.
\bibinfo{journal}{{\em Nuclear Physics A}} \bibinfo{volume}{777}:
  \bibinfo{pages}{356--394}.
ISSN \bibinfo{issn}{0375-9474}.
  \bibinfo{doi}{\doi{10.1016/j.nuclphysa.2004.06.012}}.

\bibtype{Article}%
\bibitem[Capozzi and Saviano(2022)]{Capozzi2022}
\bibinfo{author}{Capozzi F} and  \bibinfo{author}{Saviano N}
  (\bibinfo{year}{2022}), \bibinfo{month}{Feb.}
\bibinfo{title}{Neutrino flavor conversions in high-density astrophysical and
  cosmological environments}.
\bibinfo{journal}{{\em Universe}} \bibinfo{volume}{8} (\bibinfo{number}{2}):
  \bibinfo{pages}{94}.
ISSN \bibinfo{issn}{2218-1997}. \bibinfo{doi}{\doi{10.3390/universe8020094}}.

\bibtype{Article}%
\bibitem[Carlson et al.(2015)]{Carlson2015}
\bibinfo{author}{Carlson J}, \bibinfo{author}{Gandolfi S},
  \bibinfo{author}{Pederiva F}, \bibinfo{author}{Pieper SC},
  \bibinfo{author}{Schiavilla R}, \bibinfo{author}{Schmidt K} and  et al.
  (\bibinfo{year}{2015}).
\bibinfo{title}{Quantum monte carlo methods for nuclear physics}.
\bibinfo{journal}{{\em Reviews of Modern Physics}} \bibinfo{volume}{87}
  (\bibinfo{number}{3}): \bibinfo{pages}{1067--1118}.
ISSN \bibinfo{issn}{1539-0756}.
  \bibinfo{doi}{\doi{10.1103/revmodphys.87.1067}}.

\bibtype{Article}%
\bibitem[Chabanov and Rezzolla(2025)]{Chabanov2025}
\bibinfo{author}{Chabanov M} and  \bibinfo{author}{Rezzolla L}
  (\bibinfo{year}{2025}), \bibinfo{month}{Feb.}
\bibinfo{title}{Impact of bulk viscosity on the postmerger gravitational-wave
  signal from merging neutron stars}.
\bibinfo{journal}{{\em Physical Review Letters}} \bibinfo{volume}{134}
  (\bibinfo{number}{7}): \bibinfo{pages}{071402}.
ISSN \bibinfo{issn}{1079-7114}.
  \bibinfo{doi}{\doi{10.1103/physrevlett.134.071402}}.

\bibtype{Article}%
\bibitem[Chamel and Haensel(2008)]{Chamel2008}
\bibinfo{author}{Chamel N} and  \bibinfo{author}{Haensel P}
  (\bibinfo{year}{2008}), \bibinfo{month}{Dec.}
\bibinfo{title}{Physics of neutron star crusts}.
\bibinfo{journal}{{\em Living Reviews in Relativity}} \bibinfo{volume}{11}
  (\bibinfo{number}{1}).
ISSN \bibinfo{issn}{1433-8351}. \bibinfo{doi}{\doi{10.12942/lrr-2008-10}}.

\bibtype{Article}%
\bibitem[{Collins} and {Perry}(1975)]{Collins1975}
\bibinfo{author}{{Collins} JC} and  \bibinfo{author}{{Perry} MJ}
  (\bibinfo{year}{1975}), \bibinfo{month}{May}.
\bibinfo{title}{{Superdense matter: Neutrons or asymptotically free quarks?}}
\bibinfo{journal}{{\em \prl}} \bibinfo{volume}{34} (\bibinfo{number}{21}):
  \bibinfo{pages}{1353--1356}.
  \bibinfo{doi}{\doi{10.1103/PhysRevLett.34.1353}}.

\bibtype{Article}%
\bibitem[Colò(2020)]{Colo2020}
\bibinfo{author}{Colò G} (\bibinfo{year}{2020}), \bibinfo{month}{Jan.}
\bibinfo{title}{Nuclear density functional theory}.
\bibinfo{journal}{{\em Advances in Physics: X}} \bibinfo{volume}{5}
  (\bibinfo{number}{1}): \bibinfo{pages}{1740061}.
ISSN \bibinfo{issn}{2374-6149}.
  \bibinfo{doi}{\doi{10.1080/23746149.2020.1740061}}.

\bibtype{Article}%
\bibitem[{CompOSE Core Team} et al.(2022)]{Typel2022}
\bibinfo{author}{{CompOSE Core Team}}, \bibinfo{author}{{Typel, S.}},
  \bibinfo{author}{{Oertel, M.}}, \bibinfo{author}{{Klähn, T.}},
  \bibinfo{author}{{Chatterjee, D.}}, \bibinfo{author}{{Dexheimer, V.}} and  et
  al. (\bibinfo{year}{2022}).
\bibinfo{title}{Compose reference manual - version 3.01, compstar online
  supernovæ equations of state, “harmonising\ the concert of nuclear physics
  and astrophysics”, https://compose.obspm.fr}.
\bibinfo{journal}{{\em Eur. Phys. J. A}} \bibinfo{volume}{58}
  (\bibinfo{number}{11}): \bibinfo{pages}{221}.
  \bibinfo{doi}{\doi{10.1140/epja/s10050-022-00847-y}}.
\bibinfo{url}{\url{https://doi.org/10.1140/epja/s10050-022-00847-y}}.

\bibtype{Article}%
\bibitem[Cowan et al.(2021)]{Cowan2021}
\bibinfo{author}{Cowan JJ}, \bibinfo{author}{Sneden C}, \bibinfo{author}{Lawler
  JE}, \bibinfo{author}{Aprahamian A}, \bibinfo{author}{Wiescher M},
  \bibinfo{author}{Langanke K} and  et al. (\bibinfo{year}{2021}),
  \bibinfo{month}{Feb.}
\bibinfo{title}{Origin of the heaviest elements: The rapid neutron-capture
  process}.
\bibinfo{journal}{{\em Reviews of Modern Physics}} \bibinfo{volume}{93}
  (\bibinfo{number}{1}): \bibinfo{pages}{015002}.
ISSN \bibinfo{issn}{1539-0756}.
  \bibinfo{doi}{\doi{10.1103/revmodphys.93.015002}}.

\bibtype{Article}%
\bibitem[Cowperthwaite et al.(2017)]{Cowperthwaite2017}
\bibinfo{author}{Cowperthwaite PS}, \bibinfo{author}{Berger E},
  \bibinfo{author}{Villar VA}, \bibinfo{author}{Metzger BD},
  \bibinfo{author}{Nicholl M}, \bibinfo{author}{Chornock R} and  et al.
  (\bibinfo{year}{2017}), \bibinfo{month}{Oct.}
\bibinfo{title}{The electromagnetic counterpart of the binary neutron star
  merger ligo/virgo gw170817. ii. uv, optical, and near-infrared light curves
  and comparison to kilonova models}.
\bibinfo{journal}{{\em The Astrophysical Journal}} \bibinfo{volume}{848}
  (\bibinfo{number}{2}): \bibinfo{pages}{L17}.
ISSN \bibinfo{issn}{2041-8213}. \bibinfo{doi}{\doi{10.3847/2041-8213/aa8fc7}}.

\bibtype{Article}%
\bibitem[Cruz~Rojas et al.(2024)]{CruzRojas2024}
\bibinfo{author}{Cruz~Rojas J}, \bibinfo{author}{Gorda T},
  \bibinfo{author}{Hoyos C}, \bibinfo{author}{Jokela N},
  \bibinfo{author}{Järvinen M}, \bibinfo{author}{Kurkela A} and  et al.
  (\bibinfo{year}{2024}), \bibinfo{month}{Aug.}
\bibinfo{title}{Estimate for the bulk viscosity of strongly coupled quark
  matter using perturbative qcd and holography}.
\bibinfo{journal}{{\em Physical Review Letters}} \bibinfo{volume}{133}
  (\bibinfo{number}{7}): \bibinfo{pages}{071901}.
ISSN \bibinfo{issn}{1079-7114}.
  \bibinfo{doi}{\doi{10.1103/physrevlett.133.071901}}.

\bibtype{Article}%
\bibitem[Côté et al.(2018)]{Cote2018}
\bibinfo{author}{Côté B}, \bibinfo{author}{Fryer CL},
  \bibinfo{author}{Belczynski K}, \bibinfo{author}{Korobkin O},
  \bibinfo{author}{Chruślińska M}, \bibinfo{author}{Vassh N} and  et al.
  (\bibinfo{year}{2018}), \bibinfo{month}{Mar.}
\bibinfo{title}{The origin of r-process elements in the milky way}.
\bibinfo{journal}{{\em The Astrophysical Journal}} \bibinfo{volume}{855}
  (\bibinfo{number}{2}): \bibinfo{pages}{99}.
ISSN \bibinfo{issn}{1538-4357}. \bibinfo{doi}{\doi{10.3847/1538-4357/aaad67}}.

\bibtype{Article}%
\bibitem[Dessart et al.(2008)]{Dessart2008}
\bibinfo{author}{Dessart L}, \bibinfo{author}{Ott CD}, \bibinfo{author}{Burrows
  A}, \bibinfo{author}{Rosswog S} and  \bibinfo{author}{Livne E}
  (\bibinfo{year}{2008}), \bibinfo{month}{Dec.}
\bibinfo{title}{Neutrino signatures and the neutrino-driven wind in binary
  neutron star mergers}.
\bibinfo{journal}{{\em The Astrophysical Journal}} \bibinfo{volume}{690}
  (\bibinfo{number}{2}): \bibinfo{pages}{1681--1705}.
ISSN \bibinfo{issn}{1538-4357}.
  \bibinfo{doi}{\doi{10.1088/0004-637x/690/2/1681}}.

\bibtype{Book}%
\bibitem[Dickhoff and Van~Neck(2008)]{Dickhoff2008}
\bibinfo{author}{Dickhoff WH} and  \bibinfo{author}{Van~Neck D}
  (\bibinfo{year}{2008}), \bibinfo{month}{May}.
\bibinfo{title}{Many-Body Theory Exposed!: Propagator Description of Quantum
  Mechanics in Many-Body Systems}, \bibinfo{publisher}{WORLD SCIENTIFIC}.
\bibinfo{comment}{ISBN} \bibinfo{isbn}{9789812813817}.
\bibinfo{doi}{\doi{10.1142/6821}}.

\bibtype{Article}%
\bibitem[Dionysopoulou et al.(2015)]{Dionysopoulou2015}
\bibinfo{author}{Dionysopoulou K}, \bibinfo{author}{Alic D} and
  \bibinfo{author}{Rezzolla L} (\bibinfo{year}{2015}), \bibinfo{month}{Oct.}
\bibinfo{title}{General-relativistic resistive-magnetohydrodynamic simulations
  of binary neutron stars}.
\bibinfo{journal}{{\em Physical Review D}} \bibinfo{volume}{92}
  (\bibinfo{number}{8}): \bibinfo{pages}{084064}.
ISSN \bibinfo{issn}{1550-2368}.
  \bibinfo{doi}{\doi{10.1103/physrevd.92.084064}}.

\bibtype{Article}%
\bibitem[{Drago} et al.(2014)]{Drago2014}
\bibinfo{author}{{Drago} A}, \bibinfo{author}{{Lavagno} A},
  \bibinfo{author}{{Pagliara} G} and  \bibinfo{author}{{Pigato} D}
  (\bibinfo{year}{2014}), \bibinfo{month}{Dec.}
\bibinfo{title}{{Early appearance of {\ensuremath{\Delta}} isobars in neutron
  stars}}.
\bibinfo{journal}{{\em \prc}} \bibinfo{volume}{90} (\bibinfo{number}{6}),
  \bibinfo{eid}{065809}. \bibinfo{doi}{\doi{10.1103/PhysRevC.90.065809}}.

\bibtype{Article}%
\bibitem[Drischler et al.(2021)]{Drischler2021}
\bibinfo{author}{Drischler C}, \bibinfo{author}{Holt J} and
  \bibinfo{author}{Wellenhofer C} (\bibinfo{year}{2021}).
\bibinfo{title}{Chiral effective field theory and the high-density nuclear
  equation of state}.
\bibinfo{journal}{{\em Annual Review of Nuclear and Particle Science}}
  \bibinfo{volume}{71} (\bibinfo{number}{1}): \bibinfo{pages}{403--432}.
ISSN \bibinfo{issn}{1545-4134}.
  \bibinfo{doi}{\doi{10.1146/annurev-nucl-102419-041903}}.

\bibtype{Article}%
\bibitem[Duez et al.(2006{\natexlab{a}})]{Duez2006PRL}
\bibinfo{author}{Duez MD}, \bibinfo{author}{Liu YT}, \bibinfo{author}{Shapiro
  SL}, \bibinfo{author}{Shibata M} and  \bibinfo{author}{Stephens BC}
  (\bibinfo{year}{2006}{\natexlab{a}}), \bibinfo{month}{Jan.}
\bibinfo{title}{Collapse of magnetized hypermassive neutron stars in general
  relativity}.
\bibinfo{journal}{{\em Physical Review Letters}} \bibinfo{volume}{96}
  (\bibinfo{number}{3}): \bibinfo{pages}{031101}.
ISSN \bibinfo{issn}{1079-7114}.
  \bibinfo{doi}{\doi{10.1103/physrevlett.96.031101}}.

\bibtype{Article}%
\bibitem[Duez et al.(2006{\natexlab{b}})]{Duez2006PRD}
\bibinfo{author}{Duez MD}, \bibinfo{author}{Liu YT}, \bibinfo{author}{Shapiro
  SL}, \bibinfo{author}{Shibata M} and  \bibinfo{author}{Stephens BC}
  (\bibinfo{year}{2006}{\natexlab{b}}), \bibinfo{month}{May}.
\bibinfo{title}{Evolution of magnetized, differentially rotating neutron stars:
  Simulations in full general relativity}.
\bibinfo{journal}{{\em Physical Review D}} \bibinfo{volume}{73}
  (\bibinfo{number}{10}): \bibinfo{pages}{104015}.
ISSN \bibinfo{issn}{1550-2368}.
  \bibinfo{doi}{\doi{10.1103/physrevd.73.104015}}.

\bibtype{Article}%
\bibitem[Duez et al.(2020)]{Duez2020}
\bibinfo{author}{Duez MD}, \bibinfo{author}{Knight A}, \bibinfo{author}{Foucart
  F}, \bibinfo{author}{Haddadi M}, \bibinfo{author}{Jesse J},
  \bibinfo{author}{Hébert F} and  et al. (\bibinfo{year}{2020}),
  \bibinfo{month}{Nov.}
\bibinfo{title}{Comparison of momentum transport models for numerical
  relativity}.
\bibinfo{journal}{{\em Physical Review D}} \bibinfo{volume}{102}
  (\bibinfo{number}{10}): \bibinfo{pages}{104050}.
ISSN \bibinfo{issn}{2470-0029}.
  \bibinfo{doi}{\doi{10.1103/physrevd.102.104050}}.

\bibtype{Article}%
\bibitem[Epelbaum et al.(2009)]{Epelbaum2009}
\bibinfo{author}{Epelbaum E}, \bibinfo{author}{Hammer HW} and
  \bibinfo{author}{Meißner UG} (\bibinfo{year}{2009}), \bibinfo{month}{Dec.}
\bibinfo{title}{Modern theory of nuclear forces}.
\bibinfo{journal}{{\em Reviews of Modern Physics}} \bibinfo{volume}{81}
  (\bibinfo{number}{4}): \bibinfo{pages}{1773--1825}.
ISSN \bibinfo{issn}{1539-0756}.
  \bibinfo{doi}{\doi{10.1103/revmodphys.81.1773}}.

\bibtype{Misc}%
\bibitem[Evans et al.(2021)]{Evans2021}
\bibinfo{author}{Evans M}, \bibinfo{author}{Adhikari RX}, \bibinfo{author}{Afle
  C}, \bibinfo{author}{Ballmer SW}, \bibinfo{author}{Biscoveanu S},
  \bibinfo{author}{Borhanian S} and  et al. (\bibinfo{year}{2021}).
\bibinfo{title}{A horizon study for cosmic explorer: Science, observatories,
  and community}.
\bibinfo{doi}{\doi{10.48550/ARXIV.2109.09882}}.

\bibtype{Article}%
\bibitem[{Faber} and {Rasio}(2012)]{Faber2012LRR}
\bibinfo{author}{{Faber} JA} and  \bibinfo{author}{{Rasio} FA}
  (\bibinfo{year}{2012}), \bibinfo{month}{Dec.}
\bibinfo{title}{{Binary Neutron Star Mergers}}.
\bibinfo{journal}{{\em Living Reviews in Relativity}} \bibinfo{volume}{15}
  (\bibinfo{number}{1}), \bibinfo{eid}{8}.
  \bibinfo{doi}{\doi{10.12942/lrr-2012-8}}.
\eprint{1204.3858}.

\bibtype{Article}%
\bibitem[Fernández and Metzger(2013)]{Fernandez2013}
\bibinfo{author}{Fernández R} and  \bibinfo{author}{Metzger BD}
  (\bibinfo{year}{2013}), \bibinfo{month}{Aug.}
\bibinfo{title}{Delayed outflows from black hole accretion tori following
  neutron star binary coalescence}.
\bibinfo{journal}{{\em Monthly Notices of the Royal Astronomical Society}}
  \bibinfo{volume}{435} (\bibinfo{number}{1}): \bibinfo{pages}{502--517}.
ISSN \bibinfo{issn}{0035-8711}. \bibinfo{doi}{\doi{10.1093/mnras/stt1312}}.

\bibtype{Article}%
\bibitem[{Flanagan} and {Hinderer}(2008)]{Flanagan2008}
\bibinfo{author}{{Flanagan} {\'E}{\'E}} and  \bibinfo{author}{{Hinderer} T}
  (\bibinfo{year}{2008}), \bibinfo{month}{Jan.}
\bibinfo{title}{{Constraining neutron-star tidal Love numbers with
  gravitational-wave detectors}}.
\bibinfo{journal}{{\em \prd}} \bibinfo{volume}{77} (\bibinfo{number}{2}),
  \bibinfo{eid}{021502}. \bibinfo{doi}{\doi{10.1103/PhysRevD.77.021502}}.
\eprint{0709.1915}.

\bibtype{Article}%
\bibitem[Foucart(2023)]{Foucart2023}
\bibinfo{author}{Foucart F} (\bibinfo{year}{2023}), \bibinfo{month}{Feb.}
\bibinfo{title}{Neutrino transport in general relativistic neutron star merger
  simulations}.
\bibinfo{journal}{{\em Living Reviews in Computational Astrophysics}}
  \bibinfo{volume}{9} (\bibinfo{number}{1}).
ISSN \bibinfo{issn}{2365-0524}.
  \bibinfo{doi}{\doi{10.1007/s41115-023-00016-y}}.

\bibtype{Article}%
\bibitem[Foucart et al.(2016)]{Foucart2016}
\bibinfo{author}{Foucart F}, \bibinfo{author}{Haas R}, \bibinfo{author}{Duez
  MD}, \bibinfo{author}{O’Connor E}, \bibinfo{author}{Ott CD},
  \bibinfo{author}{Roberts L} and  et al. (\bibinfo{year}{2016}),
  \bibinfo{month}{Feb.}
\bibinfo{title}{Low mass binary neutron star mergers: Gravitational waves and
  neutrino emission}.
\bibinfo{journal}{{\em Physical Review D}} \bibinfo{volume}{93}
  (\bibinfo{number}{4}): \bibinfo{pages}{044019}.
ISSN \bibinfo{issn}{2470-0029}.
  \bibinfo{doi}{\doi{10.1103/physrevd.93.044019}}.

\bibtype{Article}%
\bibitem[Fraga et al.(2014)]{Fraga2014}
\bibinfo{author}{Fraga ES}, \bibinfo{author}{Kurkela A} and
  \bibinfo{author}{Vuorinen A} (\bibinfo{year}{2014}), \bibinfo{month}{Jan.}
\bibinfo{title}{Interacting quark matter equation of state for compact stars}.
\bibinfo{journal}{{\em The Astrophysical Journal}} \bibinfo{volume}{781}
  (\bibinfo{number}{2}): \bibinfo{pages}{L25}.
ISSN \bibinfo{issn}{2041-8213}.
  \bibinfo{doi}{\doi{10.1088/2041-8205/781/2/l25}}.

\bibtype{Article}%
\bibitem[Freire and Wex(2024)]{Freire2024}
\bibinfo{author}{Freire PCC} and  \bibinfo{author}{Wex N}
  (\bibinfo{year}{2024}).
\bibinfo{title}{Gravity experiments with radio pulsars}.
\bibinfo{journal}{{\em Living Reviews in Relativity}} \bibinfo{volume}{27}
  (\bibinfo{number}{1}).
ISSN \bibinfo{issn}{1433-8351}.
  \bibinfo{doi}{\doi{10.1007/s41114-024-00051-y}}.

\bibtype{Inproceedings}%
\bibitem[Froustey et al.(2024)]{Froustey2024}
\bibinfo{author}{Froustey J}, \bibinfo{author}{Richers S},
  \bibinfo{author}{Grohs E}, \bibinfo{author}{Flynn S},
  \bibinfo{author}{Foucart F}, \bibinfo{author}{Kneller J} and  et al.
  (\bibinfo{year}{2024}), \bibinfo{month}{Feb.}, \bibinfo{title}{Neutrino
  flavor transformation with moments: application to fast flavor instabilities
  in neutron star mergers}, \bibinfo{booktitle}{Proceedings of XVIII
  International Conference on Topics in Astroparticle and Underground Physics
  — PoS(TAUP2023)}, \bibinfo{series}{TAUP2023}, \bibinfo{publisher}{Sissa
  Medialab}, pp. \bibinfo{pages}{341}.

\bibtype{Article}%
\bibitem[Fujibayashi et al.(2018)]{Fujibayashi2018}
\bibinfo{author}{Fujibayashi S}, \bibinfo{author}{Kiuchi K},
  \bibinfo{author}{Nishimura N}, \bibinfo{author}{Sekiguchi Y} and
  \bibinfo{author}{Shibata M} (\bibinfo{year}{2018}).
\bibinfo{title}{Mass ejection from the remnant of a binary neutron star merger:
  Viscous-radiation hydrodynamics study}.
\bibinfo{journal}{{\em The Astrophysical Journal}} \bibinfo{volume}{860}
  (\bibinfo{number}{1}): \bibinfo{pages}{64}.
ISSN \bibinfo{issn}{1538-4357}. \bibinfo{doi}{\doi{10.3847/1538-4357/aabafd}}.

\bibtype{Article}%
\bibitem[Ghosh et al.(2024)]{Ghosh2024}
\bibinfo{author}{Ghosh S}, \bibinfo{author}{Pradhan BK} and
  \bibinfo{author}{Chatterjee D} (\bibinfo{year}{2024}), \bibinfo{month}{May}.
\bibinfo{title}{Tidal heating as a direct probe of strangeness inside neutron
  stars}.
\bibinfo{journal}{{\em Physical Review D}} \bibinfo{volume}{109}
  (\bibinfo{number}{10}): \bibinfo{pages}{103036}.
ISSN \bibinfo{issn}{2470-0029}.
  \bibinfo{doi}{\doi{10.1103/physrevd.109.103036}}.

\bibtype{Article}%
\bibitem[Gieg et al.(2025)]{Gieg2025}
\bibinfo{author}{Gieg H}, \bibinfo{author}{Ujevic M},
  \bibinfo{author}{Sedrakian A} and  \bibinfo{author}{Dietrich T}
  (\bibinfo{year}{2025}), \bibinfo{month}{Dec.}
\bibinfo{title}{Simulating binary neutron star mergers with finite-temperature
  equations of state: The influences of the slope of the symmetry energy and
  artificial heating}.
\bibinfo{journal}{{\em Physical Review D}} \bibinfo{volume}{112}
  (\bibinfo{number}{12}).
ISSN \bibinfo{issn}{2470-0029}. \bibinfo{doi}{\doi{10.1103/k4mp-ksxy}}.

\bibtype{Article}%
\bibitem[Glendenning and Kettner(2000)]{Glendenning2000}
\bibinfo{author}{Glendenning NK} and  \bibinfo{author}{Kettner C}
  (\bibinfo{year}{2000}).
\bibinfo{title}{Non-identical neutron star twins}.
\bibinfo{journal}{{\em Astronomy and Astrophysics}} \bibinfo{volume}{353}:
  \bibinfo{pages}{L9}. \bibinfo{doi}{\doi{10.48550/arXiv.astro-ph/9807155}}.
\eprint{astro-ph/9807155}.

\bibtype{Article}%
\bibitem[Gorda et al.(2021)]{Gorda2021}
\bibinfo{author}{Gorda T}, \bibinfo{author}{Kurkela A},
  \bibinfo{author}{Paatelainen R}, \bibinfo{author}{Säppi S} and
  \bibinfo{author}{Vuorinen A} (\bibinfo{year}{2021}), \bibinfo{month}{Oct.}
\bibinfo{title}{Soft interactions in cold quark matter}.
\bibinfo{journal}{{\em Physical Review Letters}} \bibinfo{volume}{127}
  (\bibinfo{number}{16}): \bibinfo{pages}{162003}.
ISSN \bibinfo{issn}{1079-7114}.
  \bibinfo{doi}{\doi{10.1103/physrevlett.127.162003}}.

\bibtype{Article}%
\bibitem[Goriely et al.(2011)]{Goriely2011}
\bibinfo{author}{Goriely S}, \bibinfo{author}{Bauswein A} and
  \bibinfo{author}{Janka HT} (\bibinfo{year}{2011}).
\bibinfo{title}{r-process nucleosynthesis in dynamically ejected matter of
  neutron star mergers}.
\bibinfo{journal}{{\em Astrophysical Journal Letters}} \bibinfo{volume}{738}
  (\bibinfo{number}{2}): \bibinfo{pages}{L32}.
  \bibinfo{doi}{\doi{10.1088/2041-8205/738/2/L32}}.
\eprint{1107.0899}.

\bibtype{Article}%
\bibitem[Hanauske et al.(2017)]{Hanauske2017}
\bibinfo{author}{Hanauske M}, \bibinfo{author}{Takami K},
  \bibinfo{author}{Bovard L}, \bibinfo{author}{Rezzolla L},
  \bibinfo{author}{Font JA}, \bibinfo{author}{Galeazzi F} and
  \bibinfo{author}{Stöcker H} (\bibinfo{year}{2017}), \bibinfo{month}{Aug.}
\bibinfo{title}{Rotational properties of hypermassive neutron stars from binary
  mergers}.
\bibinfo{journal}{{\em Physical Review D}} \bibinfo{volume}{96}
  (\bibinfo{number}{4}): \bibinfo{pages}{043004}.
ISSN \bibinfo{issn}{2470-0029}.
  \bibinfo{doi}{\doi{10.1103/physrevd.96.043004}}.

\bibtype{Article}%
\bibitem[Harutyunyan and Sedrakian(2016)]{Harutyunyan2016}
\bibinfo{author}{Harutyunyan A} and  \bibinfo{author}{Sedrakian A}
  (\bibinfo{year}{2016}), \bibinfo{month}{Aug.}
\bibinfo{title}{Electrical conductivity of a warm neutron star crust in
  magnetic fields}.
\bibinfo{journal}{{\em Physical Review C}} \bibinfo{volume}{94}
  (\bibinfo{number}{2}): \bibinfo{pages}{025805}.
ISSN \bibinfo{issn}{2469-9993}.
  \bibinfo{doi}{\doi{10.1103/physrevc.94.025805}}.

\bibtype{Article}%
\bibitem[Harutyunyan and Sedrakian(2024)]{Harutyunyan2024}
\bibinfo{author}{Harutyunyan A} and  \bibinfo{author}{Sedrakian A}
  (\bibinfo{year}{2024}), \bibinfo{month}{Nov.}
\bibinfo{title}{Thermal conductivity and thermal hall effect in dense
  electron-ion plasma}.
\bibinfo{journal}{{\em Particles}} \bibinfo{volume}{7} (\bibinfo{number}{4}):
  \bibinfo{pages}{967--983}.
ISSN \bibinfo{issn}{2571-712X}. \bibinfo{doi}{\doi{10.3390/particles7040059}}.

\bibtype{Article}%
\bibitem[Harutyunyan et al.(2018)]{Harutyunyan2018}
\bibinfo{author}{Harutyunyan A}, \bibinfo{author}{Nathanail A},
  \bibinfo{author}{Rezzolla L} and  \bibinfo{author}{Sedrakian A}
  (\bibinfo{year}{2018}), \bibinfo{month}{Nov.}
\bibinfo{title}{Electrical resistivity and hall effect in binary neutron star
  mergers}.
\bibinfo{journal}{{\em The European Physical Journal A}} \bibinfo{volume}{54}
  (\bibinfo{number}{11}).
ISSN \bibinfo{issn}{1434-601X}.
  \bibinfo{doi}{\doi{10.1140/epja/i2018-12624-1}}.

\bibtype{Article}%
\bibitem[Hernández et al.(2024)]{Hernandez2024}
\bibinfo{author}{Hernández JL}, \bibinfo{author}{Manuel C} and
  \bibinfo{author}{Tolos L} (\bibinfo{year}{2024}).
\bibinfo{title}{Damping of density oscillations from bulk viscosity in quark
  matter}.
\bibinfo{journal}{{\em Physical Review D}} \bibinfo{volume}{109}
  (\bibinfo{number}{12}): \bibinfo{pages}{123022}.
ISSN \bibinfo{issn}{2470-0029}.
  \bibinfo{doi}{\doi{10.1103/physrevd.109.123022}}.

\bibtype{Article}%
\bibitem[{Hinderer}(2008)]{Hinderer2008}
\bibinfo{author}{{Hinderer} T} (\bibinfo{year}{2008}), \bibinfo{month}{Apr.}
\bibinfo{title}{{Tidal Love Numbers of Neutron Stars}}.
\bibinfo{journal}{{\em \apj}} \bibinfo{volume}{677} (\bibinfo{number}{2}):
  \bibinfo{pages}{1216--1220}. \bibinfo{doi}{\doi{10.1086/533487}}.
\eprint{0711.2420}.

\bibtype{Article}%
\bibitem[Hinderer et al.(2016)]{Hinderer2016}
\bibinfo{author}{Hinderer T}, \bibinfo{author}{Taracchini A},
  \bibinfo{author}{Foucart F}, \bibinfo{author}{Buonanno A},
  \bibinfo{author}{Steinhoff J}, \bibinfo{author}{Duez M} and  et al.
  (\bibinfo{year}{2016}), \bibinfo{month}{May}.
\bibinfo{title}{Effects of neutron-star dynamic tides on gravitational
  waveforms within the effective-one-body approach}.
\bibinfo{journal}{{\em Physical Review Letters}} \bibinfo{volume}{116}
  (\bibinfo{number}{18}): \bibinfo{pages}{181101}.
ISSN \bibinfo{issn}{1079-7114}.
  \bibinfo{doi}{\doi{10.1103/physrevlett.116.181101}}.

\bibtype{Article}%
\bibitem[Horowitz and Berry(2008)]{Horowitz2008}
\bibinfo{author}{Horowitz CJ} and  \bibinfo{author}{Berry DK}
  (\bibinfo{year}{2008}).
\bibinfo{title}{Shear viscosity and thermal conductivity of nuclear
  “pasta”}.
\bibinfo{journal}{{\em Physical Review C}} \bibinfo{volume}{78}
  (\bibinfo{number}{3}): \bibinfo{pages}{035806}.
ISSN \bibinfo{issn}{1089-490X}.
  \bibinfo{doi}{\doi{10.1103/physrevc.78.035806}}.

\bibtype{Article}%
\bibitem[Horowitz et al.(2004)]{Horowitz2004}
\bibinfo{author}{Horowitz CJ}, \bibinfo{author}{Pérez-García MA},
  \bibinfo{author}{Carriere J}, \bibinfo{author}{Berry DK} and
  \bibinfo{author}{Piekarewicz J} (\bibinfo{year}{2004}), \bibinfo{month}{Dec.}
\bibinfo{title}{Nonuniform neutron-rich matter and coherent neutrino
  scattering}.
\bibinfo{journal}{{\em Physical Review C}} \bibinfo{volume}{70}
  (\bibinfo{number}{6}): \bibinfo{pages}{065806}.
ISSN \bibinfo{issn}{1089-490X}.
  \bibinfo{doi}{\doi{10.1103/physrevc.70.065806}}.

\bibtype{Article}%
\bibitem[Hotokezaka et al.(2011)]{Hotokezaka2011}
\bibinfo{author}{Hotokezaka K}, \bibinfo{author}{Kyutoku K},
  \bibinfo{author}{Okawa H}, \bibinfo{author}{Shibata M} and
  \bibinfo{author}{Kiuchi K} (\bibinfo{year}{2011}).
\bibinfo{title}{Binary neutron star mergers: Dependence on the nuclear equation
  of state}.
\bibinfo{journal}{{\em Physical Review D}} \bibinfo{volume}{83}
  (\bibinfo{number}{12}): \bibinfo{pages}{124008}.
ISSN \bibinfo{issn}{1550-2368}.
  \bibinfo{doi}{\doi{10.1103/physrevd.83.124008}}.

\bibtype{Article}%
\bibitem[Hotokezaka et al.(2013)]{Hotokezaka2013}
\bibinfo{author}{Hotokezaka K}, \bibinfo{author}{Kiuchi K},
  \bibinfo{author}{Kyutoku K}, \bibinfo{author}{Okawa H},
  \bibinfo{author}{Sekiguchi Yi}, \bibinfo{author}{Shibata M} and  et al.
  (\bibinfo{year}{2013}).
\bibinfo{title}{Mass ejection from the merger of binary neutron stars}.
\bibinfo{journal}{{\em Physical Review D}} \bibinfo{volume}{87}
  (\bibinfo{number}{2}): \bibinfo{pages}{024001}.
  \bibinfo{doi}{\doi{10.1103/PhysRevD.87.024001}}.
\eprint{1212.0905}.

\bibtype{Article}%
\bibitem[Hulse and Taylor(1975)]{Hulse1975}
\bibinfo{author}{Hulse RA} and  \bibinfo{author}{Taylor JH}
  (\bibinfo{year}{1975}), \bibinfo{month}{Jan.}
\bibinfo{title}{Discovery of a pulsar in a binary system}.
\bibinfo{journal}{{\em The Astrophysical Journal}} \bibinfo{volume}{195}:
  \bibinfo{pages}{L51}.
ISSN \bibinfo{issn}{1538-4357}. \bibinfo{doi}{\doi{10.1086/181708}}.

\bibtype{Article}%
\bibitem[Ivanova et al.(2013)]{Ivanova2013}
\bibinfo{author}{Ivanova N}, \bibinfo{author}{Justham S}, \bibinfo{author}{Chen
  X}, \bibinfo{author}{De~Marco O}, \bibinfo{author}{Fryer CL},
  \bibinfo{author}{Gaburov E} and  et al. (\bibinfo{year}{2013}),
  \bibinfo{month}{Feb.}
\bibinfo{title}{Common envelope evolution: where we stand and how we can move
  forward}.
\bibinfo{journal}{{\em The Astronomy and Astrophysics Review}}
  \bibinfo{volume}{21} (\bibinfo{number}{1}).
ISSN \bibinfo{issn}{1432-0754}. \bibinfo{doi}{\doi{10.1007/s00159-013-0059-2}}.

\bibtype{Article}%
\bibitem[Just et al.(2015)]{Just2015}
\bibinfo{author}{Just O}, \bibinfo{author}{Bauswein A},
  \bibinfo{author}{Pulpillo RA}, \bibinfo{author}{Goriely S} and
  \bibinfo{author}{Janka HT} (\bibinfo{year}{2015}), \bibinfo{month}{Feb.}
\bibinfo{title}{Comprehensive nucleosynthesis analysis for ejecta of compact
  binary mergers}.
\bibinfo{journal}{{\em Monthly Notices of the Royal Astronomical Society}}
  \bibinfo{volume}{448} (\bibinfo{number}{1}): \bibinfo{pages}{541--567}.
ISSN \bibinfo{issn}{0035-8711}. \bibinfo{doi}{\doi{10.1093/mnras/stv009}}.

\bibtype{Article}%
\bibitem[Kajino et al.(2019)]{Kajino2019}
\bibinfo{author}{Kajino T}, \bibinfo{author}{Aoki W},
  \bibinfo{author}{Balantekin A}, \bibinfo{author}{Diehl R},
  \bibinfo{author}{Famiano M} and  \bibinfo{author}{Mathews G}
  (\bibinfo{year}{2019}).
\bibinfo{title}{Current status of r-process nucleosynthesis}.
\bibinfo{journal}{{\em Progress in Particle and Nuclear Physics}}
  \bibinfo{volume}{107}: \bibinfo{pages}{109--166}.
ISSN \bibinfo{issn}{0146-6410}.
  \bibinfo{doi}{\doi{10.1016/j.ppnp.2019.02.008}}.

\bibtype{Article}%
\bibitem[Kasen et al.(2013)]{Kasen2013}
\bibinfo{author}{Kasen D}, \bibinfo{author}{Badnell NR} and
  \bibinfo{author}{Barnes J} (\bibinfo{year}{2013}), \bibinfo{month}{Aug.}
\bibinfo{title}{Opacities and spectra of ther-process ejecta from neutron star
  mergers}.
\bibinfo{journal}{{\em The Astrophysical Journal}} \bibinfo{volume}{774}
  (\bibinfo{number}{1}): \bibinfo{pages}{25}.
ISSN \bibinfo{issn}{1538-4357}.
  \bibinfo{doi}{\doi{10.1088/0004-637x/774/1/25}}.

\bibtype{Article}%
\bibitem[Kasen et al.(2017)]{Kasen2017}
\bibinfo{author}{Kasen D}, \bibinfo{author}{Metzger B}, \bibinfo{author}{Barnes
  J}, \bibinfo{author}{Quataert E} and  \bibinfo{author}{Ramirez-Ruiz E}
  (\bibinfo{year}{2017}).
\bibinfo{title}{Origin of the heavy elements in binary neutron-star mergers
  from a gravitational-wave event}.
\bibinfo{journal}{{\em Nature}} \bibinfo{volume}{551} (\bibinfo{number}{7678}):
  \bibinfo{pages}{80--84}. \bibinfo{doi}{\doi{10.1038/nature24453}}.
\eprint{1710.05463}.

\bibtype{Article}%
\bibitem[Kiuchi et al.(2014)]{Kiuchi2014}
\bibinfo{author}{Kiuchi K}, \bibinfo{author}{Kyutoku K},
  \bibinfo{author}{Sekiguchi Y}, \bibinfo{author}{Shibata M} and
  \bibinfo{author}{Wada T} (\bibinfo{year}{2014}), \bibinfo{month}{Aug.}
\bibinfo{title}{High resolution numerical relativity simulations for the merger
  of binary magnetized neutron stars}.
\bibinfo{journal}{{\em Physical Review D}} \bibinfo{volume}{90}
  (\bibinfo{number}{4}): \bibinfo{pages}{041502}.
ISSN \bibinfo{issn}{1550-2368}.
  \bibinfo{doi}{\doi{10.1103/physrevd.90.041502}}.

\bibtype{Article}%
\bibitem[Kiuchi et al.(2015)]{Kiuchi2015}
\bibinfo{author}{Kiuchi K}, \bibinfo{author}{Cerdá-Durán P},
  \bibinfo{author}{Kyutoku K}, \bibinfo{author}{Sekiguchi Y} and
  \bibinfo{author}{Shibata M} (\bibinfo{year}{2015}), \bibinfo{month}{Dec.}
\bibinfo{title}{Efficient magnetic-field amplification due to the
  kelvin-helmholtz instability in binary neutron star mergers}.
\bibinfo{journal}{{\em Physical Review D}} \bibinfo{volume}{92}
  (\bibinfo{number}{12}): \bibinfo{pages}{124034}.
ISSN \bibinfo{issn}{1550-2368}.
  \bibinfo{doi}{\doi{10.1103/physrevd.92.124034}}.

\bibtype{Article}%
\bibitem[Köppel et al.(2019)]{Koeppel2019}
\bibinfo{author}{Köppel S}, \bibinfo{author}{Bovard L} and
  \bibinfo{author}{Rezzolla L} (\bibinfo{year}{2019}), \bibinfo{month}{Feb.}
\bibinfo{title}{A general-relativistic determination of the threshold mass to
  prompt collapse in binary neutron star mergers}.
\bibinfo{journal}{{\em The Astrophysical Journal Letters}}
  \bibinfo{volume}{872} (\bibinfo{number}{1}): \bibinfo{pages}{L16}.
ISSN \bibinfo{issn}{2041-8213}. \bibinfo{doi}{\doi{10.3847/2041-8213/ab0210}}.

\bibtype{Article}%
\bibitem[{Lattimer}(2023)]{Lattimer2023}
\bibinfo{author}{{Lattimer} JM} (\bibinfo{year}{2023}), \bibinfo{month}{Jan.}
\bibinfo{title}{{Constraints on Nuclear Symmetry Energy Parameters}}.
\bibinfo{journal}{{\em Particles}} \bibinfo{volume}{6} (\bibinfo{number}{1}):
  \bibinfo{pages}{30--56}. \bibinfo{doi}{\doi{10.3390/particles6010003}}.
\eprint{2301.03666}.

\bibtype{Article}%
\bibitem[{Lattimer} and {Prakash}(2001)]{Lattimer2001}
\bibinfo{author}{{Lattimer} JM} and  \bibinfo{author}{{Prakash} M}
  (\bibinfo{year}{2001}), \bibinfo{month}{Mar.}
\bibinfo{title}{{Neutron Star Structure and the Equation of State}}.
\bibinfo{journal}{{\em \apj}} \bibinfo{volume}{550} (\bibinfo{number}{1}):
  \bibinfo{pages}{426--442}. \bibinfo{doi}{\doi{10.1086/319702}}.
\eprint{astro-ph/0002232}.

\bibtype{Article}%
\bibitem[Lattimer and Prakash(2005)]{Lattimer2005}
\bibinfo{author}{Lattimer JM} and  \bibinfo{author}{Prakash M}
  (\bibinfo{year}{2005}), \bibinfo{month}{Mar.}
\bibinfo{title}{Ultimate energy density of observable cold baryonic matter}.
\bibinfo{journal}{{\em Physical Review Letters}} \bibinfo{volume}{94}
  (\bibinfo{number}{11}): \bibinfo{pages}{111101}.
ISSN \bibinfo{issn}{1079-7114}.
  \bibinfo{doi}{\doi{10.1103/physrevlett.94.111101}}.

\bibtype{Article}%
\bibitem[Lee(2025)]{Lee2025}
\bibinfo{author}{Lee D} (\bibinfo{year}{2025}).
\bibinfo{title}{Lattice effective field theory simulations of nuclei}.
\bibinfo{journal}{{\em Annual Review of Nuclear and Particle Science}}
  \bibinfo{volume}{75} (\bibinfo{number}{1}): \bibinfo{pages}{109--128}.
ISSN \bibinfo{issn}{1545-4134}.
  \bibinfo{doi}{\doi{10.1146/annurev-nucl-101918-023343}}.

\bibtype{Article}%
\bibitem[{Li} and {Sedrakian}(2023)]{Li2023}
\bibinfo{author}{{Li} JJ} and  \bibinfo{author}{{Sedrakian} A}
  (\bibinfo{year}{2023}), \bibinfo{month}{Nov.}
\bibinfo{title}{{New Covariant Density Functionals of Nuclear Matter for
  Compact Star Simulations}}.
\bibinfo{journal}{{\em \apj}} \bibinfo{volume}{957} (\bibinfo{number}{1}),
  \bibinfo{eid}{41}. \bibinfo{doi}{\doi{10.3847/1538-4357/acfa73}}.
\eprint{2308.14457}.

\bibtype{Article}%
\bibitem[{Li} et al.(2018)]{Li2018}
\bibinfo{author}{{Li} JJ}, \bibinfo{author}{{Sedrakian} A} and
  \bibinfo{author}{{Weber} F} (\bibinfo{year}{2018}), \bibinfo{month}{Aug.}
\bibinfo{title}{{Competition between delta isobars and hyperons and properties
  of compact stars}}.
\bibinfo{journal}{{\em Physics Letters B}} \bibinfo{volume}{783}:
  \bibinfo{pages}{234--240}.
  \bibinfo{doi}{\doi{10.1016/j.physletb.2018.06.051}}.
\eprint{1803.03661}.

\bibtype{Article}%
\bibitem[Li et al.(2025)]{Li2025}
\bibinfo{author}{Li JJ}, \bibinfo{author}{Sedrakian A} and
  \bibinfo{author}{Alford M} (\bibinfo{year}{2025}), \bibinfo{month}{Feb.}
\bibinfo{title}{Confronting new nicer mass-radius measurements with phase
  transition in dense matter and twin compact stars}.
\bibinfo{journal}{{\em Journal of Cosmology and Astroparticle Physics}}
  \bibinfo{volume}{2025} (\bibinfo{number}{02}): \bibinfo{pages}{002}.
ISSN \bibinfo{issn}{1475-7516}.
  \bibinfo{doi}{\doi{10.1088/1475-7516/2025/02/002}}.

\bibtype{Article}%
\bibitem[Lippuner and Roberts(2015)]{Lippuner2015}
\bibinfo{author}{Lippuner J} and  \bibinfo{author}{Roberts LF}
  (\bibinfo{year}{2015}), \bibinfo{month}{Dec.}
\bibinfo{title}{r-process lanthanide production and heating rates in
  kilonovae}.
\bibinfo{journal}{{\em The Astrophysical Journal}} \bibinfo{volume}{815}
  (\bibinfo{number}{2}): \bibinfo{pages}{82}.
ISSN \bibinfo{issn}{1538-4357}.
  \bibinfo{doi}{\doi{10.1088/0004-637x/815/2/82}}.

\bibtype{Article}%
\bibitem[Lynn et al.(2019)]{Lynn2019}
\bibinfo{author}{Lynn J}, \bibinfo{author}{Tews I}, \bibinfo{author}{Gandolfi
  S} and  \bibinfo{author}{Lovato A} (\bibinfo{year}{2019}),
  \bibinfo{month}{Oct.}
\bibinfo{title}{Quantum monte carlo methods in nuclear physics: Recent
  advances}.
\bibinfo{journal}{{\em Annual Review of Nuclear and Particle Science}}
  \bibinfo{volume}{69} (\bibinfo{number}{1}): \bibinfo{pages}{279--305}.
ISSN \bibinfo{issn}{1545-4134}.
  \bibinfo{doi}{\doi{10.1146/annurev-nucl-101918-023600}}.

\bibtype{Article}%
\bibitem[Machleidt and Entem(2011)]{Machleidt2011}
\bibinfo{author}{Machleidt R} and  \bibinfo{author}{Entem D}
  (\bibinfo{year}{2011}).
\bibinfo{title}{Chiral effective field theory and nuclear forces}.
\bibinfo{journal}{{\em Physics Reports}} \bibinfo{volume}{503}
  (\bibinfo{number}{1}): \bibinfo{pages}{1--75}.
ISSN \bibinfo{issn}{0370-1573}.
  \bibinfo{doi}{\doi{10.1016/j.physrep.2011.02.001}}.

\bibtype{Article}%
\bibitem[{Margueron} et al.(2018)]{Margueron2018}
\bibinfo{author}{{Margueron} J}, \bibinfo{author}{{Hoffmann Casali} R} and
  \bibinfo{author}{{Gulminelli} F} (\bibinfo{year}{2018}),
  \bibinfo{month}{Feb.}
\bibinfo{title}{{Equation of state for dense nucleonic matter from
  metamodeling. I. Foundational aspects}}.
\bibinfo{journal}{{\em \prc}} \bibinfo{volume}{97} (\bibinfo{number}{2}),
  \bibinfo{eid}{025805}. \bibinfo{doi}{\doi{10.1103/PhysRevC.97.025805}}.
\eprint{1708.06894}.

\bibtype{Article}%
\bibitem[Martin et al.(2015)]{Martin2015}
\bibinfo{author}{Martin D}, \bibinfo{author}{Perego A},
  \bibinfo{author}{Arcones A}, \bibinfo{author}{Thielemann FK},
  \bibinfo{author}{Korobkin O} and  \bibinfo{author}{Rosswog S}
  (\bibinfo{year}{2015}), \bibinfo{month}{Oct.}
\bibinfo{title}{Neutrino-driven winds in the aftermath of a neutron star
  merger: Nucleosynthesis and electromagnetic transients}.
\bibinfo{journal}{{\em The Astrophysical Journal}} \bibinfo{volume}{813}
  (\bibinfo{number}{1}): \bibinfo{pages}{2}.
ISSN \bibinfo{issn}{1538-4357}. \bibinfo{doi}{\doi{10.1088/0004-637x/813/1/2}}.

\bibtype{Article}%
\bibitem[Metzger(2017)]{Metzger2017}
\bibinfo{author}{Metzger BD} (\bibinfo{year}{2017}).
\bibinfo{title}{Kilonovae}.
\bibinfo{journal}{{\em Living Reviews in Relativity}} \bibinfo{volume}{20}
  (\bibinfo{number}{1}): \bibinfo{pages}{3}.
  \bibinfo{doi}{\doi{10.1007/s41114-017-0006-z}}.
\eprint{1610.09381}.

\bibtype{Article}%
\bibitem[Metzger and Fernández(2014)]{Metzger2014}
\bibinfo{author}{Metzger BD} and  \bibinfo{author}{Fernández R}
  (\bibinfo{year}{2014}), \bibinfo{month}{May}.
\bibinfo{title}{Red or blue? a potential kilonova imprint of the delay until
  black hole formation following a neutron star merger}.
\bibinfo{journal}{{\em Monthly Notices of the Royal Astronomical Society}}
  \bibinfo{volume}{441} (\bibinfo{number}{4}): \bibinfo{pages}{3444--3453}.
ISSN \bibinfo{issn}{0035-8711}. \bibinfo{doi}{\doi{10.1093/mnras/stu802}}.

\bibtype{Article}%
\bibitem[Metzger et al.(2010)]{Metzger2010}
\bibinfo{author}{Metzger BD} and  et al. (\bibinfo{year}{2010}).
\bibinfo{title}{Electromagnetic counterparts of neutron star mergers}.
\bibinfo{journal}{{\em Monthly Notices of the Royal Astronomical Society}}
  \bibinfo{volume}{406}: \bibinfo{pages}{2650}.
  \bibinfo{doi}{\doi{10.1111/j.1365-2966.2010.16864.x}}.
\eprint{1001.5029}.

\bibtype{Article}%
\bibitem[{Miller} et al.(2019)]{Miller2019}
\bibinfo{author}{{Miller} MC}, \bibinfo{author}{{Lamb} FK},
  \bibinfo{author}{{Dittmann} AJ}, \bibinfo{author}{{Bogdanov} S},
  \bibinfo{author}{{Arzoumanian} Z}, \bibinfo{author}{{Gendreau} KC} and  et
  al. (\bibinfo{year}{2019}), \bibinfo{month}{Dec.}
\bibinfo{title}{{PSR J0030+0451 Mass and Radius from NICER Data and
  Implications for the Properties of Neutron Star Matter}}.
\bibinfo{journal}{{\em \apjl}} \bibinfo{volume}{887} (\bibinfo{number}{1}),
  \bibinfo{eid}{L24}. \bibinfo{doi}{\doi{10.3847/2041-8213/ab50c5}}.
\eprint{1912.05705}.

\bibtype{Article}%
\bibitem[{Miller} et al.(2021)]{Miller2021}
\bibinfo{author}{{Miller} MC}, \bibinfo{author}{{Lamb} FK},
  \bibinfo{author}{{Dittmann} AJ}, \bibinfo{author}{{Bogdanov} S},
  \bibinfo{author}{{Arzoumanian} Z}, \bibinfo{author}{{Gendreau} KC} and  et
  al. (\bibinfo{year}{2021}), \bibinfo{month}{Sep.}
\bibinfo{title}{{The Radius of PSR J0740+6620 from NICER and XMM-Newton Data}}.
\bibinfo{journal}{{\em \apjl}} \bibinfo{volume}{918} (\bibinfo{number}{2}),
  \bibinfo{eid}{L28}. \bibinfo{doi}{\doi{10.3847/2041-8213/ac089b}}.
\eprint{2105.06979}.

\bibtype{Article}%
\bibitem[Most et al.(2019)]{Most2019}
\bibinfo{author}{Most ER}, \bibinfo{author}{Papenfort LJ},
  \bibinfo{author}{Dexheimer V}, \bibinfo{author}{Hanauske M},
  \bibinfo{author}{Schramm S}, \bibinfo{author}{Stöcker H} and  et al.
  (\bibinfo{year}{2019}), \bibinfo{month}{Feb.}
\bibinfo{title}{Signatures of quark-hadron phase transitions in
  general-relativistic neutron-star mergers}.
\bibinfo{journal}{{\em Physical Review Letters}} \bibinfo{volume}{122}
  (\bibinfo{number}{6}): \bibinfo{pages}{061101}.
ISSN \bibinfo{issn}{1079-7114}.
  \bibinfo{doi}{\doi{10.1103/physrevlett.122.061101}}.

\bibtype{Article}%
\bibitem[Most et al.(2020)]{Most2020}
\bibinfo{author}{Most ER}, \bibinfo{author}{Jens~Papenfort L},
  \bibinfo{author}{Dexheimer V}, \bibinfo{author}{Hanauske M},
  \bibinfo{author}{Stoecker H} and  \bibinfo{author}{Rezzolla L}
  (\bibinfo{year}{2020}), \bibinfo{month}{Feb.}
\bibinfo{title}{On the deconfinement phase transition in neutron-star mergers}.
\bibinfo{journal}{{\em The European Physical Journal A}} \bibinfo{volume}{56}
  (\bibinfo{number}{2}).
ISSN \bibinfo{issn}{1434-601X}.
  \bibinfo{doi}{\doi{10.1140/epja/s10050-020-00073-4}}.

\bibtype{Article}%
\bibitem[Most et al.(2024)]{Most2024}
\bibinfo{author}{Most ER}, \bibinfo{author}{Haber A}, \bibinfo{author}{Harris
  SP}, \bibinfo{author}{Zhang Z}, \bibinfo{author}{Alford MG} and
  \bibinfo{author}{Noronha J} (\bibinfo{year}{2024}), \bibinfo{month}{May}.
\bibinfo{title}{Emergence of microphysical bulk viscosity in binary neutron
  star postmerger dynamics}.
\bibinfo{journal}{{\em The Astrophysical Journal Letters}}
  \bibinfo{volume}{967} (\bibinfo{number}{1}): \bibinfo{pages}{L14}.
ISSN \bibinfo{issn}{2041-8213}. \bibinfo{doi}{\doi{10.3847/2041-8213/ad454f}}.

\bibtype{Article}%
\bibitem[Mumpower et al.(2016)]{Mumpower2016}
\bibinfo{author}{Mumpower M}, \bibinfo{author}{Surman R},
  \bibinfo{author}{McLaughlin G} and  \bibinfo{author}{Aprahamian A}
  (\bibinfo{year}{2016}), \bibinfo{month}{Jan.}
\bibinfo{title}{The impact of individual nuclear properties on r-process
  nucleosynthesis}.
\bibinfo{journal}{{\em Progress in Particle and Nuclear Physics}}
  \bibinfo{volume}{86}: \bibinfo{pages}{86--126}.
ISSN \bibinfo{issn}{0146-6410}.
  \bibinfo{doi}{\doi{10.1016/j.ppnp.2015.09.001}}.

\bibtype{Article}%
\bibitem[Müther et al.(2017)]{Muether2017}
\bibinfo{author}{Müther H}, \bibinfo{author}{Sammarruca F} and
  \bibinfo{author}{Ma Z} (\bibinfo{year}{2017}), \bibinfo{month}{Mar.}
\bibinfo{title}{Relativistic effects and three-nucleon forces in nuclear matter
  and nuclei}.
\bibinfo{journal}{{\em International Journal of Modern Physics E}}
  \bibinfo{volume}{26} (\bibinfo{number}{03}): \bibinfo{pages}{1730001}.
ISSN \bibinfo{issn}{1793-6608}. \bibinfo{doi}{\doi{10.1142/s0218301317300016}}.

\bibtype{Article}%
\bibitem[Ng et al.(2025)]{Ng2025}
\bibinfo{author}{Ng HHY}, \bibinfo{author}{Musolino C}, \bibinfo{author}{Tootle
  SD} and  \bibinfo{author}{Rezzolla L} (\bibinfo{year}{2025}),
  \bibinfo{month}{May}.
\bibinfo{title}{Accurate muonic interactions in neutron star mergers and impact
  on heavy-element nucleosynthesis}.
\bibinfo{journal}{{\em The Astrophysical Journal Letters}}
  \bibinfo{volume}{985} (\bibinfo{number}{2}): \bibinfo{pages}{L36}.
ISSN \bibinfo{issn}{2041-8213}. \bibinfo{doi}{\doi{10.3847/2041-8213/add324}}.

\bibtype{Article}%
\bibitem[Oechslin et al.(2007)]{Oechslin2007}
\bibinfo{author}{Oechslin R}, \bibinfo{author}{Janka HT} and
  \bibinfo{author}{Marek A} (\bibinfo{year}{2007}), \bibinfo{month}{Mar.}
\bibinfo{title}{Relativistic neutron star merger simulations with non-zero
  temperature equations of state: I. variation of binary parameters and
  equation of state}.
\bibinfo{journal}{{\em A\&A}} \bibinfo{volume}{467} (\bibinfo{number}{2}):
  \bibinfo{pages}{395--409}.
ISSN \bibinfo{issn}{1432-0746}.
  \bibinfo{doi}{\doi{10.1051/0004-6361:20066682}}.

\bibtype{Article}%
\bibitem[{Oertel} et al.(2017)]{Oertel2017}
\bibinfo{author}{{Oertel} M}, \bibinfo{author}{{Hempel} M},
  \bibinfo{author}{{Kl{\"a}hn} T} and  \bibinfo{author}{{Typel} S}
  (\bibinfo{year}{2017}), \bibinfo{month}{Jan.}
\bibinfo{title}{{Equations of state for supernovae and compact stars}}.
\bibinfo{journal}{{\em Reviews of Modern Physics}} \bibinfo{volume}{89}
  (\bibinfo{number}{1}), \bibinfo{eid}{015007}.
  \bibinfo{doi}{\doi{10.1103/RevModPhys.89.015007}}.
\eprint{1610.03361}.

\bibtype{Article}%
\bibitem[{Oppenheimer} and {Volkoff}(1939)]{Oppenheimer1939}
\bibinfo{author}{{Oppenheimer} JR} and  \bibinfo{author}{{Volkoff} GM}
  (\bibinfo{year}{1939}), \bibinfo{month}{Feb.}
\bibinfo{title}{{On Massive Neutron Cores}}.
\bibinfo{journal}{{\em Physical Review}} \bibinfo{volume}{55}
  (\bibinfo{number}{4}): \bibinfo{pages}{374--381}.
  \bibinfo{doi}{\doi{10.1103/PhysRev.55.374}}.

\bibtype{Article}%
\bibitem[{{\"O}zel} and {Freire}(2016)]{Oezel2016}
\bibinfo{author}{{{\"O}zel} F} and  \bibinfo{author}{{Freire} P}
  (\bibinfo{year}{2016}), \bibinfo{month}{Sep.}
\bibinfo{title}{{Masses, Radii, and the Equation of State of Neutron Stars}}.
\bibinfo{journal}{{\em \araa}} \bibinfo{volume}{54} (\bibinfo{number}{1}):
  \bibinfo{pages}{401--440}.
ISSN \bibinfo{issn}{1545-4282}.
  \bibinfo{doi}{\doi{10.1146/annurev-astro-081915-023322}}.
\eprint{1603.02698}.

\bibtype{Article}%
\bibitem[Page and Reddy(2006)]{Page2006}
\bibinfo{author}{Page D} and  \bibinfo{author}{Reddy S} (\bibinfo{year}{2006}),
  \bibinfo{month}{Nov.}
\bibinfo{title}{Dense matter in compact stars: Theoretical developments and
  observational constraints}.
\bibinfo{journal}{{\em Annual Review of Nuclear and Particle Science}}
  \bibinfo{volume}{56} (\bibinfo{number}{1}): \bibinfo{pages}{327--374}.
ISSN \bibinfo{issn}{1545-4134}.
  \bibinfo{doi}{\doi{10.1146/annurev.nucl.56.080805.140600}}.

\bibtype{Article}%
\bibitem[Pani et al.(2015)]{Pani2015}
\bibinfo{author}{Pani P}, \bibinfo{author}{Gualtieri L},
  \bibinfo{author}{Maselli A} and  \bibinfo{author}{Ferrari V}
  (\bibinfo{year}{2015}).
\bibinfo{title}{Tidal deformations of a spinning compact object}.
\bibinfo{journal}{{\em Physical Review D}} \bibinfo{volume}{92}
  (\bibinfo{number}{2}): \bibinfo{pages}{024010}.
ISSN \bibinfo{issn}{1550-2368}.
  \bibinfo{doi}{\doi{10.1103/physrevd.92.024010}}.

\bibtype{Article}%
\bibitem[Pearson et al.(2018)]{Pearson2018}
\bibinfo{author}{Pearson JM}, \bibinfo{author}{Chamel N},
  \bibinfo{author}{Potekhin AY}, \bibinfo{author}{Fantina AF},
  \bibinfo{author}{Ducoin C}, \bibinfo{author}{Dutta AK} and  et al.
  (\bibinfo{year}{2018}).
\bibinfo{title}{Unified equations of state for cold non-accreting neutron stars
  with brussels-montreal functionals. i. role of symmetry energy}.
\bibinfo{journal}{{\em Monthly Notices of the Royal Astronomical Society}} ISSN
  \bibinfo{issn}{1365-2966}. \bibinfo{doi}{\doi{10.1093/mnras/sty2413}}.

\bibtype{Article}%
\bibitem[Perego et al.(2014)]{Perego2014}
\bibinfo{author}{Perego A}, \bibinfo{author}{Rosswog S},
  \bibinfo{author}{Cabezon RM}, \bibinfo{author}{Korobkin O},
  \bibinfo{author}{Kappeli R}, \bibinfo{author}{Arcones A} and  et al.
  (\bibinfo{year}{2014}), \bibinfo{month}{Aug.}
\bibinfo{title}{Neutrino-driven winds from neutron star merger remnants}.
\bibinfo{journal}{{\em Monthly Notices of the Royal Astronomical Society}}
  \bibinfo{volume}{443} (\bibinfo{number}{4}): \bibinfo{pages}{3134--3156}.
ISSN \bibinfo{issn}{1365-2966}. \bibinfo{doi}{\doi{10.1093/mnras/stu1352}}.

\bibtype{Article}%
\bibitem[Peters(1964)]{Peters1964}
\bibinfo{author}{Peters PC} (\bibinfo{year}{1964}), \bibinfo{month}{Nov.}
\bibinfo{title}{Gravitational radiation and the motion of two point masses}.
\bibinfo{journal}{{\em Physical Review}} \bibinfo{volume}{136}
  (\bibinfo{number}{4B}): \bibinfo{pages}{B1224--B1232}.
ISSN \bibinfo{issn}{0031-899X}. \bibinfo{doi}{\doi{10.1103/physrev.136.b1224}}.

\bibtype{Article}%
\bibitem[Petrosyan et al.(2026)]{Petrosyan2026}
\bibinfo{author}{Petrosyan T}, \bibinfo{author}{Harutyunyan A} and
  \bibinfo{author}{Sedrakian A} (\bibinfo{year}{2026}), \bibinfo{month}{Mar.}
\bibinfo{title}{Impact of positrons on electrical conductivity of hot and dense
  astrophysical plasma}.
\bibinfo{journal}{{\em Physical Review D}} \bibinfo{volume}{113}
  (\bibinfo{number}{6}).
ISSN \bibinfo{issn}{2470-0029}. \bibinfo{doi}{\doi{10.1103/vmjv-9822}}.

\bibtype{Article}%
\bibitem[Pian et al.(2017)]{Pian2017}
\bibinfo{author}{Pian E}, \bibinfo{author}{D’Avanzo P},
  \bibinfo{author}{Benetti S}, \bibinfo{author}{Branchesi M},
  \bibinfo{author}{Brocato E}, \bibinfo{author}{Campana S} and  et al.
  (\bibinfo{year}{2017}), \bibinfo{month}{Oct.}
\bibinfo{title}{Spectroscopic identification of r-process nucleosynthesis in a
  double neutron-star merger}.
\bibinfo{journal}{{\em Nature}} \bibinfo{volume}{551} (\bibinfo{number}{7678}):
  \bibinfo{pages}{67--70}.
ISSN \bibinfo{issn}{1476-4687}. \bibinfo{doi}{\doi{10.1038/nature24298}}.

\bibtype{Article}%
\bibitem[Price and Rosswog(2006)]{Price2006}
\bibinfo{author}{Price DJ} and  \bibinfo{author}{Rosswog S}
  (\bibinfo{year}{2006}), \bibinfo{month}{May}.
\bibinfo{title}{Producing ultrastrong magnetic fields in neutron star mergers}.
\bibinfo{journal}{{\em Science}} \bibinfo{volume}{312}
  (\bibinfo{number}{5774}): \bibinfo{pages}{719--722}.
ISSN \bibinfo{issn}{1095-9203}. \bibinfo{doi}{\doi{10.1126/science.1125201}}.

\bibtype{Article}%
\bibitem[Qian and Woosley(1996)]{Qian1996}
\bibinfo{author}{Qian Y} and  \bibinfo{author}{Woosley SE}
  (\bibinfo{year}{1996}), \bibinfo{month}{Nov.}
\bibinfo{title}{Nucleosynthesis in neutrino‐driven winds. i. the physical
  conditions}.
\bibinfo{journal}{{\em The Astrophysical Journal}} \bibinfo{volume}{471}
  (\bibinfo{number}{1}): \bibinfo{pages}{331--351}.
ISSN \bibinfo{issn}{1538-4357}. \bibinfo{doi}{\doi{10.1086/177973}}.

\bibtype{Article}%
\bibitem[Radice(2017)]{Radice2017}
\bibinfo{author}{Radice D} (\bibinfo{year}{2017}), \bibinfo{month}{Mar.}
\bibinfo{title}{General-relativistic large-eddy simulations of binary neutron
  star mergers}.
\bibinfo{journal}{{\em The Astrophysical Journal Letters}}
  \bibinfo{volume}{838} (\bibinfo{number}{1}): \bibinfo{pages}{L2}.
ISSN \bibinfo{issn}{2041-8213}. \bibinfo{doi}{\doi{10.3847/2041-8213/aa6483}}.

\bibtype{Article}%
\bibitem[Radice and Hawke(2024)]{Radice2024}
\bibinfo{author}{Radice D} and  \bibinfo{author}{Hawke I}
  (\bibinfo{year}{2024}), \bibinfo{month}{Feb.}
\bibinfo{title}{Turbulence modelling in neutron star merger simulations}.
\bibinfo{journal}{{\em Living Reviews in Computational Astrophysics}}
  \bibinfo{volume}{10} (\bibinfo{number}{1}).
ISSN \bibinfo{issn}{2365-0524}.
  \bibinfo{doi}{\doi{10.1007/s41115-023-00019-9}}.

\bibtype{Article}%
\bibitem[Radice et al.(2018{\natexlab{a}})]{Radice2018}
\bibinfo{author}{Radice D}, \bibinfo{author}{Perego A},
  \bibinfo{author}{Bernuzzi S} and  \bibinfo{author}{Zhang B}
  (\bibinfo{year}{2018}{\natexlab{a}}).
\bibinfo{title}{Long-lived remnants from binary neutron star mergers}.
\bibinfo{journal}{{\em Monthly Notices of the Royal Astronomical Society}}
  \bibinfo{volume}{481} (\bibinfo{number}{3}): \bibinfo{pages}{3670--3682}.
ISSN \bibinfo{issn}{1365-2966}. \bibinfo{doi}{\doi{10.1093/mnras/sty2531}}.

\bibtype{Article}%
\bibitem[Radice et al.(2018{\natexlab{b}})]{Radice2018a}
\bibinfo{author}{Radice D}, \bibinfo{author}{Perego A},
  \bibinfo{author}{Hotokezaka K}, \bibinfo{author}{Bernuzzi S},
  \bibinfo{author}{Fromm SA} and  \bibinfo{author}{Roberts LF}
  (\bibinfo{year}{2018}{\natexlab{b}}), \bibinfo{month}{Dec.}
\bibinfo{title}{Viscous-dynamical ejecta from binary neutron star mergers}.
\bibinfo{journal}{{\em The Astrophysical Journal Letters}}
  \bibinfo{volume}{869} (\bibinfo{number}{2}): \bibinfo{pages}{L35}.
ISSN \bibinfo{issn}{2041-8213}. \bibinfo{doi}{\doi{10.3847/2041-8213/aaf053}}.

\bibtype{Article}%
\bibitem[Radice et al.(2020)]{Radice2020}
\bibinfo{author}{Radice D} and  et al. (\bibinfo{year}{2020}).
\bibinfo{title}{Binary neutron star mergers: mass ejection, electromagnetic
  counterparts}.
\bibinfo{journal}{{\em Annual Review of Nuclear and Particle Science}}
  \bibinfo{volume}{70}: \bibinfo{pages}{95}.
  \bibinfo{doi}{\doi{10.1146/annurev-nucl-013120-114541}}.
\eprint{2002.03863}.

\bibtype{Article}%
\bibitem[Rau and Sedrakian(2023)]{Rau2023}
\bibinfo{author}{Rau PB} and  \bibinfo{author}{Sedrakian A}
  (\bibinfo{year}{2023}), \bibinfo{month}{May}.
\bibinfo{title}{Two first-order phase transitions in hybrid compact stars:
  Higher-order multiplet stars, reaction modes, and intermediate conversion
  speeds}.
\bibinfo{journal}{{\em Physical Review D}} \bibinfo{volume}{107}
  (\bibinfo{number}{10}): \bibinfo{pages}{103042}.
ISSN \bibinfo{issn}{2470-0029}.
  \bibinfo{doi}{\doi{10.1103/physrevd.107.103042}}.

\bibtype{Article}%
\bibitem[Read et al.(2009)]{Read2009}
\bibinfo{author}{Read JS}, \bibinfo{author}{Lackey BD}, \bibinfo{author}{Owen
  BJ} and  \bibinfo{author}{Friedman JL} (\bibinfo{year}{2009}).
\bibinfo{title}{Constraints on a phenomenologically parametrized neutron-star
  equation of state}.
\bibinfo{journal}{{\em Physical Review D}} \bibinfo{volume}{79}
  (\bibinfo{number}{12}): \bibinfo{pages}{124032}.
ISSN \bibinfo{issn}{1550-2368}.
  \bibinfo{doi}{\doi{10.1103/physrevd.79.124032}}.

\bibtype{Article}%
\bibitem[Reddy et al.(1999)]{Reddy1999}
\bibinfo{author}{Reddy S}, \bibinfo{author}{Prakash M},
  \bibinfo{author}{Lattimer JM} and  \bibinfo{author}{Pons JA}
  (\bibinfo{year}{1999}), \bibinfo{month}{May}.
\bibinfo{title}{Effects of strong and electromagnetic correlations on neutrino
  interactions in dense matter}.
\bibinfo{journal}{{\em Physical Review C}} \bibinfo{volume}{59}
  (\bibinfo{number}{5}): \bibinfo{pages}{2888--2918}.
ISSN \bibinfo{issn}{1089-490X}. \bibinfo{doi}{\doi{10.1103/physrevc.59.2888}}.

\bibtype{Article}%
\bibitem[Rezzolla and Takami(2016)]{Rezzolla2016}
\bibinfo{author}{Rezzolla L} and  \bibinfo{author}{Takami K}
  (\bibinfo{year}{2016}).
\bibinfo{title}{Gravitational-wave signal from binary neutron stars: A
  systematic analysis of the spectral properties}.
\bibinfo{journal}{{\em Physical Review D}} \bibinfo{volume}{93}
  (\bibinfo{number}{12}): \bibinfo{pages}{124051}.
ISSN \bibinfo{issn}{2470-0029}.
  \bibinfo{doi}{\doi{10.1103/physrevd.93.124051}}.

\bibtype{Book}%
\bibitem[Rezzolla and Zanotti(2013)]{RezzollaZanotti2013}
\bibinfo{author}{Rezzolla L} and  \bibinfo{author}{Zanotti O}
  (\bibinfo{year}{2013}).
\bibinfo{title}{Relativistic Hydrodynamics}, \bibinfo{publisher}{Oxford
  University Press}.

\bibtype{Article}%
\bibitem[Rezzolla et al.(2011)]{Rezzolla2011}
\bibinfo{author}{Rezzolla L}, \bibinfo{author}{Giacomazzo B},
  \bibinfo{author}{Baiotti L}, \bibinfo{author}{Granot J},
  \bibinfo{author}{Kouveliotou C} and  \bibinfo{author}{Aloy MA}
  (\bibinfo{year}{2011}), \bibinfo{month}{Apr.}
\bibinfo{title}{The missing link: Merging neutron stars naturally produce
  jet-like structures and can power short gamma-ray bursts}.
\bibinfo{journal}{{\em The Astrophysical Journal}} \bibinfo{volume}{732}
  (\bibinfo{number}{1}): \bibinfo{pages}{L6}.
ISSN \bibinfo{issn}{2041-8213}.
  \bibinfo{doi}{\doi{10.1088/2041-8205/732/1/l6}}.

\bibtype{inbook}%
\bibitem[Richers and Sen(2022)]{Richers2022}
\bibinfo{author}{Richers S} and  \bibinfo{author}{Sen M}
  (\bibinfo{year}{2022}).
\bibinfo{title}{Fast Flavor Transformations}, \bibinfo{publisher}{Springer
  Nature Singapore}.
\bibinfo{comment}{ISBN} \bibinfo{isbn}{9789811588181},  \bibinfo{pages}{1--17}.
\bibinfo{doi}{\doi{10.1007/978-981-15-8818-1}}.

\bibtype{Article}%
\bibitem[{Riley} et al.(2019)]{Riley2019}
\bibinfo{author}{{Riley} TE}, \bibinfo{author}{{Watts} AL},
  \bibinfo{author}{{Bogdanov} S}, \bibinfo{author}{{Ray} PS},
  \bibinfo{author}{{Ludlam} RM}, \bibinfo{author}{{Guillot} S} and  et al.
  (\bibinfo{year}{2019}), \bibinfo{month}{Dec.}
\bibinfo{title}{{A NICER View of PSR J0030+0451: Millisecond Pulsar Parameter
  Estimation}}.
\bibinfo{journal}{{\em \apjl}} \bibinfo{volume}{887} (\bibinfo{number}{1}),
  \bibinfo{eid}{L21}. \bibinfo{doi}{\doi{10.3847/2041-8213/ab481c}}.
\eprint{1912.05702}.

\bibtype{Article}%
\bibitem[{Riley} et al.(2021)]{Riley2021}
\bibinfo{author}{{Riley} TE}, \bibinfo{author}{{Watts} AL},
  \bibinfo{author}{{Ray} PS}, \bibinfo{author}{{Bogdanov} S},
  \bibinfo{author}{{Guillot} S}, \bibinfo{author}{{Morsink} SM} and  et al.
  (\bibinfo{year}{2021}), \bibinfo{month}{Sep.}
\bibinfo{title}{{A NICER View of the Massive Pulsar PSR J0740+6620 Informed by
  Radio Timing and XMM-Newton Spectroscopy}}.
\bibinfo{journal}{{\em \apjl}} \bibinfo{volume}{918} (\bibinfo{number}{2}),
  \bibinfo{eid}{L27}. \bibinfo{doi}{\doi{10.3847/2041-8213/ac0a81}}.
\eprint{2105.06980}.

\bibtype{Article}%
\bibitem[Rios et al.(2009)]{Rios2009}
\bibinfo{author}{Rios A}, \bibinfo{author}{Polls A} and
  \bibinfo{author}{Vidaña I} (\bibinfo{year}{2009}), \bibinfo{month}{Feb.}
\bibinfo{title}{Hot neutron matter from a self-consistent green’s-functions
  approach}.
\bibinfo{journal}{{\em Physical Review C}} \bibinfo{volume}{79}
  (\bibinfo{number}{2}): \bibinfo{pages}{025802}.
ISSN \bibinfo{issn}{1089-490X}.
  \bibinfo{doi}{\doi{10.1103/physrevc.79.025802}}.

\bibtype{Article}%
\bibitem[Ripley et al.(2023)]{Ripley2023}
\bibinfo{author}{Ripley JL}, \bibinfo{author}{R. AHK} and
  \bibinfo{author}{Yunes N} (\bibinfo{year}{2023}), \bibinfo{month}{Nov.}
\bibinfo{title}{Probing internal dissipative processes of neutron stars with
  gravitational waves during the inspiral of neutron star binaries}.
\bibinfo{journal}{{\em Phys. Rev. D 108 (2023) 10, 103037}}
  \bibinfo{volume}{108} (\bibinfo{number}{10}): \bibinfo{pages}{103037}.
ISSN \bibinfo{issn}{2470-0029}.
  \bibinfo{doi}{\doi{10.1103/physrevd.108.103037}}.
\eprint{2306.15633}.

\bibtype{Article}%
\bibitem[Roederer et al.(2014)]{Roederer2014}
\bibinfo{author}{Roederer IU}, \bibinfo{author}{Preston GW},
  \bibinfo{author}{Thompson IB}, \bibinfo{author}{Shectman SA},
  \bibinfo{author}{Sneden C}, \bibinfo{author}{Burley GS} and  et al.
  (\bibinfo{year}{2014}), \bibinfo{month}{May}.
\bibinfo{title}{A search for stars of very low metal abundance. vi. detailed
  abundances of 313 metal-poor stars}.
\bibinfo{journal}{{\em The Astronomical Journal}} \bibinfo{volume}{147}
  (\bibinfo{number}{6}): \bibinfo{pages}{136}.
ISSN \bibinfo{issn}{1538-3881}.
  \bibinfo{doi}{\doi{10.1088/0004-6256/147/6/136}}.

\bibtype{Article}%
\bibitem[{Rosswog}(2015)]{Rosswog2015}
\bibinfo{author}{{Rosswog} S} (\bibinfo{year}{2015}), \bibinfo{month}{Feb.}
\bibinfo{title}{{The multi-messenger picture of compact binary mergers}}.
\bibinfo{journal}{{\em International Journal of Modern Physics D}}
  \bibinfo{volume}{24} (\bibinfo{number}{5}), \bibinfo{eid}{1530012-52}.
  \bibinfo{doi}{\doi{10.1142/S0218271815300128}}.
\eprint{1501.02081}.

\bibtype{Article}%
\bibitem[Ruiz et al.(2016)]{Ruiz2016}
\bibinfo{author}{Ruiz M}, \bibinfo{author}{Lang RN},
  \bibinfo{author}{Paschalidis V} and  \bibinfo{author}{Shapiro SL}
  (\bibinfo{year}{2016}).
\bibinfo{title}{Binary neutron star mergers: A jet engine for short gamma-ray
  bursts}.
\bibinfo{journal}{{\em The Astrophysical Journal Letters}}
  \bibinfo{volume}{824} (\bibinfo{number}{1}): \bibinfo{pages}{L6}.
ISSN \bibinfo{issn}{2041-8213}.
  \bibinfo{doi}{\doi{10.3847/2041-8205/824/1/l6}}.

\bibtype{Article}%
\bibitem[Ruiz et al.(2018)]{Ruiz2018}
\bibinfo{author}{Ruiz M}, \bibinfo{author}{Shapiro SL} and
  \bibinfo{author}{Tsokaros A} (\bibinfo{year}{2018}), \bibinfo{month}{Dec.}
\bibinfo{title}{Jet launching from binary black hole-neutron star mergers:
  Dependence on black hole spin, binary mass ratio, and magnetic field
  orientation}.
\bibinfo{journal}{{\em Physical Review D}} \bibinfo{volume}{98}
  (\bibinfo{number}{12}): \bibinfo{pages}{123017}.
ISSN \bibinfo{issn}{2470-0029}.
  \bibinfo{doi}{\doi{10.1103/physrevd.98.123017}}.

\bibtype{Article}%
\bibitem[Salmi et al.(2024)]{Salmi2024}
\bibinfo{author}{Salmi T}, \bibinfo{author}{Choudhury D}, \bibinfo{author}{Kini
  Y}, \bibinfo{author}{Riley TE}, \bibinfo{author}{Vinciguerra S},
  \bibinfo{author}{Watts AL} and  et al. (\bibinfo{year}{2024}),
  \bibinfo{month}{Oct.}
\bibinfo{title}{The radius of the high-mass pulsar psr j0740+6620 with 3.6 yr
  of nicer data}.
\bibinfo{journal}{{\em The Astrophysical Journal}} \bibinfo{volume}{974}
  (\bibinfo{number}{2}): \bibinfo{pages}{294}.
ISSN \bibinfo{issn}{1538-4357}. \bibinfo{doi}{\doi{10.3847/1538-4357/ad5f1f}}.

\bibtype{inbook}%
\bibitem[Schmitt and Shternin(2018)]{Schmitt2018}
\bibinfo{author}{Schmitt A} and  \bibinfo{author}{Shternin P}
  (\bibinfo{year}{2018}).
\bibinfo{title}{Reaction Rates and Transport in Neutron Stars},
  \bibinfo{publisher}{Springer International Publishing}.
\bibinfo{comment}{ISBN} \bibinfo{isbn}{9783319976167},
  \bibinfo{pages}{455--574}.
\bibinfo{doi}{\doi{10.1007/978-3-319-97616-7_9}}.

\bibtype{Article}%
\bibitem[{Sedrakian}(2007)]{Sedrakian2007PrPNP}
\bibinfo{author}{{Sedrakian} A} (\bibinfo{year}{2007}), \bibinfo{month}{Jan.}
\bibinfo{title}{{The physics of dense hadronic matter and compact stars}}.
\bibinfo{journal}{{\em Progress in Particle and Nuclear Physics}}
  \bibinfo{volume}{58} (\bibinfo{number}{1}): \bibinfo{pages}{168--246}.
  \bibinfo{doi}{\doi{10.1016/j.ppnp.2006.02.002}}.
\eprint{nucl-th/0601086}.

\bibtype{Article}%
\bibitem[Sedrakian(2023)]{Sedrakian2023a}
\bibinfo{author}{Sedrakian A} (\bibinfo{year}{2023}).
\bibinfo{title}{Impact of multiple phase transitions in dense qcd on compact
  stars}.
\bibinfo{journal}{{\em Particles}} \bibinfo{volume}{6} (\bibinfo{number}{3}):
  \bibinfo{pages}{713--730}.
ISSN \bibinfo{issn}{2571-712X}. \bibinfo{doi}{\doi{10.3390/particles6030044}}.

\bibtype{Article}%
\bibitem[Sedrakian(2024)]{Sedrakian2024}
\bibinfo{author}{Sedrakian A} (\bibinfo{year}{2024}), \bibinfo{month}{Oct.}
\bibinfo{title}{Short-range correlations and urca process in neutron stars}.
\bibinfo{journal}{{\em Physical Review Letters}} \bibinfo{volume}{133}
  (\bibinfo{number}{17}): \bibinfo{pages}{171401}.
ISSN \bibinfo{issn}{1079-7114}.
  \bibinfo{doi}{\doi{10.1103/physrevlett.133.171401}}.

\bibtype{Article}%
\bibitem[{Sedrakian} and {Harutyunyan}(2021)]{Sedrakian2021}
\bibinfo{author}{{Sedrakian} A} and  \bibinfo{author}{{Harutyunyan} A}
  (\bibinfo{year}{2021}), \bibinfo{month}{Oct.}
\bibinfo{title}{{Equation of State and Composition of Proto-Neutron Stars and
  Merger Remnants with Hyperons}}.
\bibinfo{journal}{{\em Universe}} \bibinfo{volume}{7} (\bibinfo{number}{10}),
  \bibinfo{eid}{382}. \bibinfo{doi}{\doi{10.3390/universe7100382}}.
\eprint{2109.01919}.

\bibtype{Article}%
\bibitem[{Sedrakian} and {Harutyunyan}(2022)]{Sedrakian2022}
\bibinfo{author}{{Sedrakian} A} and  \bibinfo{author}{{Harutyunyan} A}
  (\bibinfo{year}{2022}), \bibinfo{month}{Jul.}
\bibinfo{title}{{Delta-resonances and hyperons in proto-neutron stars and
  merger remnants}}.
\bibinfo{journal}{{\em European Physical Journal A}} \bibinfo{volume}{58}
  (\bibinfo{number}{7}), \bibinfo{eid}{137}.
  \bibinfo{doi}{\doi{10.1140/epja/s10050-022-00792-w}}.
\eprint{2202.12083}.

\bibtype{Article}%
\bibitem[{Sedrakian} et al.(2020)]{Sedrakian2020}
\bibinfo{author}{{Sedrakian} A}, \bibinfo{author}{{Weber} F} and
  \bibinfo{author}{{Li} JJ} (\bibinfo{year}{2020}), \bibinfo{month}{Aug.}
\bibinfo{title}{{Confronting GW190814 with hyperonization in dense matter and
  hypernuclear compact stars}}.
\bibinfo{journal}{{\em \prd}} \bibinfo{volume}{102} (\bibinfo{number}{4}),
  \bibinfo{eid}{041301}. \bibinfo{doi}{\doi{10.1103/PhysRevD.102.041301}}.
\eprint{2007.09683}.

\bibtype{Article}%
\bibitem[{Sedrakian} et al.(2023)]{Sedrakian2023}
\bibinfo{author}{{Sedrakian} A}, \bibinfo{author}{{Li} JJ} and
  \bibinfo{author}{{Weber} F} (\bibinfo{year}{2023}), \bibinfo{month}{Jul.}
\bibinfo{title}{{Heavy baryons in compact stars}}.
\bibinfo{journal}{{\em Progress in Particle and Nuclear Physics}}
  \bibinfo{volume}{131}, \bibinfo{eid}{104041}.
  \bibinfo{doi}{\doi{10.1016/j.ppnp.2023.104041}}.
\eprint{2212.01086}.

\bibtype{Article}%
\bibitem[Sekiguchi et al.(2015)]{Sekiguchi2015}
\bibinfo{author}{Sekiguchi Y}, \bibinfo{author}{Kiuchi K},
  \bibinfo{author}{Kyutoku K} and  \bibinfo{author}{Shibata M}
  (\bibinfo{year}{2015}), \bibinfo{month}{Mar.}
\bibinfo{title}{Dynamical mass ejection from binary neutron star mergers:
  Radiation-hydrodynamics study in general relativity}.
\bibinfo{journal}{{\em Physical Review D}} \bibinfo{volume}{91}
  (\bibinfo{number}{6}): \bibinfo{pages}{064059}.
ISSN \bibinfo{issn}{1550-2368}.
  \bibinfo{doi}{\doi{10.1103/physrevd.91.064059}}.

\bibtype{Article}%
\bibitem[Siegel and Metzger(2017)]{Siegel2017}
\bibinfo{author}{Siegel DM} and  \bibinfo{author}{Metzger BD}
  (\bibinfo{year}{2017}), \bibinfo{month}{Dec.}
\bibinfo{title}{Three-dimensional general-relativistic magnetohydrodynamic
  simulations of remnant accretion disks from neutron star mergers: Outflows
  and r -process nucleosynthesis}.
\bibinfo{journal}{{\em Physical Review Letters}} \bibinfo{volume}{119}
  (\bibinfo{number}{23}): \bibinfo{pages}{231102}.
ISSN \bibinfo{issn}{1079-7114}.
  \bibinfo{doi}{\doi{10.1103/physrevlett.119.231102}}.

\bibtype{Article}%
\bibitem[Siegel et al.(2014)]{Siegel2014}
\bibinfo{author}{Siegel DM}, \bibinfo{author}{Ciolfi R} and
  \bibinfo{author}{Rezzolla L} (\bibinfo{year}{2014}), \bibinfo{month}{Mar.}
\bibinfo{title}{Magnetically driven winds from differentially rotating neutron
  stars and x-ray afterglows of short gamma-ray bursts}.
\bibinfo{journal}{{\em The Astrophysical Journal}} \bibinfo{volume}{785}
  (\bibinfo{number}{1}): \bibinfo{pages}{L6}.
ISSN \bibinfo{issn}{2041-8213}.
  \bibinfo{doi}{\doi{10.1088/2041-8205/785/1/l6}}.

\bibtype{Article}%
\bibitem[Takami et al.(2015)]{Takami2015}
\bibinfo{author}{Takami K}, \bibinfo{author}{Rezzolla L} and
  \bibinfo{author}{Baiotti L} (\bibinfo{year}{2015}), \bibinfo{month}{Mar.}
\bibinfo{title}{Spectral properties of the post-merger gravitational-wave
  signal from binary neutron stars}.
\bibinfo{journal}{{\em Physical Review D}} \bibinfo{volume}{91}
  (\bibinfo{number}{6}): \bibinfo{pages}{064001}.
ISSN \bibinfo{issn}{1550-2368}.
  \bibinfo{doi}{\doi{10.1103/physrevd.91.064001}}.

\bibtype{Article}%
\bibitem[Tanaka and Hotokezaka(2013)]{Tanaka2013}
\bibinfo{author}{Tanaka M} and  \bibinfo{author}{Hotokezaka K}
  (\bibinfo{year}{2013}).
\bibinfo{title}{Radiative transfer simulations of neutron star merger ejecta}.
\bibinfo{journal}{{\em The Astrophysical Journal}} \bibinfo{volume}{775}
  (\bibinfo{number}{2}): \bibinfo{pages}{113}.
ISSN \bibinfo{issn}{1538-4357}.
  \bibinfo{doi}{\doi{10.1088/0004-637x/775/2/113}}.

\bibtype{Article}%
\bibitem[Taylor and Weisberg(1982)]{Taylor1982}
\bibinfo{author}{Taylor JH} and  \bibinfo{author}{Weisberg JM}
  (\bibinfo{year}{1982}), \bibinfo{month}{Feb.}
\bibinfo{title}{A new test of general relativity - gravitational radiation and
  the binary pulsar psr 1913+16}.
\bibinfo{journal}{{\em The Astrophysical Journal}} \bibinfo{volume}{253}:
  \bibinfo{pages}{908}.
ISSN \bibinfo{issn}{1538-4357}. \bibinfo{doi}{\doi{10.1086/159690}}.

\bibtype{Article}%
\bibitem[Tews et al.(2018)]{Tews2018}
\bibinfo{author}{Tews I}, \bibinfo{author}{Margueron J} and
  \bibinfo{author}{Reddy S} (\bibinfo{year}{2018}).
\bibinfo{title}{Confronting gravitational-wave observations with nuclear
  physics constraints}.
\bibinfo{journal}{{\em Physical Review C}} \bibinfo{volume}{98}:
  \bibinfo{pages}{045804}. \bibinfo{doi}{\doi{10.1103/PhysRevC.98.045804}}.
\eprint{1804.02783}.

\bibtype{Article}%
\bibitem[{Tolman}(1939)]{Tolman1939}
\bibinfo{author}{{Tolman} RC} (\bibinfo{year}{1939}), \bibinfo{month}{Feb.}
\bibinfo{title}{{Static Solutions of Einstein's Field Equations for Spheres of
  Fluid}}.
\bibinfo{journal}{{\em Physical Review}} \bibinfo{volume}{55}
  (\bibinfo{number}{4}): \bibinfo{pages}{364--373}.
  \bibinfo{doi}{\doi{10.1103/PhysRev.55.364}}.

\bibtype{Article}%
\bibitem[Tolos and Fabbietti(2020)]{Tolos2020}
\bibinfo{author}{Tolos L} and  \bibinfo{author}{Fabbietti L}
  (\bibinfo{year}{2020}), \bibinfo{month}{May}.
\bibinfo{title}{Strangeness in nuclei and neutron stars}.
\bibinfo{journal}{{\em Progress in Particle and Nuclear Physics}}
  \bibinfo{volume}{112}: \bibinfo{pages}{103770}.
ISSN \bibinfo{issn}{0146-6410}.
  \bibinfo{doi}{\doi{10.1016/j.ppnp.2020.103770}}.

\bibtype{Article}%
\bibitem[Tootle et al.(2021)]{Tootle2021}
\bibinfo{author}{Tootle SD}, \bibinfo{author}{Papenfort LJ},
  \bibinfo{author}{Most ER} and  \bibinfo{author}{Rezzolla L}
  (\bibinfo{year}{2021}), \bibinfo{month}{Nov.}
\bibinfo{title}{Quasi-universal behavior of the threshold mass in unequal-mass,
  spinning binary neutron star mergers}.
\bibinfo{journal}{{\em The Astrophysical Journal Letters}}
  \bibinfo{volume}{922} (\bibinfo{number}{1}): \bibinfo{pages}{L19}.
ISSN \bibinfo{issn}{2041-8213}. \bibinfo{doi}{\doi{10.3847/2041-8213/ac350d}}.

\bibtype{Article}%
\bibitem[{Truran} et al.(2002)]{Truran2002}
\bibinfo{author}{{Truran} JW}, \bibinfo{author}{{Cowan} JJ},
  \bibinfo{author}{{Pilachowski} CA} and  \bibinfo{author}{{Sneden} C}
  (\bibinfo{year}{2002}), \bibinfo{month}{Dec.}
\bibinfo{title}{{Probing the Neutron-Capture Nucleosynthesis History of
  Galactic Matter}}.
\bibinfo{journal}{{\em \pasp}} \bibinfo{volume}{114} (\bibinfo{number}{802}):
  \bibinfo{pages}{1293--1308}. \bibinfo{doi}{\doi{10.1086/344585}}.
\eprint{astro-ph/0209308}.

\bibtype{Article}%
\bibitem[Tsiopelas et al.(2024)]{Tsiopelas2024}
\bibinfo{author}{Tsiopelas S}, \bibinfo{author}{Sedrakian A} and
  \bibinfo{author}{Oertel M} (\bibinfo{year}{2024}).
\bibinfo{title}{Finite-temperature equations of state of compact stars with
  hyperons: three-dimensional tables}.
\bibinfo{journal}{{\em The European Physical Journal A}} \bibinfo{volume}{60}
  (\bibinfo{number}{6}).
ISSN \bibinfo{issn}{1434-601X}.
  \bibinfo{doi}{\doi{10.1140/epja/s10050-024-01351-1}}.

\bibtype{Article}%
\bibitem[{Typel}(2018)]{Typel2018}
\bibinfo{author}{{Typel} S} (\bibinfo{year}{2018}), \bibinfo{month}{Feb.}
\bibinfo{title}{{Relativistic Mean-Field Models with Different Parametrizations
  of Density Dependent Couplings}}.
\bibinfo{journal}{{\em Particles}} \bibinfo{volume}{1} (\bibinfo{number}{1}):
  \bibinfo{pages}{2}. \bibinfo{doi}{\doi{10.3390/particles1010002}}.

\bibtype{Article}%
\bibitem[{Typel} et al.(2010)]{Typel2010}
\bibinfo{author}{{Typel} S}, \bibinfo{author}{{R{\"o}pke} G},
  \bibinfo{author}{{Kl{\"a}hn} T}, \bibinfo{author}{{Blaschke} D} and
  \bibinfo{author}{{Wolter} HH} (\bibinfo{year}{2010}), \bibinfo{month}{Jan.}
\bibinfo{title}{{Composition and thermodynamics of nuclear matter with light
  clusters}}.
\bibinfo{journal}{{\em \prc}} \bibinfo{volume}{81} (\bibinfo{number}{1}),
  \bibinfo{eid}{015803}. \bibinfo{doi}{\doi{10.1103/PhysRevC.81.015803}}.
\eprint{0908.2344}.

\bibtype{Article}%
\bibitem[{Vida{\~n}a}(2018)]{Vidana2018}
\bibinfo{author}{{Vida{\~n}a} I} (\bibinfo{year}{2018}), \bibinfo{month}{Sep.}
\bibinfo{title}{{Hyperons: the strange ingredients of the nuclear equation of
  state}}.
\bibinfo{journal}{{\em Proceedings of the Royal Society of London Series A}}
  \bibinfo{volume}{474} (\bibinfo{number}{2217}), \bibinfo{eid}{20180145}.
  \bibinfo{doi}{\doi{10.1098/rspa.2018.0145}}.
\eprint{1803.00504}.

\bibtype{Article}%
\bibitem[Vigna-Gómez et al.(2018)]{VignaGomez2018}
\bibinfo{author}{Vigna-Gómez A}, \bibinfo{author}{Neijssel CJ},
  \bibinfo{author}{Stevenson S}, \bibinfo{author}{Barrett JW},
  \bibinfo{author}{Belczynski K}, \bibinfo{author}{Justham S} and  et al.
  (\bibinfo{year}{2018}).
\bibinfo{title}{On the formation history of galactic double neutron stars}.
\bibinfo{journal}{{\em Monthly Notices of the Royal Astronomical Society}}
  \bibinfo{volume}{481} (\bibinfo{number}{3}): \bibinfo{pages}{4009--4029}.
ISSN \bibinfo{issn}{1365-2966}. \bibinfo{doi}{\doi{10.1093/mnras/sty2463}}.

\bibtype{Article}%
\bibitem[Villar et al.(2017)]{Villar2017}
\bibinfo{author}{Villar VA}, \bibinfo{author}{Guillochon J},
  \bibinfo{author}{Berger E}, \bibinfo{author}{Metzger BD},
  \bibinfo{author}{Cowperthwaite PS}, \bibinfo{author}{Nicholl M} and  et al.
  (\bibinfo{year}{2017}), \bibinfo{month}{Dec.}
\bibinfo{title}{The combined ultraviolet, optical, and near-infrared light
  curves of the kilonova associated with the binary neutron star merger
  gw170817: Unified data set, analytic models, and physical implications}.
\bibinfo{journal}{{\em The Astrophysical Journal Letters}}
  \bibinfo{volume}{851} (\bibinfo{number}{1}): \bibinfo{pages}{L21}.
ISSN \bibinfo{issn}{2041-8213}. \bibinfo{doi}{\doi{10.3847/2041-8213/aa9c84}}.

\bibtype{Article}%
\bibitem[Wanajo et al.(2014)]{Wanajo2014}
\bibinfo{author}{Wanajo S}, \bibinfo{author}{Sekiguchi Y},
  \bibinfo{author}{Nishimura N}, \bibinfo{author}{Kiuchi K},
  \bibinfo{author}{Kyutoku K} and  \bibinfo{author}{Shibata M}
  (\bibinfo{year}{2014}).
\bibinfo{title}{Production of all the r -process nuclides in the dynamical
  ejecta of neutron star mergers}.
\bibinfo{journal}{{\em The Astrophysical Journal}} \bibinfo{volume}{789}
  (\bibinfo{number}{2}): \bibinfo{pages}{L39}.
ISSN \bibinfo{issn}{2041-8213}.
  \bibinfo{doi}{\doi{10.1088/2041-8205/789/2/l39}}.

\bibtype{Article}%
\bibitem[Weber et al.(2007)]{Weber2007}
\bibinfo{author}{Weber F}, \bibinfo{author}{Negreiros R},
  \bibinfo{author}{Rosenfield P} and  \bibinfo{author}{Stejner M}
  (\bibinfo{year}{2007}).
\bibinfo{title}{Pulsars as astrophysical laboratories for nuclear and particle
  physics}.
\bibinfo{journal}{{\em Progress in Particle and Nuclear Physics}}
  \bibinfo{volume}{59} (\bibinfo{number}{1}): \bibinfo{pages}{94--113}.
ISSN \bibinfo{issn}{0146-6410}.
  \bibinfo{doi}{\doi{10.1016/j.ppnp.2006.12.008}}.

\bibtype{Article}%
\bibitem[Weih et al.(2020)]{Weih2020}
\bibinfo{author}{Weih LR}, \bibinfo{author}{Hanauske M} and
  \bibinfo{author}{Rezzolla L} (\bibinfo{year}{2020}), \bibinfo{month}{Apr.}
\bibinfo{title}{Postmerger gravitational-wave signatures of phase transitions
  in binary mergers}.
\bibinfo{journal}{{\em Physical Review Letters}} \bibinfo{volume}{124}
  (\bibinfo{number}{17}): \bibinfo{pages}{171103}.
ISSN \bibinfo{issn}{1079-7114}.
  \bibinfo{doi}{\doi{10.1103/physrevlett.124.171103}}.

\bibtype{Article}%
\bibitem[Wu and Tamborra(2017)]{Wu2017}
\bibinfo{author}{Wu MR} and  \bibinfo{author}{Tamborra I}
  (\bibinfo{year}{2017}), \bibinfo{month}{May}.
\bibinfo{title}{Fast neutrino conversions: Ubiquitous in compact binary merger
  remnants}.
\bibinfo{journal}{{\em Physical Review D}} \bibinfo{volume}{95}
  (\bibinfo{number}{10}): \bibinfo{pages}{103007}.
ISSN \bibinfo{issn}{2470-0029}.
  \bibinfo{doi}{\doi{10.1103/physrevd.95.103007}}.

\bibtype{Article}%
\bibitem[{Yagi} and {Yunes}(2013)]{Yagi2013}
\bibinfo{author}{{Yagi} K} and  \bibinfo{author}{{Yunes} N}
  (\bibinfo{year}{2013}), \bibinfo{month}{Jul.}
\bibinfo{title}{{I-Love-Q: Unexpected Universal Relations for Neutron Stars and
  Quark Stars}}.
\bibinfo{journal}{{\em Science}} \bibinfo{volume}{341}
  (\bibinfo{number}{6144}): \bibinfo{pages}{365--368}.
  \bibinfo{doi}{\doi{10.1126/science.1236462}}.
\eprint{1302.4499}.

\bibtype{Article}%
\bibitem[Zdunik and Haensel(2013)]{Zdunik2013}
\bibinfo{author}{Zdunik JL} and  \bibinfo{author}{Haensel P}
  (\bibinfo{year}{2013}).
\bibinfo{title}{{Maximum mass of neutron stars and strange neutron-star
  cores}}.
\bibinfo{journal}{{\em A\&A}} \bibinfo{volume}{551}: \bibinfo{pages}{A61}.
  \bibinfo{doi}{\doi{10.1051/0004-6361/201220697}}.
\eprint{1211.1231}.

\end{thebibliography*}

\end{document}